\PassOptionsToPackage{dvipsnames,table,xcdraw}{xcolor}
\documentclass[acmtog]{acmart}
\acmSubmissionID{835}
\acmJournal{TOG}

\usepackage{multicol}
\usepackage{multirow}

\usepackage{pgfplotstable}
\usepackage{pgffor}
\usepackage{ifthen}
\usepackage{booktabs} %
\usepackage{makecell}
\usepackage{listings}
\usepackage{graphicx}
\usepackage{calc}
\usepackage{wrapfig}
\usepackage{afterpage}

\usepackage{pgfplots}
\pgfplotsset{compat=1.18} %

\usepackage{iftex} %

\newcommand{\minipageimage}[3][0.12]{%
    \begin{minipage}[t]{#1\linewidth}
        \centering 
        \small #3\par
        \if\relax\detokenize{#2}\relax
        \else
            \includegraphics[width=\linewidth]{#2}
        \fi
    \end{minipage}\hfill%
}

\usepackage{soul}
\usepackage{multirow}
\usepackage{xspace}

\newcommand{\editprompt}[1]{\footnotesize{Edit prompt: \textit{#1}}}

\definecolor{graytext}{RGB}{130,130,130}

\newcommand{\hao}[1]{\textcolor{blue}{[\textbf{HAO:} #1]}}

\newcommand{\ignorethis}[1]{}

\newcommand{\new}[1]{{#1}}

\def\methodName{Prox$\cdot$E\xspace} %
\def\ProxE{Prox$\cdot$E}

\def\pass{$\color{Green}{\checkmark}$}

\newbox\jsavebox

\definecolor{count_color}{rgb}{0.87, 0.63, 0.3}
\definecolor{attribute_color}{rgb}{0.73, 0.33, 0.95}
\definecolor{spatial_color}{rgb}{0, 0.8, 1.0}
\definecolor{warp_color}{rgb}{0.2, 0.4, 1.0}

\newcommand{\counttext}[1]{\textbf{\textcolor{count_color}{#1}}}
\newcommand{\attrtext}[1]{\textbf{\textcolor{attribute_color}{#1}}}
\newcommand{\warptext}[1]{\textbf{\textcolor{warp_color}{#1}}}

\newcommand{\minihead}[1]{\medskip \noindent \textbf{#1}}

\lstdefinestyle{promptstyle}{
  basicstyle=\ttfamily\tiny,
  breaklines=true,
  breakatwhitespace=true,
  breakautoindent=false,
  breakindent=0pt,
  postbreak=,
  columns=fullflexible,
  keepspaces=true,
  frame=none,
  xleftmargin=0pt,
  xrightmargin=0pt,
  aboveskip=4pt,
  belowskip=4pt,
}

\copyrightyear{2026}
\acmYear{2026}
\setcopyright{cc}
\setcctype{by}
\acmConference[SIGGRAPH Conference Papers '26]{Special Interest Group on Computer Graphics and Interactive Techniques Conference Conference Papers}{July 19--23, 2026}{Los Angeles, CA, USA}
\acmBooktitle{Special Interest Group on Computer Graphics and Interactive Techniques Conference Conference Papers (SIGGRAPH Conference Papers '26), July 19--23, 2026, Los Angeles, CA, USA}
\acmDOI{10.1145/3799902.3811141}
\acmISBN{979-8-4007-2554-8/2026/07}


\begin{document}

\title{\ProxE: Fine-Grained 3D Shape Editing via Primitive-Based Abstractions
}

\author{Etai Sella}\authornote{Denotes equal contribution}
\affiliation{%
  \institution{Tel Aviv University}
  \country{Israel}
}
\email{etaisella@gmail.com}
\orcid{0009-0002-2119-9808}

\author{Hao Phung}\authornotemark[1]
\affiliation{%
  \institution{Cornell University}
  \country{USA}
}
\email{htp26@cornell.edu}
\orcid{0000-0002-6834-4139}

\author{Nitay Amiel}
\affiliation{%
  \institution{Technion - Israel Institute of Technology}
  \country{Israel}
}
\email{nitay.amiel@campus.technion.ac.il}
\orcid{0000-0000-0000-0000}

\author{Or Litany}
\affiliation{%
  \institution{Technion - Israel Institute of Technology}
  \country{Israel}
}
\email{or.litany@gmail.com}
\orcid{0000-0001-6700-7379}

\author{Or Patashnik}
\affiliation{%
  \institution{Tel Aviv University}
  \country{Israel}
}
\email{orpatashnik@gmail.com}
\orcid{0000-0001-7757-6137}

\author{Hadar Averbuch-Elor}
\affiliation{%
  \institution{Cornell University}
  \city{New York}
  \state{New York}
  \country{USA}
}
\email{hadarelor@cornell.edu}
\orcid{0000-0003-3476-0940}

\begin{abstract}
Text-based 2D image editing models have recently reached an impressive level of maturity, motivating a growing body of work that heavily depends on these models to drive 3D edits. While effective for appearance-based modifications, such 2D-centric 3D editing pipelines often struggle with fine-grained 3D editing, where localized structural changes must be applied while strictly preserving an object’s overall identity. To address this limitation, we propose \ProxE{}, a training-free framework that enables fine-grained 3D control through an explicit, primitive-based geometric abstraction. Our framework first abstracts an input 3D shape into a compact set of geometric primitives. A pretrained vision–language model (VLM) then edits this abstraction to specify primitive-level changes. These structural edits are subsequently used to guide a 3D generative model, enabling fine-grained, localized modifications while preserving unchanged regions of the original shape. Through extensive experiments, we demonstrate that our method consistently balances identity preservation, shape quality, and instruction fidelity more effectively than various existing approaches, including 2D-based 3D editors and training-based methods.

\end{abstract}

\begin{teaserfigure}
\centering
\includegraphics[width=0.999\textwidth,trim={0 0.6cm 0 0},clip]{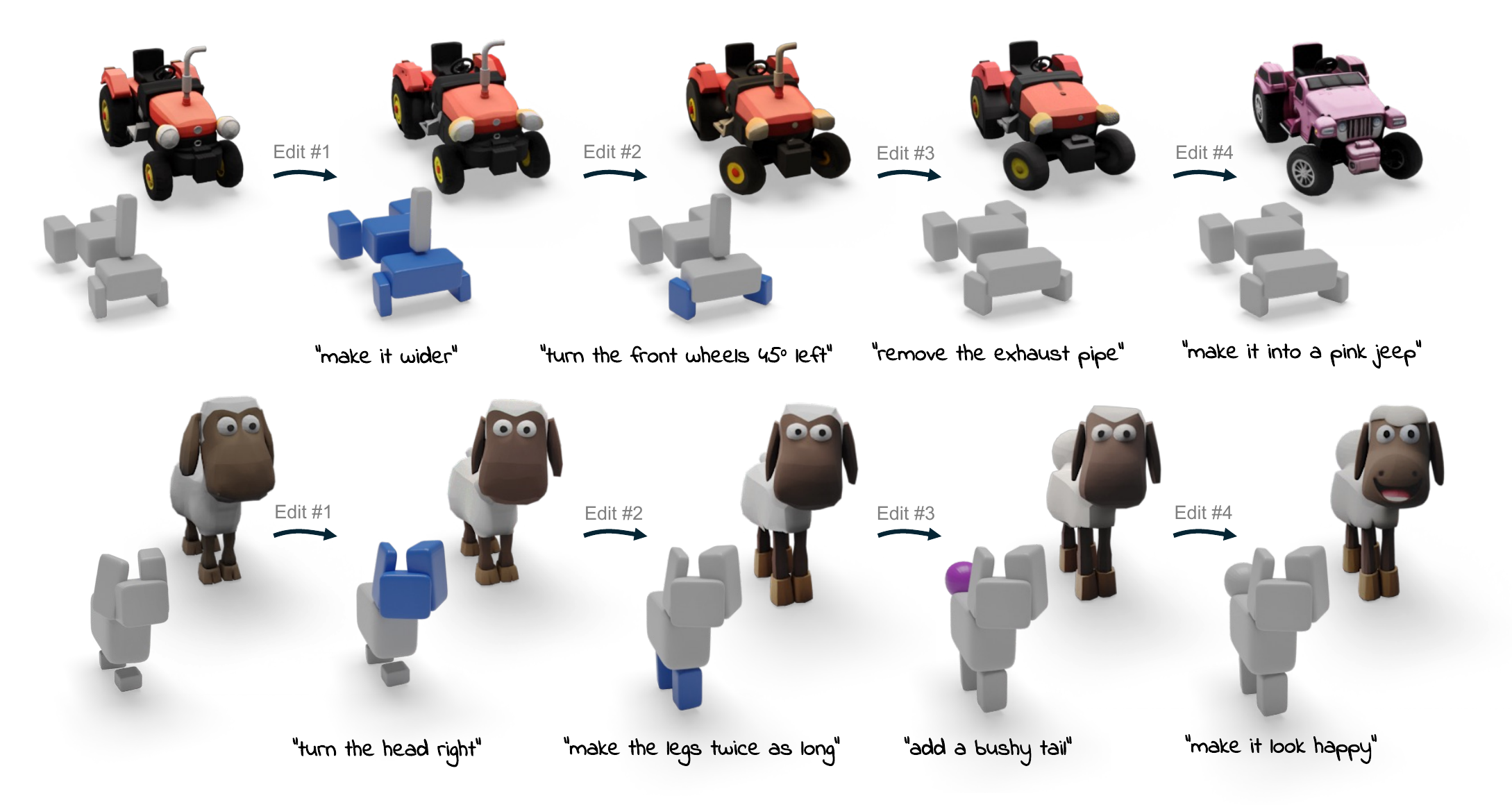}
\vspace{-20pt}
\Description{Teaser figure in a grid: rows of 3D objects with primitive-based abstractions. Middle and lower rows show edited proxy shapes, with original geometry and edited or added super-quadric parts highlighted in blue and purple, alongside text prompts for each edit, illustrating global, parametric, part, and appearance edits.}%
\captionof{figure}{We introduce \ProxE, a training-free 3D editing framework that operates on a primitive-based geometric abstraction. By editing this proxy representation (second and bottom rows; edited primitives shown in \textcolor{blue}{blue}, added ones shown in \textcolor{violet}{purple}) and using it to guide 3D generation, \ProxE{} enables precise, fine-grained edits while preserving the object's identity.
As illustrated above, our method supports a wide range of text-guided edits, spanning global and localized geometric transformations (edits 1 and 2) including parametric edits (edits involving a numeric parameter, i.e. edit 2), addition and removal of object parts (edit 3), and stylistic appearance-based modifications (edit 4).}%
\label{fig:teaser}
\end{teaserfigure}

\maketitle

\section{Introduction}
\label{sec:intro}

Recent years have seen a Cambrian explosion of methods capable of generating novel 3D shapes directly from text. However, practical 3D creation workflows are far more often defined by the need for fine-grained modification rather than wholesale generation. Designers often seek to precisely alter existing geometry—such as lengthening a table’s legs by a fixed factor, introducing gentle ornamental details along a teapot's spout, or turning a vehicle's wheels as illustrated in Figure \ref{fig:teaser}—while strictly preserving the object’s overall identity. However, achieving this level of localized control remains challenging for modern 3D editors, which struggle to faithfully execute such fine-grained edits. 

\begin{figure}
    \small 
    \centering
    \includegraphics[width=0.95\linewidth]{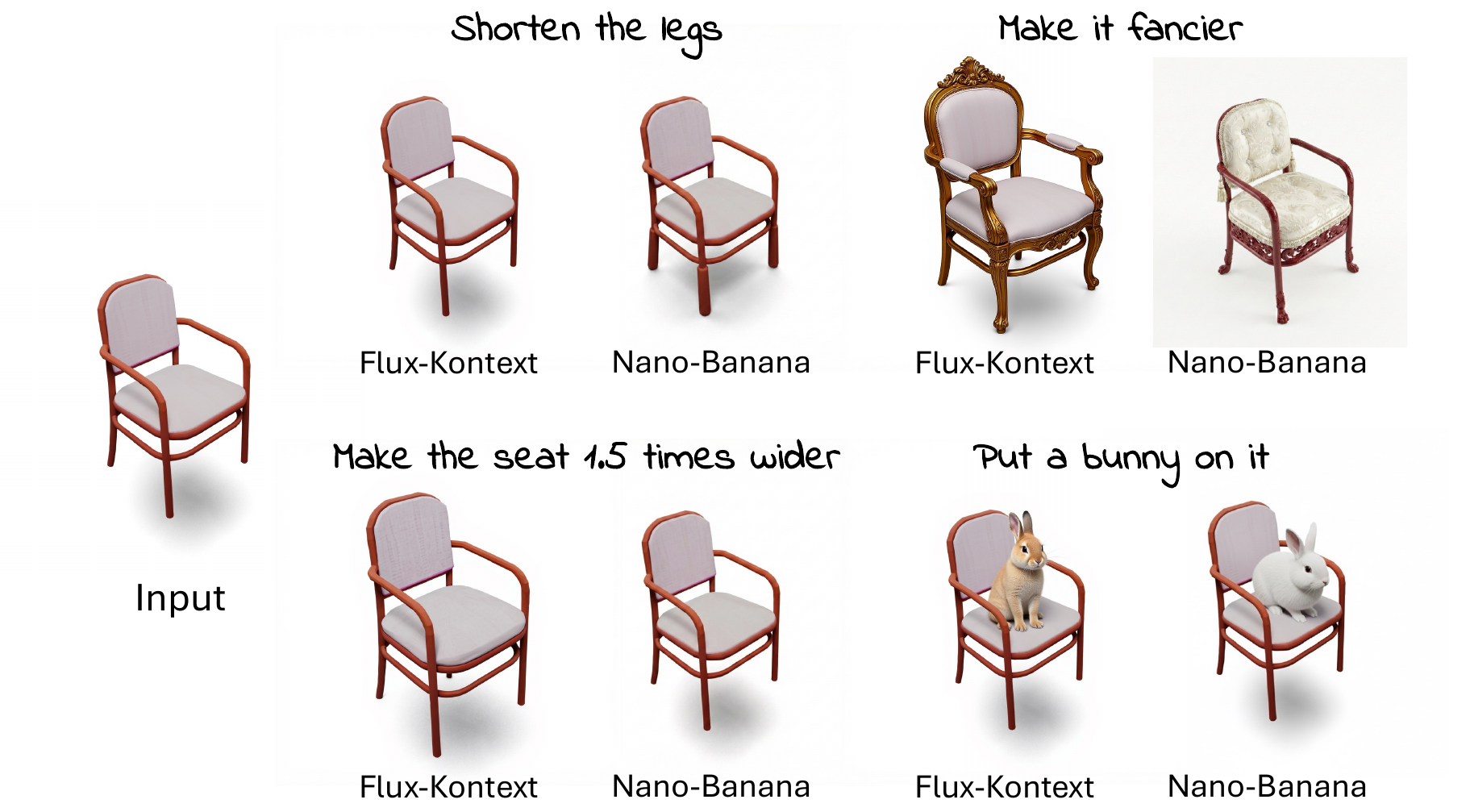}
    \vspace{-5pt}
    \caption{Editing 3D objects with 2D generative models. Given an input image of a chair (left), state-of-the-art open and closed source image editors (Flux-Kontext~\cite{labs2025flux1kontextflowmatching} and Nano-Banana~\cite{GoogleDeepMind2025NanoBanana}) successfully perform appearance-based edits and semantic insertions (right). In contrast, they struggle with fine-grained geometric instructions that require metric reasoning about existing structure (center). These failures reveal a fundamental mismatch between the capabilities of pixel-based editors and the requirements of of fine-grained, controllable 3D editing.
    }
    \label{fig:motivation}
\end{figure}

The remarkable capabilities of modern 2D generative models have recently given rise to a 3D editing paradigm that shifts much of the editing work to image-based generators, treating 3D structure primarily as a scaffold for multi-view synthesis and aggregation~\cite{li2025voxhammer, ye2025nano3d, xia2025towards, gilo2026instructmix2mix}. Implicit in this paradigm are two key assumptions: (i) that a pretrained 2D diffusion model can produce a semantically and geometrically correct edit of the underlying 3D asset from a projected view, and (ii) that a single, or a small set of, edited views is sufficient to faithfully propagate the modification to the full 3D shape. For fine-grained 3D edits, where success hinges on subtle geometric or localized stylistic nuances, the validity of these assumptions remains unclear. 

Consider the image edits produced by two state-of-the-art 2D image editing model, as illustrated in Figure \ref{fig:motivation}. While these models readily succeed at introducing new semantic content—(e.g., placing a bunny on the chair) or appearance-based modifications, they are challenged by instructions targeting fine-grained manipulations of existing geometry. Such \emph{structural} edits require an explicit understanding of metric properties in 3D space, which is largely absent from pixel-based diffusion models. This gap challenges the assumption that pretrained 2D models can reliably support fine-grained edits of a 3D asset from projected views, highlighting a fundamental mismatch between image-based editors and the requirements of controllable 3D editing.

In this work, we propose a training-free 3D editing framework that bridges image-based editors and fine-grained 3D controllability through an explicit, primitive-based shape abstraction. Rather than operating directly in pixel space, our framework decomposes an input 3D asset into a small set of interpretable geometric primitives. A pretrained vision–language model (VLM) operates on this abstraction to specify precise primitive-level edits, which are then used to guide a state-of-the-art 3D generative model~\cite{xiang2025structured} in producing fine-grained, localized modifications of the underlying 3D shape. By exposing structure and metric relationships explicitly, our framework unlocks the ability of VLMs to reason about fine-grained 3D edits that are difficult to express in pixel space alone.

To accurately guide the 3D generative process, we introduce a proxy-induced denoising strategy that uses the primitive-based edits to determine where geometry should be preserved, transformed, or newly synthesized, and enforces these constraints directly in the latent space of a 3D diffusion model. At its core, our strategy employs a blending mechanism that composites inverted latent representations originating from both the input shape and the edited proxy, allowing the generative process to follow the specified structural edits while preserving the identity of the original shape. 
Finally, after generating a coherent edited structure, we refine appearance using 2D image editors, leveraging their strong visual priors to apply stylistic modifications and output a high-quality 3D shape. %

We conduct extensive experiments comparing our method against a broad range of 3D editing paradigms, including training-based 3D editors and image-based 3D editing approaches. We evaluate performance using metrics that quantify preservation of structural identity, quality of the generated shapes, and fidelity to the edit text prompt. Our results demonstrate that our approach achieves a superior balance among these criteria, enabling precise 3D edits while reliably preserving the original shape’s identity.

\section{Related Work}
\label{sec:related_works}

\subsection{Text-Guided 3D Editing}
Text-guided 3D manipulation has evolved rapidly, with recent surveys categorizing approaches into stylization, generation, and editing~\cite{chao2023text, zhu2026survey}. Despite the remarkable fidelity of text-to-3D generative models, extending these capabilities to precise structural editing remains an unresolved challenge.

\paragraph{Stylization, Deformation, and Optimization.}
Early works focused on stylizing geometry without topological changes. Methods like Text2Mesh~\cite{michel2022text2mesh} and Tango~\cite{chen2022tango} leverage CLIP or depth-to-image diffusion for color and displacement optimization, with~\cite{chung20243dstyleglip} adding part-level control. This paradigm extends to implicit fields and 3D Gaussian Splatting via Score Distillation Sampling (SDS)~\cite{haque2023instruct, poole2022dreamfusion, palandra2024gsedit, chen2023fantasia3d, chen2024dge}. While effective for generation, optimization-based methods like Vox-E~\cite{sella2023vox}, DreamEditor~\cite{zhuang2023dreameditor}, and TIP-Editor~\cite{zhuang2024tip}, often struggle to balance metric precision with identity preservation. Prior supervised methods~\cite{huang2022ladis, achlioptas2022changeit3d} required paired data and lack generalization. Similarly, explicit deformation techniques~\cite{gao2023textdeformer, yang2025genvdm, meng2025text2vdm} achieve high-fidelity sculpting but  lacking in ability to alter functional topology.

\paragraph{Latent and Lifting Approaches.}
To accelerate editing, recent works operate in latent spaces~\cite{edelstein2025sharp, chen2024shap} but often suffer from encoding information loss. Alternatively, lifting approaches~\cite{ye2025nano3d,xia2025towards, bar2025editp23,gilo2026instructmix2mix} integrate 2D editing priors directly into 3D flow-matching or multi-view diffusion. However, these methods rely on the \textit{geometric validity} of 2D input derived from pixel space. Consequently, they often fail to execute precise metric instructions (e.g., ``widen seat 1.5x'') and struggle to reconcile spatial transformations with the original identity. Our framework uses explicit primitives to provide the view-agnostic, metric guidance these methods lack.

\paragraph{Auxiliary Control.}
To ensure spatial control, methods utilize masks \cite{barda2025instant3dit, weber2024nerfiller, erkoc2025preditor3d} or bounding boxes~\cite{xiang2025structured, li2025voxhammer}. While effective for local edits, these pre-defined constraints limit global flexibility. While SpaceControl~\cite{fedele2026spacecontrol} utilizes superquadrics to constrain the generated content, it relies on the user to manually define these geometric guides. Our VLM agent bridges this gap by automatically generating precise spatial constraints for both local and global edits.

\subsection{Primitive-Based Abstractions}
Explicit primitives offer interpretable ``building blocks'' for 3D reasoning. While automated decomposition methods exist~\cite{tulsiani2017learning, paschalidou2019superquadrics, paschalidou2021neural, fedele2025superdec, ye2025primitiveanything}, these approaches have not demonstrated how to leverage such coarse geometric handles to drive detailed structural editing, such as component addition or removal, in high-fidelity assets. Unlike pixel-aligned triplanes~\cite{kathare2025, bilecen2025}, primitives are semantically interpretable. Recent differentiable rendering works have improved fidelity~\cite{held20253d, govindarajan2025radiant}, yet their focus remains on representation rather than manipulation. Hybrid approaches~\cite{hao2020dualsdf, hu2024cns, liu2023exim} couple coarse proxies with implicit functions, but are largely limited to deformation. Crucially, they lack the capacity for explicit topological edits—such as adding a handle—which our neuro-symbolic approach uniquely enables by treating primitives as flexible volumetric guides rather than rigid constraints.

\subsection{VLMs for 3D Generation}
Bridging the modality gap between Vision-Language Models (VLMs) and native 3D representations presents a fundamental challenge. Direct generation methods~\cite{siddiqui2024meshgpt, wang2024llama, fang2025meshllm} burden models with topological consistency by outputting dense tokens. Procedural code methods~\cite{sun20253d, lu2025ll3m, man2025videocad, avetisyan2024scenescript} rely on \textit{blind execution}, forcing the model to simulate transformations without visual feedback, often yielding incoherent geometry. In contrast, we propose that parametric primitives serve as a token-efficient vocabulary, allowing the VLM to act as a spatial reasoning agent that manipulates structure with visual verification.

\section{Method}
\label{sec:method}

\begin{figure*}
    \centering
    \includegraphics[width=0.99\linewidth]{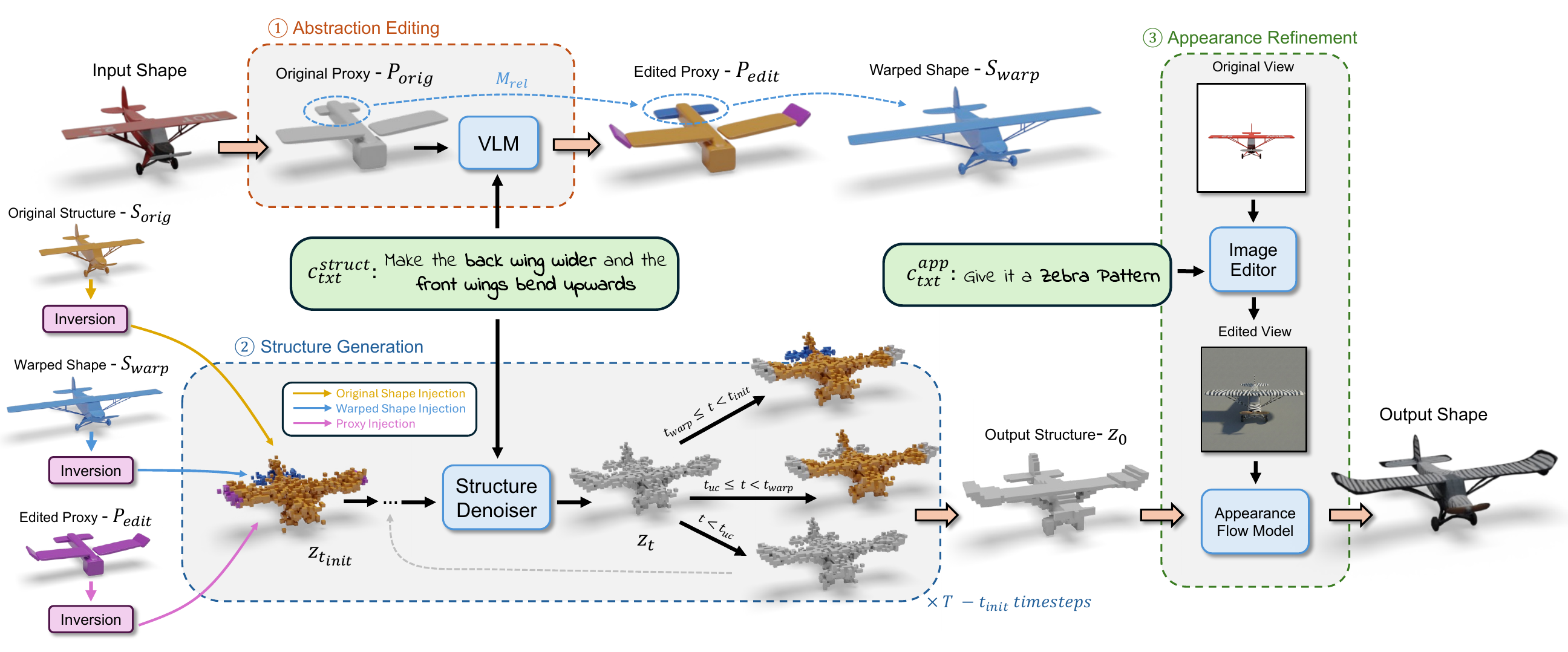}
    \vspace{-5pt}
    \caption{\textbf{An overview of our approach.} Given an input 3D shape and a text prompt, we first edit a primitive-based abstraction using a vision–language model to specify structure-aware modifications (Section \ref{sec:editing_abstraction}). These edits guide a proxy-induced denoising process by blending inverted latents from the original structure (yellow), warped shape (blue) and edited proxy (purple) to generate an updated structure while preserving object identity (Section \ref{sec:structure_editing}). Finally, we refine appearance using 2D image editors to apply stylistic changes and produce the final edited 3D shape (Section \ref{sec:appearance}).
    }
    \label{fig:structure_editing}
\end{figure*}

Given an input 3D shape $\mathcal{S}_{orig}$ and a text-based editing instruction $c_{\text{txt}}$, our method generates an edited 3D shape $\mathcal{S}_{edit}$ by extracting and editing a primitive-based proxy representation. We extract this proxy by abstracting the input shape and editing the abstraction using a vision-language model (Sec. \ref{sec:editing_abstraction}). We then use the edited abstraction to modify the structure (Sec. \ref{sec:structure_editing}) and appearance (Sec. \ref{sec:appearance}) of the original 3D shape. See Figure \ref{fig:structure_editing} for an overview.

\smallskip
\paragraph{Prompt Parsing} As a precursor to our method, we employ an LLM to parse the text instruction $c_{\text{txt}}$ into two separate textual descriptions: an instruction prompt specifying \textit{structural} edits $c_{\text{txt}}^{struct}$ and an instruction prompt specifying \textit{appearance}-based edits $c_\text{txt}^{app}$; see Figure \ref{fig:structure_editing} for an example of each. 
The full system prompt and additional details are provided in the supplementary material.

\subsection{Background}
\label{sec:background}
We begin by providing background on the primitive-based representation we adopt and the 3D generative backbone model.

\smallskip
\paragraph{Superquadrics.} We utilize superquadrics (SQs) \cite{barr1981superquadrics,pentland1986parts,paschalidou2019superquadrics} as the geometric primitives composing our proxy shapes. A superquadric surface is defined as the set of points $(x, y, z)$ satisfying the implicit equation:

\begin{equation}
    f(x, y, z; \lambda) = \left( \left| \frac{x}{a_1} \right|^{\frac{2}{\epsilon_2}} + \left| \frac{y}{a_2} \right|^{\frac{2}{\epsilon_2}} \right)^{\frac{\epsilon_2}{\epsilon_1}} + \left| \frac{z}{a_3} \right|^{\frac{2}{\epsilon_1}} = 1,
\end{equation}
where $\lambda$ represents the set of shape parameters. Specifically, we parameterize each primitive $q$ using 11 parameters: scale $\mathbf{a} = [a_1, a_2, a_3] \in \mathbb{R}_{>0}^3$, shape exponents $\mathbf{\epsilon} = [\epsilon_1, \epsilon_2] \in \mathbb{R}_{>0}^2$, translation $\mathbf{t} \in \mathbb{R}^3$, and rotation $\mathbf{r} \in \mathbb{R}^3$.

\smallskip
\paragraph{TRELLIS.} We build upon TRELLIS \cite{xiang2025structured}, a cascaded framework that first generates a \textit{Sparse Structure} ($64^3$ occupancy grid) and subsequently synthesizes appearance features in a \textit{Structured Latent} (SLAT) space. Both phases employ rectified flow transformers on latent voxel grids to predict 3D data from noise, conditioned on text or images. The resulting features are finally decoded into diverse formats such as Gaussian Splats or textured meshes.

\subsection{Editing Abstractions with a Vision-Language Model}
\label{sec:editing_abstraction}
To perform structural manipulation, we first abstract the geometry of $\mathcal{S}_{orig}$ into a discrete, parametric representation (top left in Figure~\ref{fig:structure_editing}). We sample the surface to obtain a dense point cloud, which is processed by \textbf{SuperDec}~\cite{fedele2025superdec} to decompose the shape into a set of superquadrics, forming the original proxy shape $\mathcal{P}_{orig}$. 
Critically, our framework is designed to be robust to the approximation errors inherent in this decomposition. We treat the resulting primitives not as rigid boundary conditions, but as coarse volumetric guides that condition the generative process. This allows us to leverage the strong shape priors of the 3D diffusion model to compensate for discretization artifacts or minor misalignments in the proxy, synthesizing high-fidelity geometry even when the initial primitive fit is imperfect.

We then employ a VLM as an editing agent (Abstraction Editing part in Figure~\ref{fig:structure_editing}). 
To facilitate visual reasoning, we assign each primitive in $\mathcal{P}_{orig}$ a unique index and a distinct color, and group the parameters of all primitives (scale, position, rotation, shape, and the assigned color code) into a structured JSON file.
The agent is provided with a multi-modal context comprising: (1) a composite image containing four orthogonal views (front, back, left, right) of the colored proxy $\mathcal{P}_{orig}$; (2) a reference rendering of the original shape $\mathcal{S}_{orig}$; (3) the structured JSON containing the primitives' parameters; and (4) the structural editing instruction $c_{\text{txt}}^{struct}$.

Crucially, the inclusion of color codes in the JSON enables the VLM to ground visual features (e.g., identifying a ``red'' leg in the render) to specific symbolic entries in the parameter list. 
The VLM is instructed to manipulate the primitive parameters (e.g., updating scale, orientation, or position) or modify the list structure (adding/deleting primitives) within the JSON to satisfy $c_{\text{txt}}^{struct}$, while adhering to a strict principle of minimal intervention to preserve the object's identity (see supplementary material for the exact prompt). We prompt the VLM to output a chain-of-thought reasoning process, first localizing relevant primitives and planning the edit, followed by the output of the fully updated JSON file.

To ensure robustness, we implement a visual verification loop. 
Upon receiving the edited JSON, we render the new proxy $\mathcal{P}_{edit}$ from the same four viewpoints %
and feed these renders back to the VLM alongside the conversation history. The model is asked to verify if the geometric changes satisfy the instruction $c_{\text{txt}}^{struct}$. If the edit is deemed insufficient or erroneous, the VLM generates a refined JSON. This verification strategy benefits from the inherent simplicity of the proxy representation. While VLMs often exhibit inconsistent judgment when evaluating detailed, textured meshes (a limitation we discuss in Sec.~\ref{sec:experiments}), the proxy offers a clean, unambiguous visualization of the object's structure. By verifying the edit on this color-coded abstraction, the agent can reliably confirm that the geometric constraints are met before the downstream generation process begins. This iterative process continues until the edit is verified or a maximum number of iterations is reached.

\subsection{Structural Editing via an Edited Abstraction}
\label{sec:structure_editing}

Having obtained the edited proxy $\mathcal{P}_{edit}$, our goal is to generate the detailed edited 3D shape $\mathcal{S}_{edit}$. 
We begin by constructing an approximated edited shape by warping the original one, and denote it by $\mathcal{S}_{warp}$. Then, given $\mathcal{S}_{orig}$, $\mathcal{S}_{warp}$, and $\mathcal{P}_{edit}$ we generate $\mathcal{S}_{edit}$ by employing the structure diffusion model of TRELLIS. 
We apply DDIM inversion to each of these shapes to timestep $t_{init}$, initialize a denoising process with the inverted $\mathcal{P}_{edit}$ at $t_{init}$, and store the intermediate inverted latent grids of $\mathcal{S}_{orig}$ and $\mathcal{S}_{warp}$.

Both the construction of $\mathcal{S}_{warp}$ and our denoising process rely on localizing regions that should remain unchanged, regions that should be modified, and regions that should be added or removed.
To achieve this localization, we classify the primitives in $\mathcal{P}_{edit}$ relative to $\mathcal{P}_{orig}$ into three categories: \textit{unchanged} ($\mathcal{Q}_{uc}$), \textit{edited} ($\mathcal{Q}_{ed}$), and \textit{new} ($\mathcal{Q}_{new}$), with the latter encompassing both added and deleted elements. The volumetric union of primitives in each category defines a corresponding 3D spatial mask—$\mathcal{M}_{uc}$, $\mathcal{M}_{ed}$, and $\mathcal{M}_{new}$.

Next, we first describe how do we obtain $\mathcal{S}_{warp}$, and then we describe our denoising process strategy.

\subsubsection{Constructing Warped Shape.}
We first note that the pose parameters ($\mathbf{t}, \mathbf{r}, \mathbf{a}$) of Superquadrics allow us to construct a local-to-world transformation matrix $M \in \mathbb{R}^{4 \times 4}$ for each primitive, composed as $M = T R S$. Here, $T$, $R$, and $S$ correspond to translation, rotation, and non-uniform scaling, respectively. This matrix $M$ defines the bounding volume and orientation of the primitive but ignores the curvature parameters $\epsilon$. 
Consequently, given two corresponding Superquadrics, their respective matrices can be used to define a relative affine transformation (specifically translation, rotation, and scaling, without shear) that maps the coordinate frame of one primitive to the other. To construct $\mathcal{S}_{warp}$, we process each edited primitive pair $(q^{(i)}_{orig}, q^{(i)}_{edit}) \in \mathcal{Q}_{ed}$ individually. First, we compute the specific relative transformation $M^{(i)}_{rel} = M^{(i)}_{edit} (M^{(i)}_{orig})^{-1}$ and apply it to the vertices of $\mathcal{S}_{orig}$ to generate a corresponding warped reference shape $\mathcal{S}_{warp}$. Note that this $\mathcal{S}_{warp}$ is not the final edited shape, since it does not taking into account primitives that were added or removed.

\subsubsection{Proxy-Induced Denoising Process.}
The denoising process is initialized with $z_{t_{init}}$, the inverted latent grid of $\mathcal{P}_{edit}$ composited with the inverted latent grids of $\mathcal{S}_{orig}$ and $\mathcal{S}_{warp}$, and is conditioned on the structural text prompt $c_{\text{txt}}^{struct}$. \new{We determine the spatial masks $\mathcal{M}_{uc}$, $\mathcal{M}_{ed}$, and $\mathcal{M}_{new}$ by voxelizing the relevant sources: $\mathcal{M}_{uc}$ is obtained from the unchanged geometry of the original shape, while $\mathcal{M}_{ed}$ and $\mathcal{M}_{new}$ are obtained from the edited proxy.} At each denoising step, \new{we first apply the denoiser update $z_{t+1} \rightarrow z_t$, and then override voxels in each masked region with their assigned reference latent.} We blend the evolving noisy latent grid with the inverted latent grids of $\mathcal{S}_{orig}$ and $\mathcal{S}_{warp}$ to preserve the details of the original shape. Next, we describe how each spatial region defined by the masks is processed at each timestep.

\paragraph{$\mathcal{M}_{uc}$: Original Shape Injection}
In regions where the proxy remains unchanged, our objective is strict preservation of the original geometry. To achieve this, we utilize $z^{orig}_{t}$, the inverted latent grids of $\mathcal{S}_{orig}$.
For all voxels $v \in \mathcal{M}_{uc}$, we replace the generated latent with the reference latent: $z_t[v] \leftarrow z^{orig}_{t}[v]$. \new{That is, after the denoiser predicts the next latent, we explicitly overwrite all voxels in $\mathcal{M}_{uc}$ with the corresponding inverted latent values from the original shape.} This injection is applied from $t_{init}$ down to a late timestep $t_{uc}$ (close to 0), ensuring the original structure is perfectly retained.

\paragraph{$\mathcal{M}_{new}$: New Regions} For regions corresponding to added or deleted primitives ($\mathcal{Q}_{new}$), we enforce the coarse structure specified by the edited proxy. Since the denoising process is initialized from the inverted proxy, voxels within $\mathcal{M}_{new}$ are taken directly from the evolving latent grid $z_t$. %

\paragraph{$\mathcal{M}_{edit}$: Edited Regions Injection}
For primitives that underwent affine transformations ($q^{(i)}_{edit} \in \mathcal{Q}_{ed}$), we aim to preserve surface details while adhering to the new geometric pose.
We utilize the inverted latent grids of $\mathcal{S}_{warp}$, denoted $z^{warp}_{t}$, and inject them into the volumetric region defined by the edited primitive $q^{(i)}_{edit}$. \new{Concretely, after each denoiser step, we overwrite voxels $v \in \mathcal{M}_{ed}$ with the corresponding warped inverted latents, i.e., $z_t[v] \leftarrow z^{warp}_{t}[v]$.} 
This injection is applied from $t_{init}$ down to an intermediate timestep $t_{warp}$ (where $t_{uc} < t_{warp} < t_{init}$). This strategy effectively relocates the original surface details to their new positions piece-by-piece, allowing the subsequent denoising steps (from $t_{warp}$ to $0$) to seamlessly stitch the distinct warped regions into a coherent global structure.

\subsection{Appearance Refinement} 
\label{sec:appearance}

Given the edited sparse structure produced in the previous stage, our final goal is to synthesize fine-grained details and appearance features. We leverage the decoupled architecture of TRELLIS, which allows us to transition from text-based guidance used for structure generation to image-based guidance for appearance. Image-based guidance enables us to exploit the capabilities of state-of-the-art 2D image editing models for high-quality appearance edits.

We begin by rendering a view $V_{orig}$ of the original shape $\mathcal{S}_{orig}$ and editing it according to the appearance instruction $c_{\text{txt}}^{app}$ using a 2D image editor, resulting in an edited view $V_{edited}$. We then invert the SLAT features of the original shape $\mathcal{S}_{orig}$, conditioning on $V_{orig}$, to obtain the latent SLAT features ${z^{app}_{t}}$ of the original shape.
Next, we initialize a denoising process from Gaussian noise $z_T$ and apply the appearance diffusion model of TRELLIS. The edited view $V_{edited}$ is used as the conditioning signal, and we apply a blending strategy similar to the one used during the structure generation stage.

Specifically, we use the masks $\mathcal{M}_{uc}$ and $\mathcal{M}_{edit}$  to blend the evolving noisy SLAT features $z_t$ with features from $z^{app}_{t}$. For regions in $\mathcal{M}_{uc}$, we directly copy the corresponding features from $z^{app}_{t}$ and inject them into the same voxel locations. For a voxel $v \in \mathcal{M}_{edit}$, we compute its pre-edit location $v' = (M^{(i)}_{rel})^{-1} v$, retrieve the feature from $z^{app}_{t}$ at $v'$, and inject it into the current denoising step. 
The denoising process proceeds from $t = T$ down to a threshold $t_{app}$, which controls the tradeoff between preserving the original appearance and allowing the edited appearance to emerge.
In practice, $t_{app}$ is set using a binary policy: when an appearance edit is applied, we set $t_{app}$ close to $T$; when no appearance edit is requested, we use a smaller fixed value to preserve the original appearance.

\section{Experiments}
\label{sec:experiments}
We present our main results in this section. Additional details, experiments and ablations, \new{including scene editing results and a runtime comparison,} are provided in the supplementary material.

\subsection{Datasets}
\label{subsec:data}

We use ShapeTalk \cite{achlioptas2023shapetalk} for qualitative and quantitative evaluation. ShapeTalk contains pairs of ShapeNet shapes accompanied by human-written descriptions of their differences. The dataset provides easy and hard splits, where hard pairs exhibit smaller geometric differences and therefore require more fine-grained edits. As our task focuses on precise textual shape editing, these hard samples are particularly well suited for evaluation. Following prior work \cite{sella2025blended,achlioptas2023shapetalk}, we report quantitative results on the Chair, Table, and Lamp categories, randomly sampling 200 ``hard'' pairs per category.

\new{To demonstrate generalization beyond ShapeNet objects, we also present qualitative results on Edit3D-bench \cite{li2025voxhammer}, a recently proposed benchmark made up of 100 high quality 3D objects paired with multiple localized editing prompts. }

\subsection{Evaluation Metrics}
\label{subsec:metrics}

\subsubsection{Identity Preservation.} We use the following metrics to evaluate to what extent the identity of the shape is preserved:

\paragraph{localized-Geometric Distance (l-GD)}
Following prior work~\cite{achlioptas2022changeit3d}, we use l-GD to measure the Chamfer distance between points outside the edited region and their counterparts in the input shape. Unlike GD on the full shape, this metric doesn't penalize differences in regions that are supposed to be modified. %

\paragraph{LPIPS and DINO-I} To quantify how well the edited objects preserve the visual characteristics of the input shapes, we calculate LPIPS and DINO-I scores between their rendered images and the source ones, following prior work~\cite{li2025voxhammer}.

\subsubsection{Quality.} To evaluate the quality of the results, we use the following:

\paragraph{Fréchet Point Distance (FPD) and Fréchet Inception Distance (FID)} FPD \cite{shu20193d} measures the distributional divergence between input point cloud and output point cloud features sampled from the generated objects using a pretrained PointNet \cite{nichol2022point}, while FID \cite{heusel2017gans} compares the distribution of rendered images between input and edited shapes. Since LPIPS, DINO-I, FID are computed based on textured renderings of the edited objects, point-based methods are omitted from this evaluation.

\subsubsection{Edit Fidelity.} We evaluate how well the generated results align with the text prompt using the following metrics:

\paragraph{CLIP Similarity (CLIP)} evaluates the edit quality by measuring the cosine similarity between the textual description of the desired edited output and a rendered view of the output shape.  %

\paragraph{VQAScore (VQA)} We measure VQAScore ~\cite{lin2024evaluating} by presenting a VLM Qwen2.5-VL-7B-Instruct \cite{bai2025qwen2} with rendered images of the original and edited objects along with the input prompt, and using the following prompt: \textit{``Image 1 is the original and Image 2 is the edited version. Does the change from Image 1 to Image 2 reflect the text [input prompt]? Answer Yes or No.''} The probability assigned to \textit{``Yes''} is used as the final score. \new{We enhance the reliability of the metric by incorporating Chain-of-Thought (CoT) reasoning into the model's generation process; additional details are provided in the supplementary.} 
\ignorethis{
We also conduct experiments on SenseNova-SI\cite{sensenova-si}, which has been demonstrated as effective for spatial reasoning. 
}

\subsection{Baselines}
\label{subsec:baseline}
We compare against a broad set of baselines representative of various 3D editing paradigms (see the supplementary for additional details): 

\paragraph{Training-based 3D Editors}. By contrast to our training-free approach, prior work considering the task of fine-grained 3D editing were typically supervised over samples belonging to one of the categories in the ShapeTalk dataset. Specifically, we compare against ChangeIt3D \cite{achlioptas2022changeit3d} and BlendedPC \cite{sella2025blended}, which predict point cloud coordinates and hence cannot directly represent detailed mesh topologies, and Spice-E~\cite{sella2024spice}, which builds upon the Shape-E~\cite{jun2023shap} backbone. 

\paragraph{Single-view 2D Editing-based 3D Editors}. We compare against  VoxHammer~\cite{li2025voxhammer} and  TRELLIS~\cite{xiang2025structured}. VoxHammer builds upon TRELLIS,  performing localized 3D editing given a user-provided 3D mask describing the edit region.
Therefore, performance for this baseline is reported only over test samples containing localized prompts, with the segmentation masks extracted using PointNet~\cite{qi2017pointnet}.
Furthermore, we also report performance over TRELLIS, the backbone generative model we build upon in this work. This baseline is constructed by editing a rendered view of the input object using FLUX Kontext~\cite{labs2025flux1kontextflowmatching}, and conditioning the sampling process on the edited image.

\paragraph{Multi-view 2D Editing-based 3D Editors}. We also compare against EditP23 \cite{bar2025editp23}, which jointly edits multiple rendered views and reconstructs the 3D shape from the edited images.

\begin{table}[t]
\caption{\textbf{Quantitative Comparison}. We evaluate our method against a wide range of baselines. Point-cloud based editors are shown on top (first two rows). Note that these methods operate directly on the input point cloud, giving them an inherent advantage on point-based metrics, while being less directly comparable on other metrics. %
}
\centering
\footnotesize %
\setlength{\tabcolsep}{0pt} %

\begin{tabular*}{\linewidth}{@{\extracolsep{\fill}} l ccc cc cc @{}}
\toprule
& \multicolumn{3}{c}{Identity Preservation} & \multicolumn{2}{c}{3D Quality} & \multicolumn{2}{c}{Edit Fidelity} \\
\cmidrule(lr){2-4} \cmidrule(lr){5-6} \cmidrule(lr){7-8}
Model & l-GD$\downarrow$ & LPIPS$\downarrow$ & DINO-I$\uparrow$ & PFD$\downarrow$ & FID$\downarrow$ & CLIP$\uparrow$ & \new{VQA}$\uparrow$ \\
\midrule
ChangeIt3D & 0.02 & -- & -- & 30.58 & -- & 0.21 & \new{0.49} \\
BlendedPC  & \textbf{0.01} & -- & -- & \textbf{7.81} & -- & 0.23 & \new{0.51} \\
Spice-E    & 0.03 & 0.15 & 0.86 & 11.79 & 56.56 & 0.27 & \new{0.62} \\
EditP23    & 0.03 & 0.16 & 0.82 & 39.58 & 54.13 & \textbf{0.28} & \new{0.58} \\
VoxHammer  & \textbf{0.01} & 0.14 & 0.86 & 12.89 & 52.45 & 0.27 & \new{0.55} \\
TRELLIS    & 0.02 & 0.15 & 0.91 & 16.43 & 36.64 & \textbf{0.28} & \new{0.65} \\
Ours       & 0.02 & \textbf{0.10} & \textbf{0.92} & 11.34 & \textbf{32.60} & \textbf{0.28} & \new{\textbf{0.71}} \\
\bottomrule
\end{tabular*}
\label{tab:metrics}
\end{table}

\subsection{Comparisons} 
\label{subsec:results} 
As shown in Table \ref{tab:metrics}, our method achieves the best overall performance across most metrics. For identity preservation, our method obtains the best LPIPS and DINO-I scores, while VoxHammer and BlendedPC achieve slightly lower l-GD values. This is somewhat expected as both methods rely on explicit 3D edit masks that are aligned with the edit region. While this setup favors identity preservation metrics, it also limits the expressive flexibility of the edits, as reflected in their lower VQA scores.
As for 3D quality, our method achieves the lowest FID while attaining lower PFD compared to other 2D editing-based methods. BlendedPC yields a lower PFD value, largely due to its training objective which explicitly preserves the input point cloud outside the edit region, giving it a built-in advantage on this metric. However, this again comes at the expense of edit expressiveness, as reflected in its lower edit fidelity scores.

Notably, our method achieves the highest VQA score, highlighting superior edit fidelity. Beyond automated metrics, we also conduct pairwise comparisons against individual baselines using both VLM-based evaluation and human raters (details in the supplementary material). In a user study with 44 participants, our method achieves the highest win rates against all competitors by a significant margin, with TRELLIS coming in closest with a \new{21.2\% win rate on edit quality.} %

\new{We present qualitative comparisons in Figure \ref{fig:comp}. The examples highlight improved edit fidelity in part modification (top row), part generation (middle row), and global edits (bottom row), while maintaining stronger identity preservation. Qualitative results on ShapeTalk abd Edit3D-Bench are presented in Figure \ref{fig:results_edit3dbench} and Figure \ref{fig:results}, respectively. These results further demonstrate that our method generalizes to diverse shapes and handles both major structural modifications and subtle edits.}

\begin{table}[t]
\caption{\textbf{Ablation Study}. We evaluate the impact of the different inverted latents in our blending mechanism (first three rows, colored according to Figure \ref{fig:structure_editing}) and our appearance refinement module (fourth row).}
\centering
\setlength{\tabcolsep}{0pt} %
\small %

\begin{tabular*}{\linewidth}{@{\extracolsep{\fill}} l @{\hspace{6pt}} cccc @{\hspace{12pt}} ccc cc cc @{}}
\toprule
& \multicolumn{4}{c}{Components} & \multicolumn{3}{c}{Identity Preservation} & \multicolumn{2}{c}{3D Quality} & \multicolumn{2}{c}{Edit Fidelity} \\
\cmidrule(lr){2-5} \cmidrule(lr){6-8} \cmidrule(lr){9-10} \cmidrule(lr){11-12}

& \attrtext{$\blacksquare$} & \counttext{$\blacksquare$} & \warptext{$\blacksquare$} & App & l-GD$\downarrow$ & LPIPS$\downarrow$ & DINO-I$\uparrow$ & P-FID$\downarrow$ & FID$\downarrow$ & CLIP$\uparrow$ & \new{VQA}$\uparrow$ \\
\midrule

$\mathcal{P}_{edit}$ only & \pass{} & & & \pass{} & 0.03 & 0.13 & 0.87 & 21.35 & 49.74 & 0.27 & \new{0.64} \\
w/o $\mathcal{P}_{edit}$  & & \pass{} & \pass{} & \pass{} & 0.05 & \textbf{0.08} & \textbf{0.93} & \textbf{9.13} & \textbf{27.00} & \textbf{0.28} & \new{0.63} \\
w/o $\mathcal{S}_{warp}$  & \pass{} & \pass{} & & \pass{} & \textbf{0.02} & 0.11 & 0.90 & 14.30 & 43.20 & 0.27 & \new{0.65} \\
w/o App                  & \pass{} & \pass{} & \pass{} & & \textbf{0.02} & 0.11 & 0.91 & 10.92 & 34.38 & \textbf{0.28} & \new{0.70} \\
Ours                     & \pass{} & \pass{} & \pass{} & \pass{} & \textbf{0.02} & 0.10 & 0.92 & 11.34 & 32.60 & \textbf{0.28} & \new{\textbf{0.71}} \\

\bottomrule
\end{tabular*}
\vspace{-8pt}

\label{tab:ablations_tab}
\end{table}

\begin{figure}[t]
    \centering
    \setlength{\tabcolsep}{0.5pt} %
    
    \begin{tabular}{cccccc}

        \multicolumn{6}{c}{\editprompt{"The lamp has nubs that touch the ground, just below the rectangular base."}} 
        \\

        \includegraphics[width=0.16\linewidth]{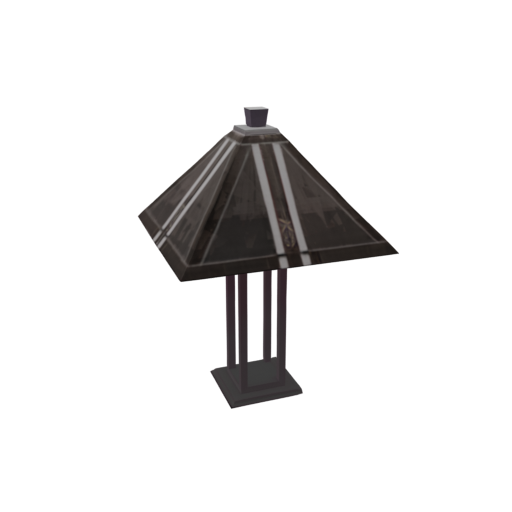} &
        \includegraphics[width=0.16\linewidth]{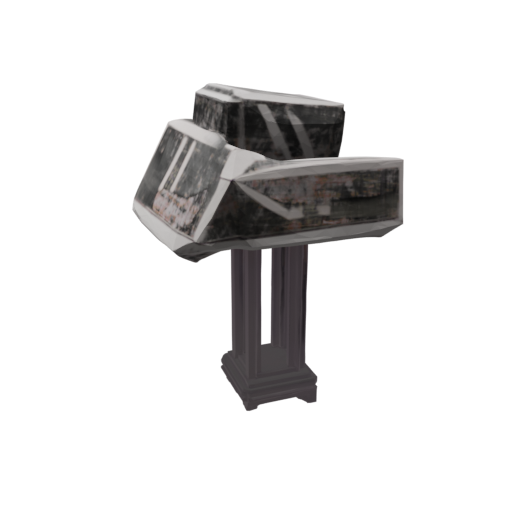} &
        \includegraphics[width=0.16\linewidth]{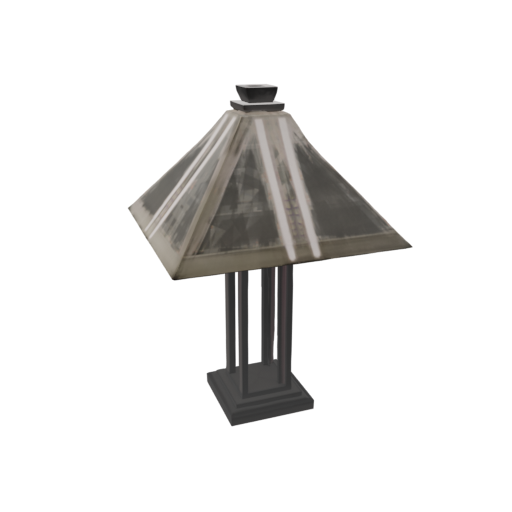} &
        \includegraphics[width=0.16\linewidth]{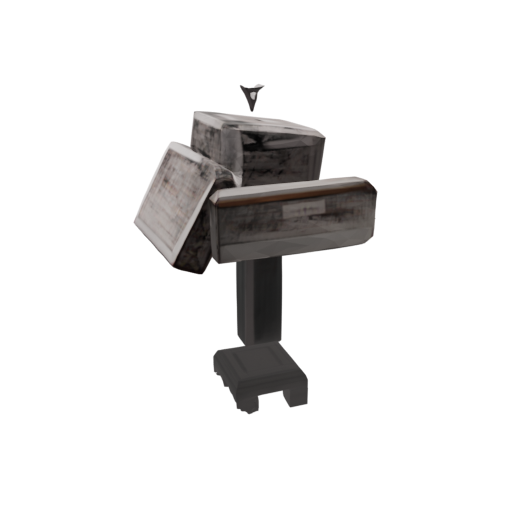} &
        \includegraphics[width=0.16\linewidth]{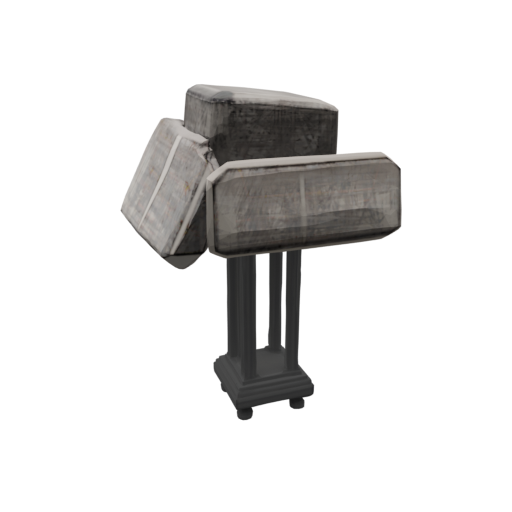} &
        \includegraphics[width=0.16\linewidth]{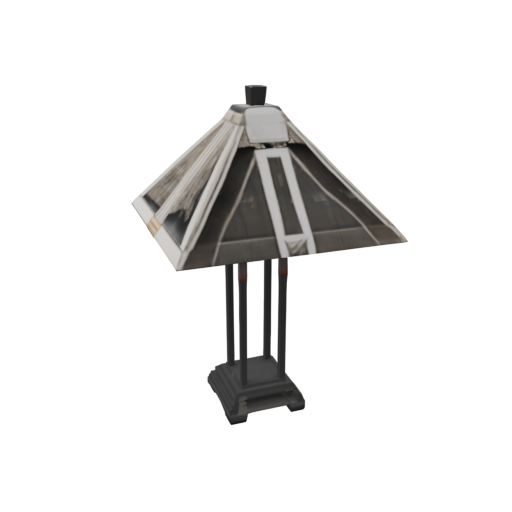} 
        \\

        \multicolumn{6}{c}{\editprompt{"The table top has round discs as design."}} 
        \\

        \includegraphics[width=0.16\linewidth]{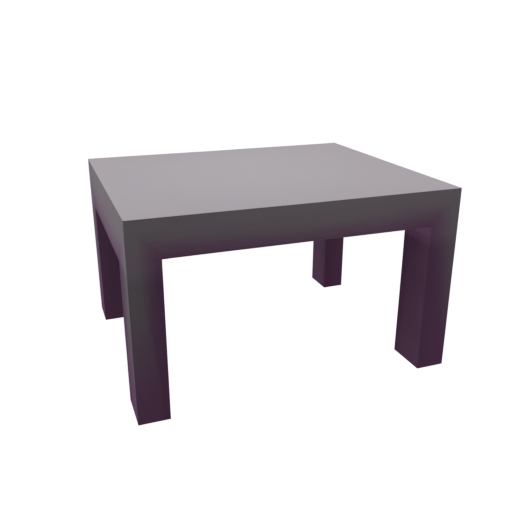} &
        \includegraphics[width=0.16\linewidth]{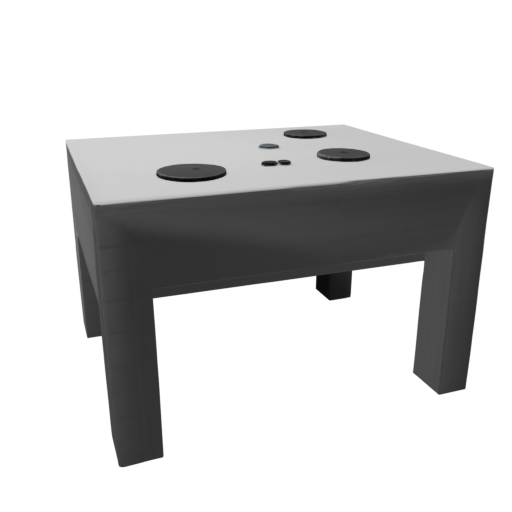} &
        \includegraphics[width=0.16\linewidth]{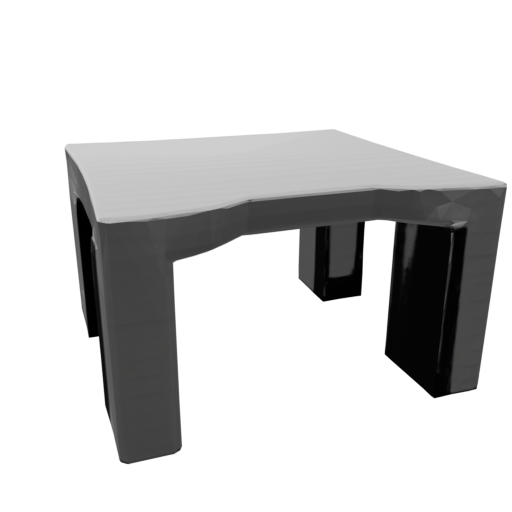} &
        \includegraphics[width=0.16\linewidth]{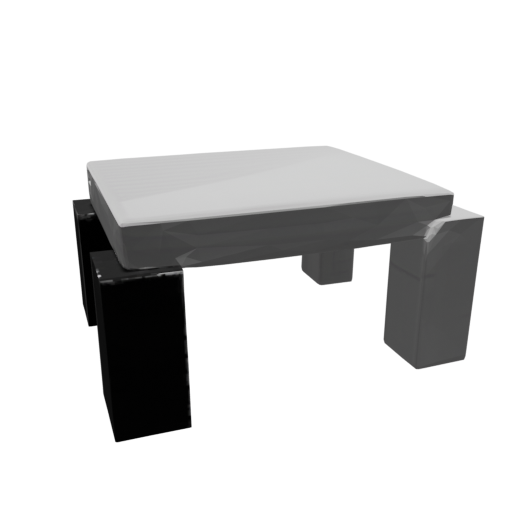} &
        \includegraphics[width=0.16\linewidth]{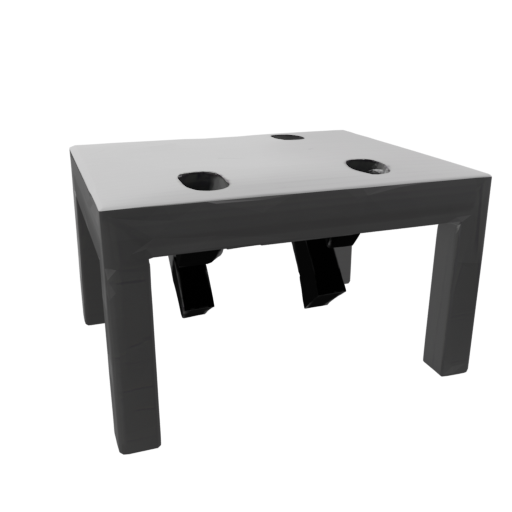} &
        \includegraphics[width=0.16\linewidth]{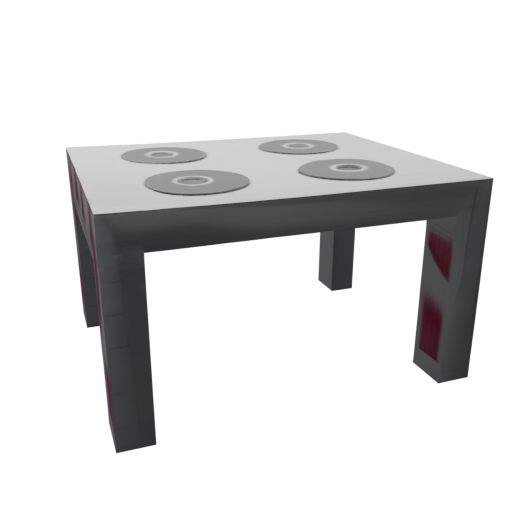} 
        \\

        \multicolumn{6}{c}{\editprompt{"The backrest of the chair is not squared."}} 
        \\

        \includegraphics[width=0.16\linewidth]{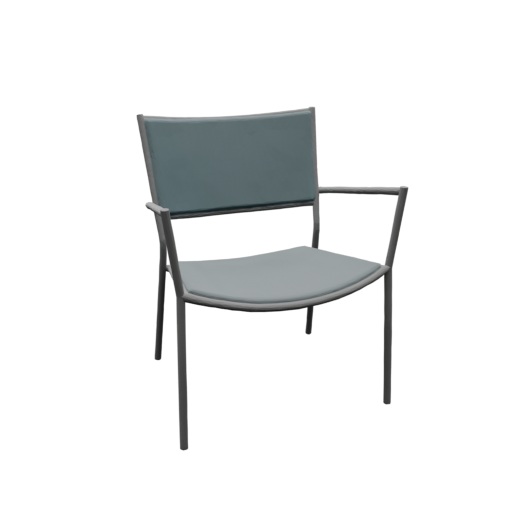} &
        \includegraphics[width=0.16\linewidth]{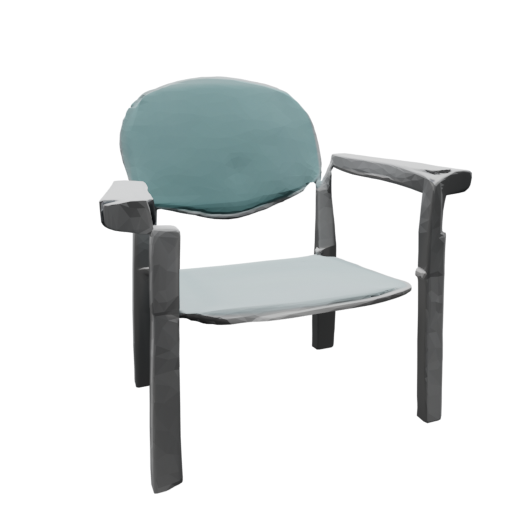} &
        \includegraphics[width=0.16\linewidth]{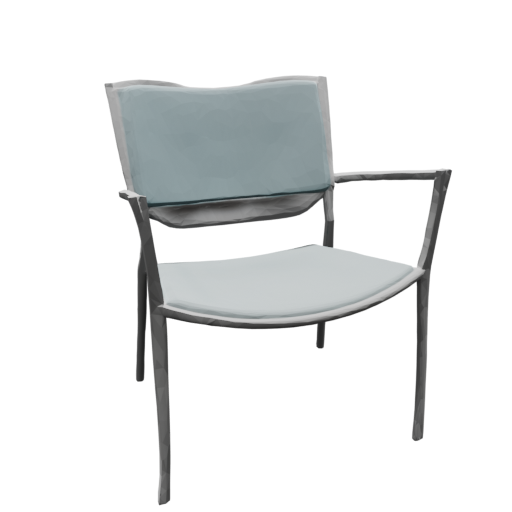} &
        \includegraphics[width=0.16\linewidth]{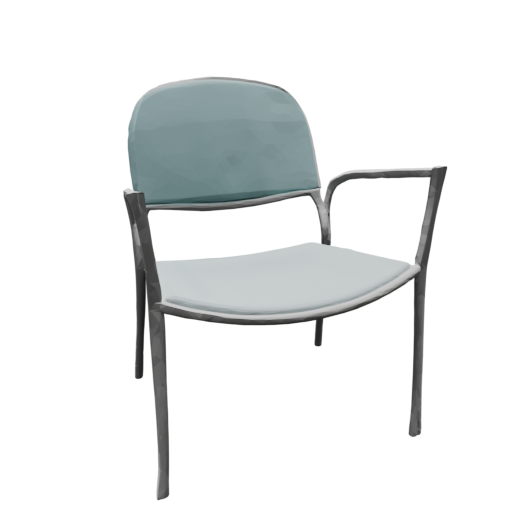} &
        \includegraphics[width=0.16\linewidth]{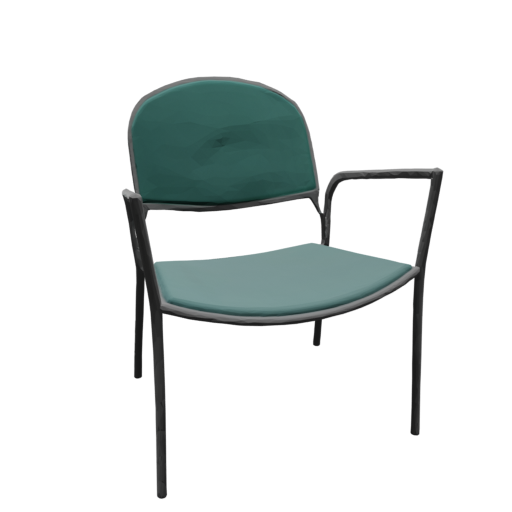} &
        \includegraphics[width=0.16\linewidth]{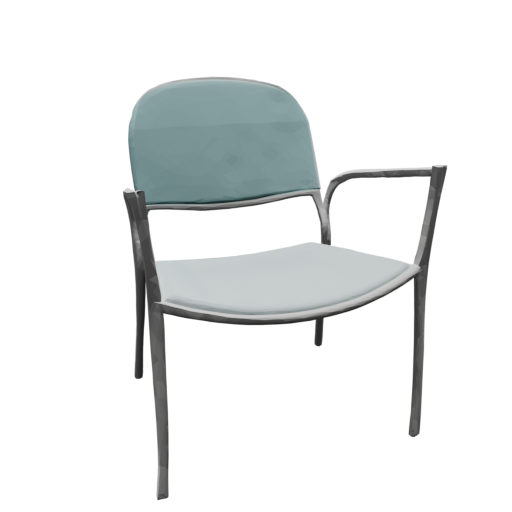} 
        \\

        \footnotesize{Input} &
        \footnotesize{$\mathcal{P}_{edit}$ only} &
        \footnotesize{w/o $\mathcal{P}_{edit}$} &
        \footnotesize{w/o $\mathcal{S}_{warp}$} &
        \footnotesize{w/o App} &
        \footnotesize{Ours} 
        \\[2mm] %

    \end{tabular}
    \label{tab:ablations}
    \vspace{-15pt}
    \caption{\textbf{Qualitative Ablation Results}. As detailed in Section \ref{subsec:abl}, we compare against several ablated variants of our model. Our full model best achieves fine-grained, identity-preserving edits, as illustrated above. }
    \vspace{-11pt}
    \label{fig:ablation_results}
    
\end{figure}

\subsection{Ablations}
\label{subsec:abl}

We conduct an ablation study analyzing the effect of the different inverted latents in our blending mechanism as well as our appearance refinement module. Specifically, we evaluate the following ablated model variants. (i) ``\textbf{$\mathcal{P}_{edit}$ only}'' uses only the edited proxy $\mathcal{P}_{edit}$ produced by the VLM as guidance in the denoising process.
(ii) ``\textbf{w/o $\mathcal{P}_{edit}$}'' excludes the edited proxy, conditioning solely on the original structure $\mathcal{S}_{org}$ and warped shape $\mathcal{S}_{warp}$.
(iii) ``\textbf{w/o $\mathcal{S}_{warp}$}'' omits the warped shape and uses only the original structure $\mathcal{S}_{org}$ and edited proxy $\mathcal{P}_{edit}$.
(iv) ``\textbf{w/o App}'' disables appearance refinement by using the original rendered image of the input object, rather than the edited image, along with TRELLIS's standard appearance flow model without any form of injection.

 Ablation results are reported in Table 3; several results are provided in Figure \ref{fig:ablation_results}. As illustrated in the table, all sources of inverted latents are important for achieving fine-grained, identity preserving edits. In particular, we can observe that indeed the ``$\mathcal{P}_{edit}$ only'' variant fails to preserve the identity of the original object, yielding significantly lower identity preservation scores in comparison to our approach. Conversely, ``w/o $\mathcal{P}_{edit}$'' achieves the best identity preservation, but at the cost of reduced edit fidelity, as reflected by the lower VQA score. This demonstrates that the proxy latents are essential for achieving precise edits. This is particularly important for insertion and deletion of new regions, as illustrated in Figure \ref{fig:ablation_results} (top row) where the lamp "nubs" cannot be added in this ablated variant. Without proxy latent injection, the denoising process is dominated by the input structure, making it difficult to introduce new parts or remove existing ones. As a result, the generated shapes remain overly similar to the input and often fail to fully realize the textual edit. Moreover, our appearance refinement module yields improvements across all metrics, underscoring the importance of handling appearance-based modifications in addition to structure.

Altogether, these experiments show that each component contributes to overall performance, yielding the most favorable trade-off among evaluated metrics.

\subsection{Limitations}
\label{subsec:limit}
While our framework demonstrates robust editing capabilities, its performance is contingent upon the granularity and semantic accuracy of the initial primitive decomposition. Since our VLM agent operates strictly on the proxy $\mathcal{P}_{orig}$, it can only manipulate geometry that is explicitly disentangled as a distinct primitive.

\begin{figure}[t]
    \centering
    
    \newlength{\imgheight}
    \setlength{\imgheight}{2.2cm} 
    
    {\editprompt{"The chair's back has no spindles"}} \par\vspace{1pt}
    
    \includegraphics[width=0.24\linewidth, height=\imgheight, trim=2cm 2cm 2cm 2cm, clip]{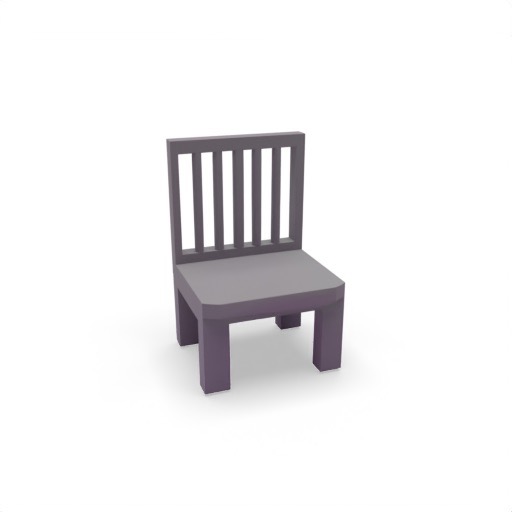}\hfill
    \includegraphics[width=0.24\linewidth, height=\imgheight, trim=2cm 2cm 2cm 2cm, clip]{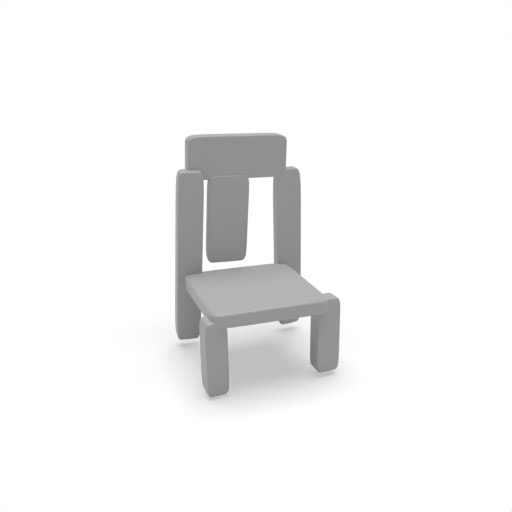}\hfill
    \includegraphics[width=0.24\linewidth, height=\imgheight, trim=2cm 2cm 2cm 2cm, clip]{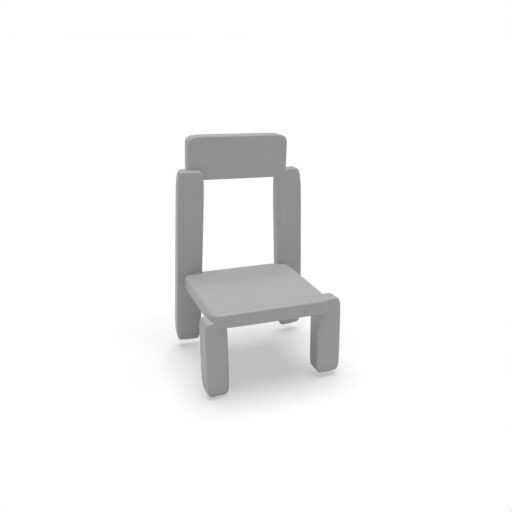}\hfill
    \includegraphics[width=0.24\linewidth, height=\imgheight, trim=2cm 2cm 2cm 2cm, clip]{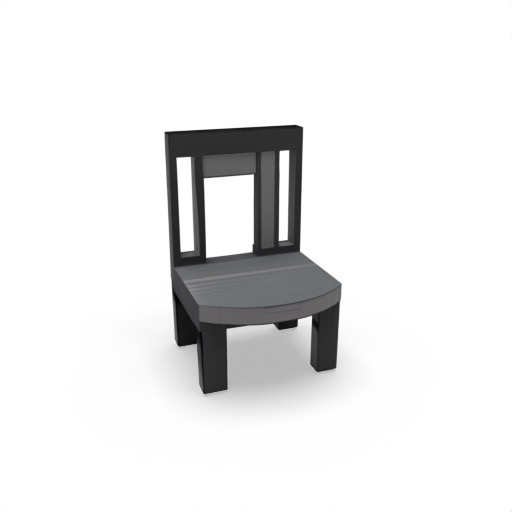}

    {\editprompt{"The lamp's shade is larger"}} \par\vspace{1pt}
    
     \includegraphics[width=0.24\linewidth, height=\imgheight, trim=4cm 4cm 4cm 4cm, clip]{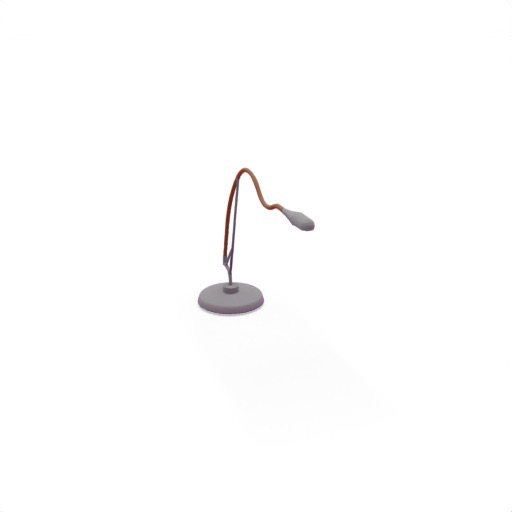}\hfill
    \includegraphics[width=0.24\linewidth, height=\imgheight, trim=4cm 4cm 4cm 4cm, clip]{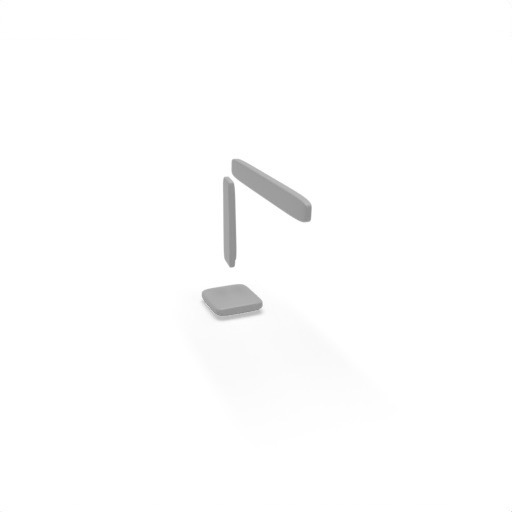}\hfill
    \includegraphics[width=0.24\linewidth, height=\imgheight, trim=4cm 4cm 4cm 4cm, clip]{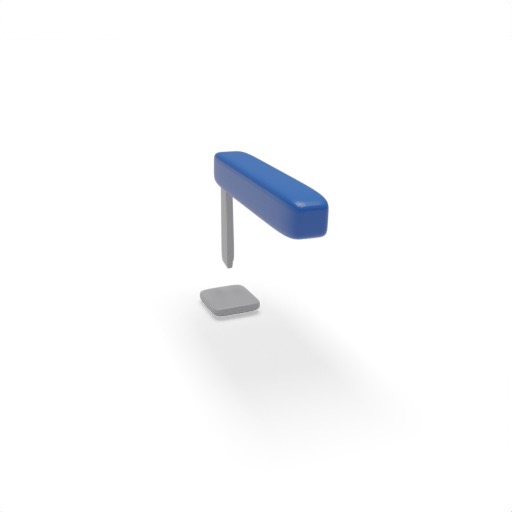}\hfill
    \includegraphics[width=0.24\linewidth, height=\imgheight, trim=4cm 4cm 4cm 4cm, clip]{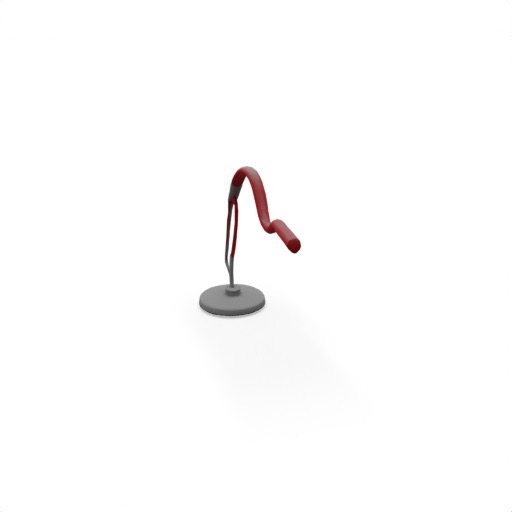}

    \makebox[0.24\linewidth][c]{\scriptsize Original Shape}\hfill
    \makebox[0.24\linewidth][c]{\scriptsize Original Proxy}\hfill
    \makebox[0.24\linewidth][c]{\scriptsize Edited Proxy}\hfill
    \makebox[0.24\linewidth][c]{\scriptsize Ours}
\vspace{-8pt}
    \caption{Limitation examples, illustrating how our method is constrained by the granularity and semantic accuracy of the initial primitive decomposition. %
    }
    \label{fig:limitations}
    \vspace{-8pt}
\end{figure}

When the abstraction module incorrectly merges distinct components into a single primitive, fine-grained control is lost. In Figure~\ref{fig:limitations} (top row), the algorithm failed to isolate all spindles of the chair's backrest—the central spindles were correctly segmented, but the side spindles were absorbed into the frame primitive. As a result, ``remove the spindles'' only partially succeeded. Similarly, in the bottom example the algorithm merged the lamp's handle and shade into one primitive, causing ``enlarge the shade'' to inadvertently distort the handle. However, our framework is agnostic to the decomposition backbone. As more expressive 3D decomposition methods emerge, our pipeline will directly benefit from improved granularity and semantic disentanglement without architectural modifications.

Additionally, our framework relies on strong spatial reasoning and instruction following capabilities from the editing agent as our abstraction editing task is highly demanding. In practice, we observe substantial performance differences across current VLMs, with only the most capable models reliably supporting the pipeline. Nevertheless, given the rapid progress in general-purpose VLMs, we expect our framework to naturally benefit from continued improvements in reasoning and instruction-following abilities.

\section{Conclusion}
\label{sec:conclusion}
We presented a training-free editing approach centered on a primitive-based geometric abstraction. This abstraction serves as a controllable proxy through which precise,  structural edits can be specified while preserving object identity. To translate these edits into high-quality 3D shapes, we proposed a novel denoising strategy that guides a 3D diffusion model using blended latent representations derived from both the input shape and the edited proxy. Altogether, our approach demonstrates how explicit geometric abstractions can bridge image and language-based reasoning and generative 3D models to support fine-grained shape editing. Looking forward, this paradigm opens new opportunities for scalable and controllable generation in more complex and dynamic 3D settings.

\begin{acks}
This work was supported by the Israel Science Foundation (grants no. 2492/20 and 1473/24), Len Blavatnik and the Blavatnik family foundation, and the U.S-Israel Binational Science Foundation (application no. 2022363).
\end{acks}
\clearpage
\bibliographystyle{ACM-Reference-Format}
\bibliography{main,blended}

@String(CVPR= {IEEE Conf. Comput. Vis. Pattern Recog.})

@String(TOG= {ACM Trans. Graph.})

@String(ICLR = {Int. Conf. Learn. Represent.})

@String(AAAI = {AAAI})

@String(CVPR  = {CVPR})

@String(TOG   = {ACM TOG})

@String(ICLR  = {ICLR})

@article{chen2022tango,
  title={Tango: Text-driven photorealistic and robust 3d stylization via lighting decomposition},
  author={Chen, Yongwei and Chen, Rui and Lei, Jiabao and Zhang, Yabin and Jia, Kui},
  journal={Advances in Neural Information Processing Systems},
  volume={35},
  pages={30923--30936},
  year={2022}
}

@article{jun2023shap,
  title={Shap-e: Generating conditional 3d implicit functions},
  author={Jun, Heewoo and Nichol, Alex},
  journal={arXiv preprint arXiv:2305.02463},
  year={2023}
}

@article{huang2022ladis,
  title={LADIS: Language disentanglement for 3D shape editing},
  author={Huang, Ian and Achlioptas, Panos and Zhang, Tianyi and Tulyakov, Sergey and Sung, Minhyuk and Guibas, Leonidas},
  journal={arXiv preprint arXiv:2212.05011},
  year={2022}
}

@article{zhuang2024tip,
  title={Tip-editor: An accurate 3d editor following both text-prompts and image-prompts},
  author={Zhuang, Jingyu and Kang, Di and Cao, Yan-Pei and Li, Guanbin and Lin, Liang and Shan, Ying},
  journal={ACM Transactions on Graphics (TOG)},
  volume={43},
  number={4},
  pages={1--12},
  year={2024},
  publisher={ACM New York, NY, USA}
}

@inproceedings{gao2023textdeformer,
  title={Textdeformer: Geometry manipulation using text guidance},
  author={Gao, William and Aigerman, Noam and Groueix, Thibault and Kim, Vova and Hanocka, Rana},
  booktitle={ACM SIGGRAPH 2023 Conference Proceedings},
  pages={1--11},
  year={2023}
}

@inproceedings{michel2022text2mesh,
  title={Text2mesh: Text-driven neural stylization for meshes},
  author={Michel, Oscar and Bar-On, Roi and Liu, Richard and Benaim, Sagie and Hanocka, Rana},
  booktitle={Proceedings of the IEEE/CVF Conference on Computer Vision and Pattern Recognition},
  pages={13492--13502},
  year={2022}
}

@article{liu2023exim,
  title={EXIM: A Hybrid Explicit-Implicit Representation for Text-Guided 3D Shape Generation},
  author={Liu, Zhengzhe and Hu, Jingyu and Hui, Ka-Hei and Qi, Xiaojuan and Cohen-Or, Daniel and Fu, Chi-Wing},
  journal={ACM Transactions on Graphics (TOG)},
  volume={42},
  number={6},
  pages={1--12},
  year={2023},
  publisher={ACM New York, NY, USA}
}

@inproceedings{hu2024cns,
  title={Cns-edit: 3d shape editing via coupled neural shape optimization},
  author={Hu, Jingyu and Hui, Ka-Hei and Liu, Zhengzhe and Zhang, Hao and Fu, Chi-Wing},
  booktitle={ACM SIGGRAPH 2024 Conference Papers},
  pages={1--12},
  year={2024}
}

@inproceedings{hao2020dualsdf,
  title={Dualsdf: Semantic shape manipulation using a two-level representation},
  author={Hao, Zekun and Averbuch-Elor, Hadar and Snavely, Noah and Belongie, Serge},
  booktitle={Proceedings of the IEEE/CVF Conference on Computer Vision and Pattern Recognition},
  pages={7631--7641},
  year={2020}
}

@inproceedings{qi2017pointnet,
  title={Pointnet: Deep learning on point sets for 3d classification and segmentation},
  author={Qi, Charles R and Su, Hao and Mo, Kaichun and Guibas, Leonidas J},
  booktitle={Proceedings of the IEEE conference on computer vision and pattern recognition},
  pages={652--660},
  year={2017}
}

@inproceedings{sella2023vox,
  title={Vox-e: Text-guided voxel editing of 3d objects},
  author={Sella, Etai and Fiebelman, Gal and Hedman, Peter and Averbuch-Elor, Hadar},
  booktitle={Proceedings of the IEEE/CVF International Conference on Computer Vision},
  pages={430--440},
  year={2023}
}

@article{poole2022dreamfusion,
  title={Dreamfusion: Text-to-3d using 2d diffusion},
  author={Poole, Ben and Jain, Ajay and Barron, Jonathan T and Mildenhall, Ben},
  journal={arXiv preprint arXiv:2209.14988},
  year={2022}
}

@article{heusel2017gans,
  title={Gans trained by a two time-scale update rule converge to a local nash equilibrium},
  author={Heusel, Martin and Ramsauer, Hubert and Unterthiner, Thomas and Nessler, Bernhard and Hochreiter, Sepp},
  journal={Advances in neural information processing systems},
  volume={30},
  year={2017}
}

@inproceedings{achlioptas2022changeit3d,
  title={ChangeIt3D: Languageassisted 3d shape edits and deformations},
  author={Achlioptas, Panos and Huang, Ian and Sung, Minhyuk and Tulyakov, Sergey and Guibas, Leonidas},
  booktitle={Conference on Computer Vision and Pattern Recognition (CVPR)},
  volume={2},
  number={5},
  pages={6},
  year={2022}
}

@inproceedings{shu20193d,
  title={3d point cloud generative adversarial network based on tree structured graph convolutions},
  author={Shu, Dong Wook and Park, Sung Woo and Kwon, Junseok},
  booktitle={Proceedings of the IEEE/CVF international conference on computer vision},
  pages={3859--3868},
  year={2019}
}

@inproceedings{sella2024spice,
  title={Spice{\textperiodcentered} E: Structural Priors in 3D Diffusion using Cross-Entity Attention},
  author={Sella, Etai and Fiebelman, Gal and Atia, Noam and Averbuch-Elor, Hadar},
  booktitle={ACM SIGGRAPH 2024 Conference Papers},
  pages={1--11},
  year={2024}
}

@inproceedings{xiang2025structured,
  title={Structured 3d latents for scalable and versatile 3d generation},
  author={Xiang, Jianfeng and Lv, Zelong and Xu, Sicheng and Deng, Yu and Wang, Ruicheng and Zhang, Bowen and Chen, Dong and Tong, Xin and Yang, Jiaolong},
  booktitle={Proceedings of the Computer Vision and Pattern Recognition Conference},
  pages={21469--21480},
  year={2025}
}

@misc{GoogleDeepMind2025NanoBanana,
  author = {{Google DeepMind}},
  title = {Introducing {Gemini} 2.5 {Flash Image}, our state-of-the-art image generation and editing model},
  howpublished = {\url{https://developers.googleblog.com/en/introducing-gemini-2-5-flash-image/}},
  month = aug,
  year = {2025},
  note = {Accessed: 2025-11-13}
}

@article{fedele2025superdec,
  title={Superdec: 3d scene decomposition with superquadric primitives},
  author={Fedele, Elisabetta and Sun, Boyang and Guibas, Leonidas and Pollefeys, Marc and Engelmann, Francis},
  journal={arXiv preprint arXiv:2504.00992},
  year={2025}
}

@inproceedings{fedele2026spacecontrol,
  title={SpaceControl: Introducing Test-Time Spatial Control to 3D Generative Modeling},
  author={Fedele, Elisabetta and Engelmann, Francis and Huang, Ian and Litany, Or and Pollefeys, Marc and Guibas, Leonidas},
  booktitle={International Conference on Learning Representations (ICLR)},
  year={2026}
}

@inproceedings{gilo2026instructmix2mix,
      title={InstructMix2Mix: Consistent Sparse-View Editing Through Multi-View Model Personalization}, 
      author={Daniel Gilo and Or Litany},
      booktitle={Proceedings of the IEEE/CVF Conference on Computer Vision and Pattern Recognition (CVPR)},
      year={2026}
}

@inproceedings{paschalidou2019superquadrics,
  title={Superquadrics revisited: Learning 3d shape parsing beyond cuboids},
  author={Paschalidou, Despoina and Ulusoy, Ali Osman and Geiger, Andreas},
  booktitle={Proceedings of the IEEE/CVF conference on computer vision and pattern recognition},
  pages={10344--10353},
  year={2019}
}

@article{li2025voxhammer,
  title={Voxhammer: Training-free precise and coherent 3d editing in native 3d space},
  author={Li, Lin and Huang, Zehuan and Feng, Haoran and Zhuang, Gengxiong and Chen, Rui and Guo, Chunchao and Sheng, Lu},
  journal={arXiv preprint arXiv:2508.19247},
  year={2025}
}

@article{ye2025nano3d,
  title={NANO3D: A Training-Free Approach for Efficient 3D Editing Without Masks},
  author={Ye, Junliang and Xie, Shenghao and Zhao, Ruowen and Wang, Zhengyi and Yan, Hongyu and Zu, Wenqiang and Ma, Lei and Zhu, Jun},
  journal={arXiv preprint arXiv:2510.15019},
  year={2025}
}

@article{xia2025towards,
  title={Towards Scalable and Consistent 3D Editing},
  author={Xia, Ruihao and Tang, Yang and Zhou, Pan},
  journal={arXiv preprint arXiv:2510.02994},
  year={2025}
}

@misc{labs2025flux1kontextflowmatching,
      title={FLUX.1 Kontext: Flow Matching for In-Context Image Generation and Editing in Latent Space}, 
      author={Black Forest Labs and Stephen Batifol and Andreas Blattmann and Frederic Boesel and Saksham Consul and Cyril Diagne and Tim Dockhorn and Jack English and Zion English and Patrick Esser and Sumith Kulal and Kyle Lacey and Yam Levi and Cheng Li and Dominik Lorenz and Jonas Müller and Dustin Podell and Robin Rombach and Harry Saini and Axel Sauer and Luke Smith},
      year={2025},
      eprint={2506.15742},
      archivePrefix={arXiv},
      primaryClass={cs.GR},
      url={https://arxiv.org/abs/2506.15742}, 
}

@inproceedings{sella2025blended,
  title={Blended Point Cloud Diffusion for Localized Text-guided Shape Editing},
  author={Sella, Etai and Atia, Noam and Mokady, Ron and Averbuch-Elor, Hadar},
  booktitle={Proceedings of the IEEE/CVF International Conference on Computer Vision},
  pages={19119--19129},
  year={2025}
}

@inproceedings{lin2024evaluating,
  title={Evaluating text-to-visual generation with image-to-text generation},
  author={Lin, Zhiqiu and Pathak, Deepak and Li, Baiqi and Li, Jiayao and Xia, Xide and Neubig, Graham and Zhang, Pengchuan and Ramanan, Deva},
  booktitle={European Conference on Computer Vision},
  pages={366--384},
  year={2024},
  organization={Springer}
}

@article{bar2025editp23,
  title={EditP23: 3D Editing via Propagation of Image Prompts to Multi-View},
  author={Bar-On, Roi and Cohen-Bar, Dana and Cohen-Or, Daniel},
  journal={arXiv preprint arXiv:2506.20652},
  year={2025}
}

@article{barr1981superquadrics,
  title={Superquadrics and angle-preserving transformations},
  author={Barr, Alan H},
  journal={IEEE Computer graphics and Applications},
  volume={1},
  number={01},
  pages={11--23},
  year={1981},
  publisher={IEEE Computer Society}
}

@inproceedings{pentland1986parts,
  title={Parts: Structured Descriptions of Shape.},
  author={Pentland, Alex},
  booktitle={AAAI},
  pages={695--701},
  year={1986}
}

@article{qi2016pointnet,
  title={PointNet: Deep Learning on Point Sets for 3D Classification and Segmentation},
  author={Qi, Charles R and Su, Hao and Mo, Kaichun and Guibas, Leonidas J},
  journal={arXiv preprint arXiv:1612.00593},
  year={2016}
}

@article{xu2024instantmesh,
  title={InstantMesh: Efficient 3D Mesh Generation from a Single Image with Sparse-view Large Reconstruction Models},
  author={Xu, Jiale and Cheng, Weihao and Gao, Yiming and Wang, Xintao and Gao, Shenghua and Shan, Ying},
  journal={arXiv preprint arXiv:2404.07191},
  year={2024}
}

@article{sensenova-si,
  title = {Scaling Spatial Intelligence with Multimodal Foundation Models},
  author = {Cai, Zhongang and Wang, Ruisi and Gu, Chenyang and Pu, Fanyi and Xu, Junxiang and Wang, Yubo and Yin, Wanqi and Yang, Zhitao and Wei, Chen and Sun, Qingping and Zhou, Tongxi and Li, Jiaqi and Pang, Hui En and Qian, Oscar and Wei, Yukun and Lin, Zhiqian and Shi, Xuanke and Deng, Kewang and Han, Xiaoyang and Chen, Zukai and Fan, Xiangyu and Deng, Hanming and Lu, Lewei and Pan, Liang and Li, Bo and Liu, Ziwei and Wang, Quan and Lin, Dahua and Yang, Lei},
  journal = {arXiv preprint arXiv:2511.13719},
  year = {2025}
}

@article{bai2025qwen2,
  title={Qwen2. 5-vl technical report},
  author={Bai, Shuai and Chen, Keqin and Liu, Xuejing and Wang, Jialin and Ge, Wenbin and Song, Sibo and Dang, Kai and Wang, Peng and Wang, Shijie and Tang, Jun and others},
  journal={arXiv preprint arXiv:2502.13923},
  year={2025}
}

@article{zhu2026survey,
    title={A survey on 3D editing based on NeRF and 3DGS},
    author={Zhu, Chen-Yang and Liu, Xin-Yao and Xu, Kai and Yi, Ren-Jiao},
    journal={Frontiers of Computer Science},
    year={2026},
    publisher={Springer}
}

@article{chao2023text,
  title={Text-guided image-and-shape editing and generation: A short survey},
  author={Chao, Cheng-Kang Ted and Gingold, Yotam},
  journal={arXiv preprint arXiv:2304.09244},
  year={2023}
}

@article{chung20243dstyleglip,
  title={3dstyleglip: Part-tailored text-guided 3d neural stylization},
  author={Chung, SeungJeh and Park, JooHyun and Kang, HyeongYeop},
  journal={arXiv preprint arXiv:2404.02634},
  year={2024}
}

@inproceedings{yang2025genvdm,
  title={GenVDM: Generating Vector Displacement Maps From a Single Image},
  author={Yang, Yuezhi and Chen, Qimin and Kim, Vladimir G and Chaudhuri, Siddhartha and Huang, Qixing and Chen, Zhiqin},
  booktitle={Proceedings of the Computer Vision and Pattern Recognition Conference},
  pages={26618--26629},
  year={2025}
}

@article{meng2025text2vdm,
  title={Text2VDM: Text to Vector Displacement Maps for Expressive and Interactive 3D Sculpting},
  author={Meng, Hengyu and Wang, Duotun and Shao, Zhijing and Liu, Ligang and Wang, Zeyu},
  journal={arXiv preprint arXiv:2502.20045},
  year={2025}
}

@inproceedings{haque2023instruct,
  title={Instruct-nerf2nerf: Editing 3d scenes with instructions},
  author={Haque, Ayaan and Tancik, Matthew and Efros, Alexei A and Holynski, Aleksander and Kanazawa, Angjoo},
  booktitle={Proceedings of the IEEE/CVF international conference on computer vision},
  pages={19740--19750},
  year={2023}
}

@inproceedings{chen2024dge,
  title={Dge: Direct gaussian 3d editing by consistent multi-view editing},
  author={Chen, Minghao and Laina, Iro and Vedaldi, Andrea},
  booktitle={European Conference on Computer Vision},
  pages={74--92},
  year={2024},
  organization={Springer}
}

@inproceedings{tulsiani2017learning,
  title={Learning shape abstractions by assembling volumetric primitives},
  author={Tulsiani, Shubham and Su, Hao and Guibas, Leonidas J and Efros, Alexei A and Malik, Jitendra},
  booktitle={Proceedings of the IEEE Conference on Computer Vision and Pattern Recognition},
  pages={2635--2643},
  year={2017}
}

@inproceedings{paschalidou2021neural,
  title={Neural parts: Learning expressive 3d shape abstractions with invertible neural networks},
  author={Paschalidou, Despoina and Katharopoulos, Angelos and Geiger, Andreas and Fidler, Sanja},
  booktitle={Proceedings of the IEEE/CVF Conference on Computer Vision and Pattern Recognition},
  pages={3204--3215},
  year={2021}
}

@inproceedings{held20253d,
  title={3D convex splatting: Radiance field rendering with 3D smooth convexes},
  author={Held, Jan and Vandeghen, Renaud and Hamdi, Abdullah and Deliege, Adrien and Cioppa, Anthony and Giancola, Silvio and Vedaldi, Andrea and Ghanem, Bernard and Van Droogenbroeck, Marc},
  booktitle={Proceedings of the Computer Vision and Pattern Recognition Conference},
  year={2025}
}

@article{govindarajan2025radiant,
  title={Radiant foam: Real-time differentiable ray tracing},
  author={Govindarajan, Shrisudhan and Rebain, Daniel and Yi, Kwang Moo and Tagliasacchi, Andrea},
  journal={arXiv preprint arXiv:2502.01157},
  year={2025}
}

@misc{ye2025primitiveanything,
    title={PrimitiveAnything: Human-Crafted 3D Primitive Assembly Generation with Auto-Regressive Transformer}, 
    author={Jingwen Ye and Yuze He and Yanning Zhou and Yiqin Zhu and Kaiwen Xiao and Yong-Jin Liu and Wei Yang and Xiao Han},
    year={2025},
    eprint={2505.04622},
    archivePrefix={arXiv},
    primaryClass={cs.GR}
}

@inproceedings{avetisyan2024scenescript,
  title={Scenescript: Reconstructing scenes with an autoregressive structured language model},
  author={Avetisyan, Armen and Xie, Christopher and Howard-Jenkins, Henry and Yang, Tsun-Yi and Aroudj, Samir and Patra, Suvam and Zhang, Fuyang and Frost, Duncan and Holland, Luke and Orme, Campbell and others},
  booktitle={European Conference on Computer Vision},
  pages={247--263},
  year={2024},
  organization={Springer}
}

@inproceedings{siddiqui2024meshgpt,
  title={Meshgpt: Generating triangle meshes with decoder-only transformers},
  author={Siddiqui, Yawar and Alliegro, Antonio and Artemov, Alexey and Tommasi, Tatiana and Sirigatti, Daniele and Rosov, Vladislav and Dai, Angela and Nie{\ss}ner, Matthias},
  booktitle={Proceedings of the IEEE/CVF conference on computer vision and pattern recognition},
  pages={19615--19625},
  year={2024}
}

@article{wang2024llama,
  title={Llama-mesh: Unifying 3d mesh generation with language models},
  author={Wang, Zhengyi and Lorraine, Jonathan and Wang, Yikai and Su, Hang and Zhu, Jun and Fidler, Sanja and Zeng, Xiaohui},
  journal={arXiv preprint arXiv:2411.09595},
  year={2024}
}

@inproceedings{fang2025meshllm,
  title={Meshllm: Empowering large language models to progressively understand and generate 3d mesh},
  author={Fang, Shuangkang and Shen, I and Wang, Yufeng and Tsai, Yi-Hsuan and Yang, Yi and Zhou, Shuchang and Ding, Wenrui and Igarashi, Takeo and Yang, Ming-Hsuan and others},
  booktitle={Proceedings of the IEEE/CVF International Conference on Computer Vision},
  pages={14061--14072},
  year={2025}
}

@inproceedings{achlioptas2023shapetalk,
  title={ShapeTalk: A language dataset and framework for 3d shape edits and deformations},
  author={Achlioptas, Panos and Huang, Ian and Sung, Minhyuk and Tulyakov, Sergey and Guibas, Leonidas},
  booktitle={Proceedings of the IEEE/CVF conference on computer vision and pattern recognition},
  pages={12685--12694},
  year={2023}
}

@inproceedings{sun20253d,
  title={3d-gpt: Procedural 3d modeling with large language models},
  author={Sun, Chunyi and Han, Junlin and Deng, Weijian and Wang, Xinlong and Qin, Zishan and Gould, Stephen},
  booktitle={2025 International Conference on 3D Vision (3DV)},
  pages={1253--1263},
  year={2025},
  organization={IEEE}
}

@article{lu2025ll3m,
  title={Ll3m: Large language 3d modelers},
  author={Lu, Sining and Chen, Guan and Dinh, Nam Anh and Lang, Itai and Holtzman, Ari and Hanocka, Rana},
  journal={arXiv preprint arXiv:2508.08228},
  year={2025}
}

@inproceedings{man2025videocad,
  title={VideoCAD: A Dataset and Model for Learning Long-Horizon 3D CAD UI Interactions from Video},
  author={Man, Brandon and Nehme, Ghadi and Alam, Md Ferdous and Ahmed, Faez},
  booktitle={The Thirty-ninth Annual Conference on Neural Information Processing Systems Datasets and Benchmarks Track},
  year={2025}
}

@article{palandra2024gsedit,
  title={Gsedit: Efficient text-guided editing of 3d objects via gaussian splatting},
  author={Palandra, Francesco and Sanchietti, Andrea and Baieri, Daniele and Rodola, Emanuele},
  journal={arXiv preprint arXiv:2403.05154},
  year={2024}
}

@inproceedings{chen2023fantasia3d,
  title={Fantasia3d: Disentangling geometry and appearance for high-quality text-to-3d content creation},
  author={Chen, Rui and Chen, Yongwei and Jiao, Ningxin and Jia, Kui},
  booktitle={Proceedings of the IEEE/CVF international conference on computer vision},
  pages={22246--22256},
  year={2023}
}

@inproceedings{zhuang2023dreameditor,
  title={Dreameditor: Text-driven 3d scene editing with neural fields},
  author={Zhuang, Jingyu and Wang, Chen and Lin, Liang and Liu, Lingjie and Li, Guanbin},
  booktitle={SIGGRAPH Asia 2023 Conference Papers},
  pages={1--10},
  year={2023}
}

@inproceedings{edelstein2025sharp,
  title={Sharp-It: A Multi-view to Multi-view Diffusion Model for 3D Synthesis and Manipulation},
  author={Edelstein, Yiftach and Patashnik, Or and Cohen-Bar, Dana and Zelnik-Manor, Lihi},
  booktitle={Proceedings of the Computer Vision and Pattern Recognition Conference},
  pages={21458--21468},
  year={2025}
}

@inproceedings{chen2024shap,
  title={Shap-editor: Instruction-guided latent 3d editing in seconds},
  author={Chen, Minghao and Xie, Junyu and Laina, Iro and Vedaldi, Andrea},
  booktitle={Proceedings of the IEEE/CVF conference on computer vision and pattern recognition},
  pages={26456--26466},
  year={2024}
}

@misc{barda2025instant3dit,
      title={Instant3dit: Multiview Inpainting for Fast Editing of 3D Objects}, 
      author={Amir Barda and Matheus Gadelha and Vladimir G. Kim and Noam Aigerman and Amit H. Bermano and Thibault Groueix},
      journal = {IEEE Conference on Computer Vision and Pattern Recognition (CVPR)},
      year = {2025}, 
}

@InProceedings{weber2024nerfiller,
    author    = {Weber, Ethan and Holynski, Aleksander and Jampani, Varun and Saxena, Saurabh and Snavely, Noah and Kar, Abhishek and Kanazawa, Angjoo},
    title     = {NeRFiller: Completing Scenes via Generative 3D Inpainting},
    booktitle = {Proceedings of the IEEE/CVF Conference on Computer Vision and Pattern Recognition (CVPR)},
    month     = {June},
    year      = {2024},
    pages     = {20731-20741}
}

@InProceedings{erkoc2025preditor3d,
    author    = {Erko\c{c}, Ziya and G\"umeli, Can and Wang, Chaoyang and Nie{\ss}ner, Matthias and Dai, Angela and Wonka, Peter and Lee, Hsin-Ying and Zhuang, Peiye},
    title     = {PrEditor3D: Fast and Precise 3D Shape Editing},
    booktitle = {Proceedings of the Computer Vision and Pattern Recognition Conference (CVPR)},
    month     = {June},
    year      = {2025},
    pages     = {640-649}
}

@misc{kathare2025,
      title={Instructive3D: Editing Large Reconstruction Models with Text Instructions}, 
      author={Kunal Kathare and Ankit Dhiman and K Vikas Gowda and Siddharth Aravindan and Shubham Monga and Basavaraja Shanthappa Vandrotti and Lokesh R Boregowda},
      year={2025},
      eprint={2501.04374},
      archivePrefix={arXiv},
      primaryClass={cs.CV},
      url={https://arxiv.org/abs/2501.04374}, 
}

@misc{bilecen2025,
      title={Reference-Based 3D-Aware Image Editing with Triplanes}, 
      author={Bahri Batuhan Bilecen and Yigit Yalin and Ning Yu and Aysegul Dundar},
      year={2024},
      eprint={2404.03632},
      archivePrefix={arXiv},
      primaryClass={cs.CV}
}

@article{nichol2022point,
  title={Point-e: A system for generating 3d point clouds from complex prompts},
  author={Nichol, Alex and Jun, Heewoo and Dhariwal, Prafulla and Mishkin, Pamela and Chen, Mark},
  journal={arXiv preprint arXiv:2212.08751},
  year={2022}
}
\clearpage
\begin{figure*}[t]
    \centering
    \setlength{\tabcolsep}{0pt} 
    
    \begin{tabular*}{\linewidth}{@{\extracolsep{\fill}}cccccccc}
        
        \multicolumn{8}{c}{\editprompt{``The target has a shorter shade''}}
        \vspace{6pt}
        \\
        
        \includegraphics[width=0.122\linewidth, trim=4.3cm 2.0cm 4.3cm 4.0cm, clip]{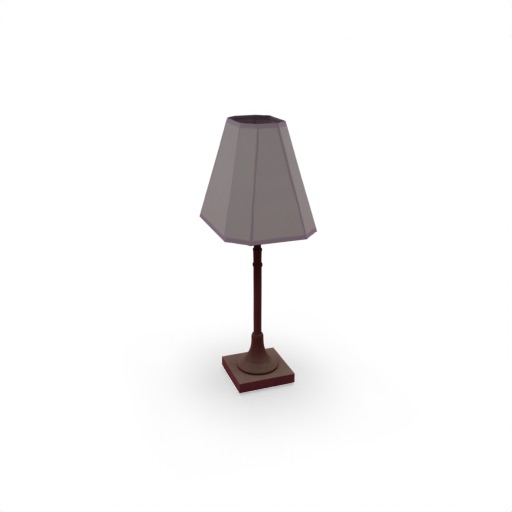} &
        \includegraphics[width=0.122\linewidth, trim=6.8cm 4.5cm 6.8cm 6.5cm, clip]{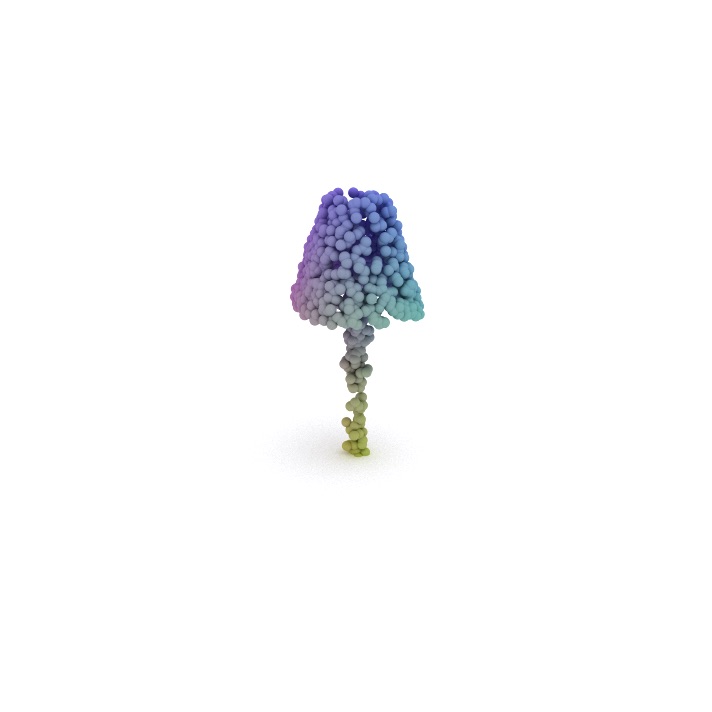} &
        \includegraphics[width=0.122\linewidth, trim=6.8cm 4.5cm 6.8cm 6.5cm, clip]{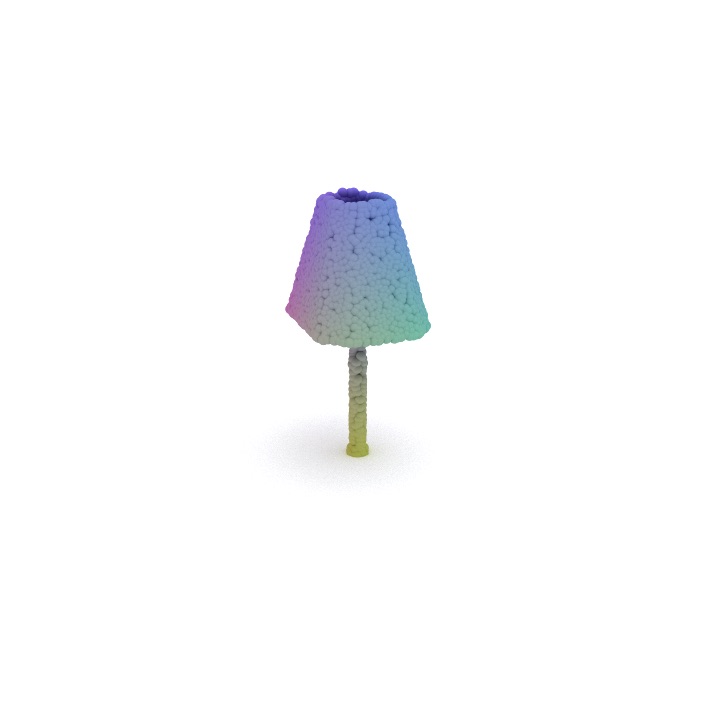} &
        \includegraphics[width=0.122\linewidth, trim=4.3cm 2.0cm 4.3cm 4.0cm, clip]{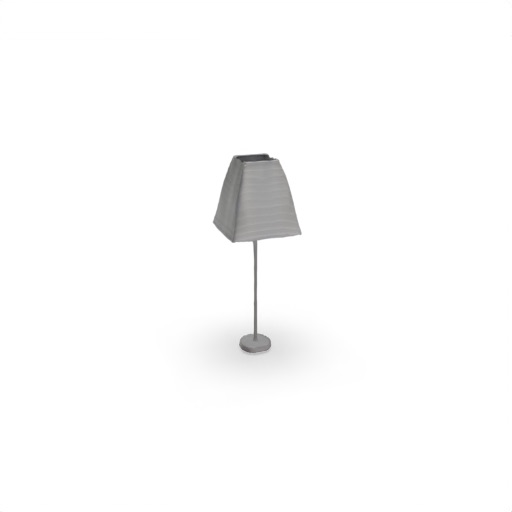} &
        
        \includegraphics[width=0.122\linewidth, trim=4.3cm 2.0cm 4.3cm 4.0cm, clip]{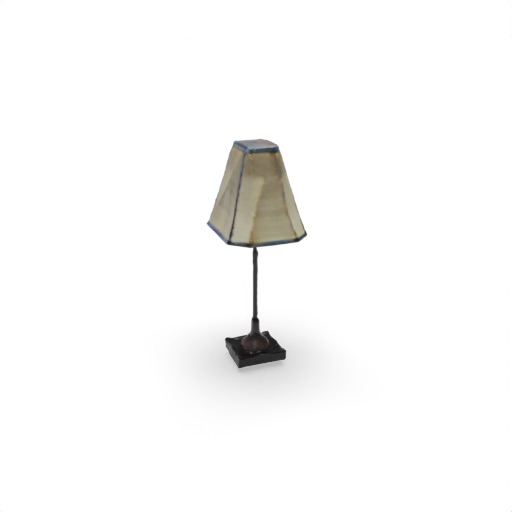} & 
        \includegraphics[width=0.122\linewidth, trim=4.3cm 2.0cm 4.3cm 4.0cm, clip]{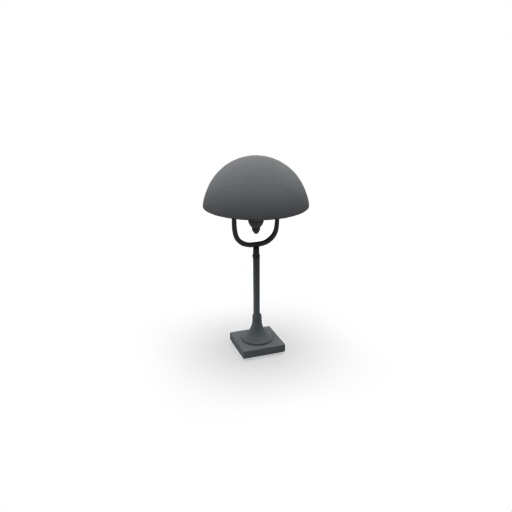} & 
        \includegraphics[width=0.122\linewidth, trim=4.3cm 2.0cm 4.3cm 4.0cm, clip]{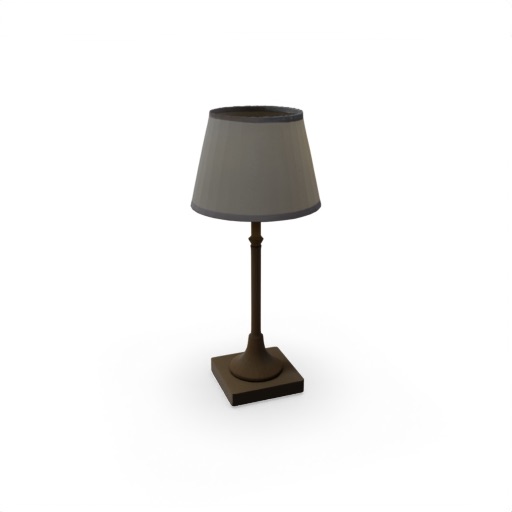} & 
        \includegraphics[width=0.122\linewidth, trim=4.3cm 2.0cm 4.3cm 4.0cm, clip]{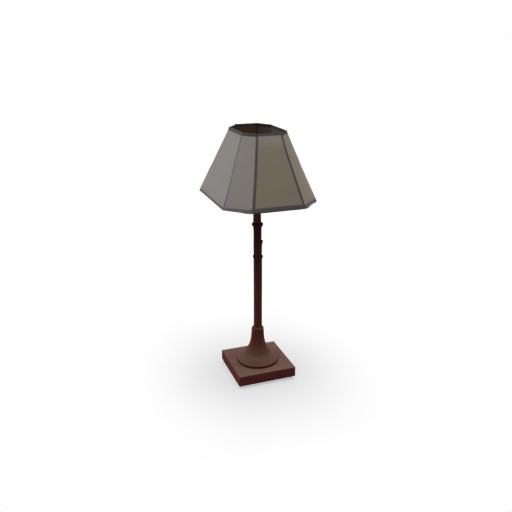} 
        \vspace{2pt}
        \\

        \multicolumn{8}{c}{\editprompt{``The target has a footrest''}}
        \vspace{6pt}
        \\
        
        \includegraphics[width=0.122\linewidth, trim=4.3cm 2.0cm 4.3cm 4.0cm, clip]{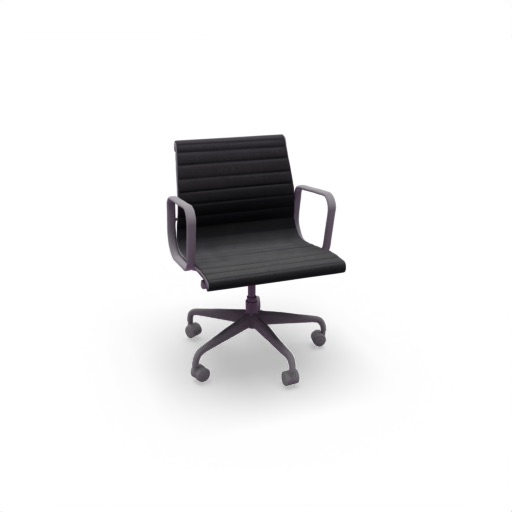} &
        \includegraphics[width=0.122\linewidth, trim=6.8cm 4.5cm 6.8cm 6.5cm, clip]{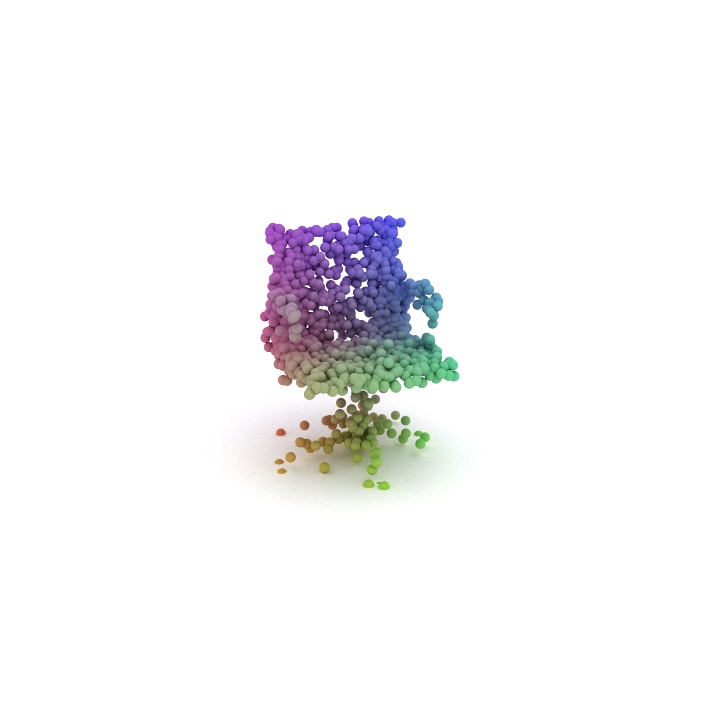} &
        \includegraphics[width=0.122\linewidth, trim=6.8cm 4.5cm 6.8cm 6.5cm, clip]{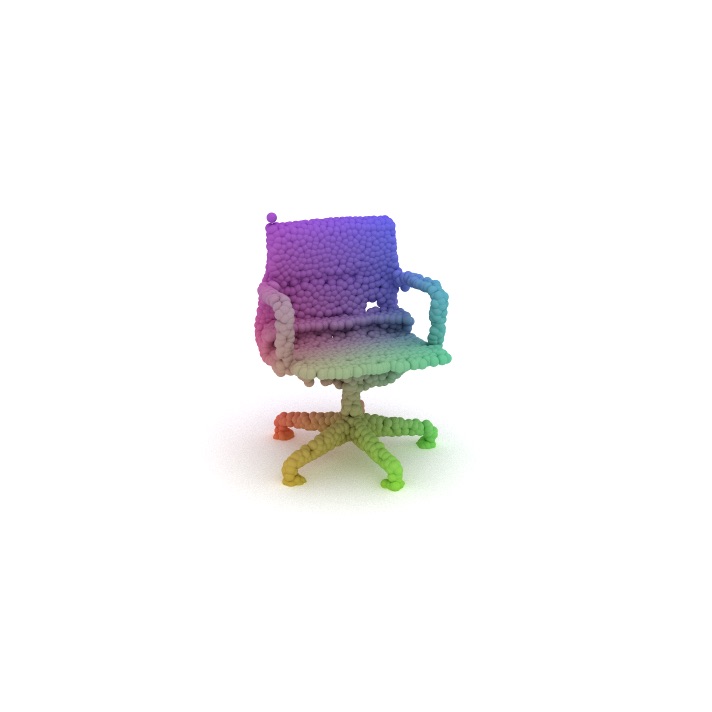} &
        \includegraphics[width=0.122\linewidth, trim=4.3cm 2.0cm 4.3cm 4.0cm, clip]{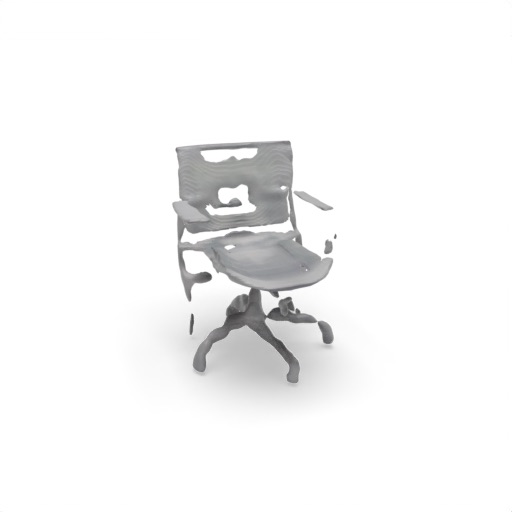} &
        
        \includegraphics[width=0.122\linewidth, trim=4.3cm 2.0cm 4.3cm 4.0cm, clip]{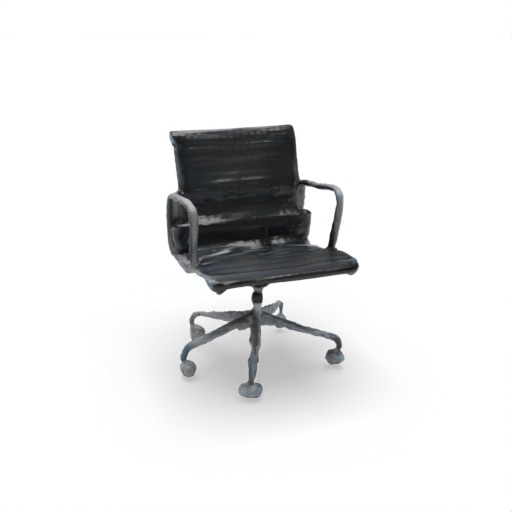} & 
        \includegraphics[width=0.122\linewidth, trim=4.3cm 2.0cm 4.3cm 4.0cm, clip]{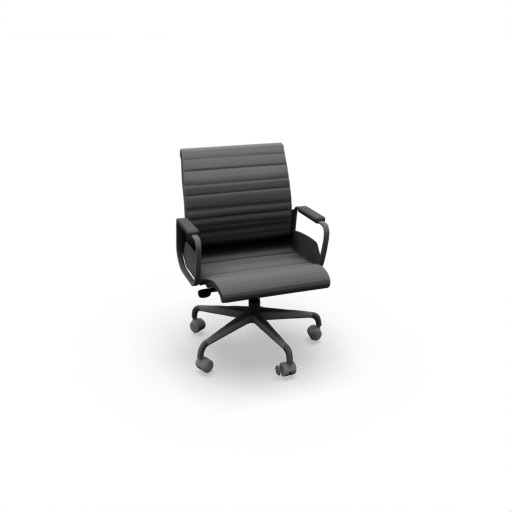} & 
        \includegraphics[width=0.122\linewidth, trim=4.3cm 2.0cm 4.3cm 4.0cm, clip]{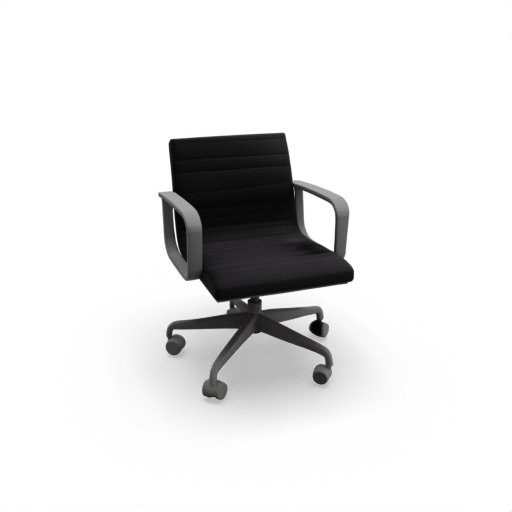} & 
        \includegraphics[width=0.122\linewidth, trim=4.3cm 2.0cm 4.3cm 4.0cm, clip]{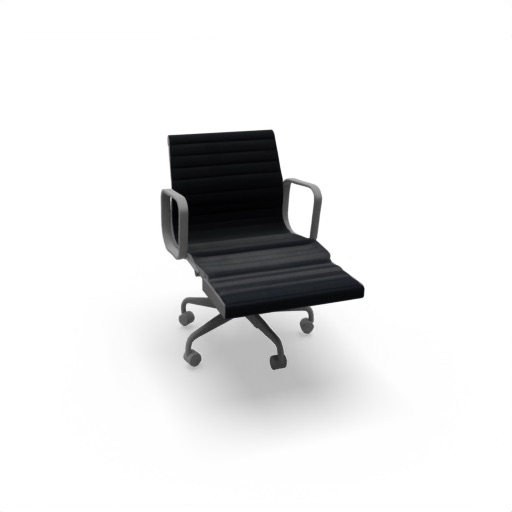} 
        \vspace{2pt}
        \\

        \multicolumn{8}{c}{\editprompt{``The target table is shorter''}}
        \\
        
        \includegraphics[width=0.122\linewidth, trim=3.0cm 2.0cm 3.0cm 4.0cm, clip]{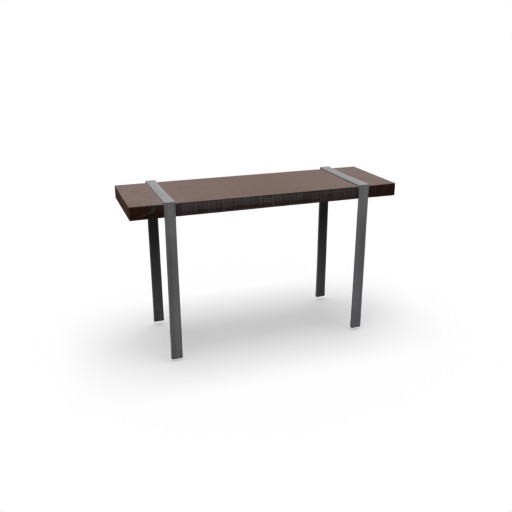} &
        \includegraphics[width=0.122\linewidth, trim=5.0cm 4.0cm 5.0cm 6.0cm, clip]{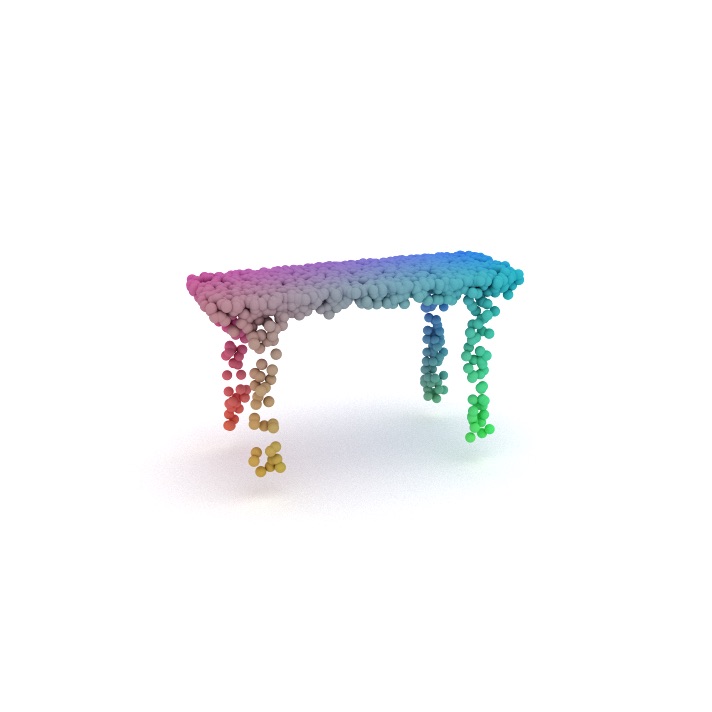} &
        \includegraphics[width=0.122\linewidth, trim=5.0cm 4.0cm 5.0cm 6.0cm, clip]{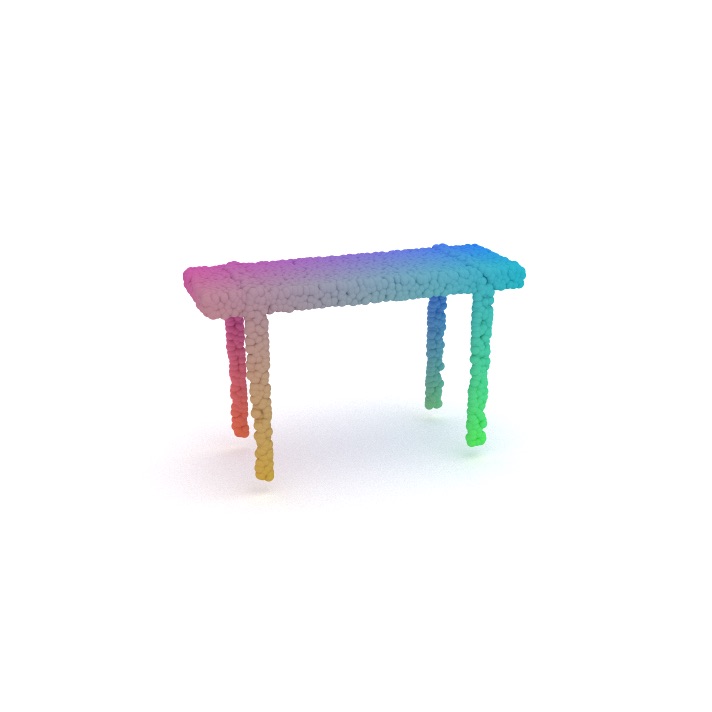} &
        \includegraphics[width=0.122\linewidth, trim=3.0cm 2.0cm 3.0cm 4.0cm, clip]{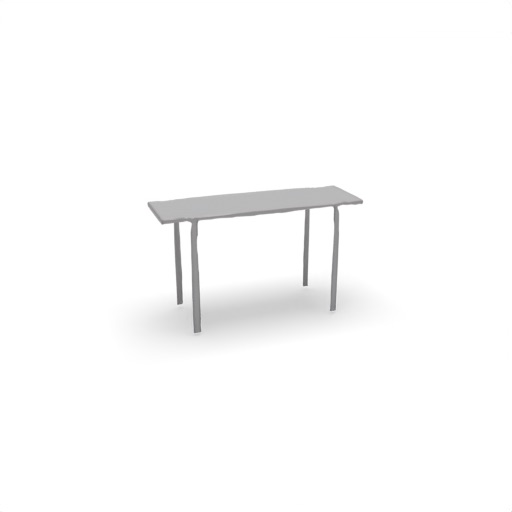} &
        
        \includegraphics[width=0.122\linewidth, trim=3.0cm 2.0cm 3.0cm 4.0cm, clip]{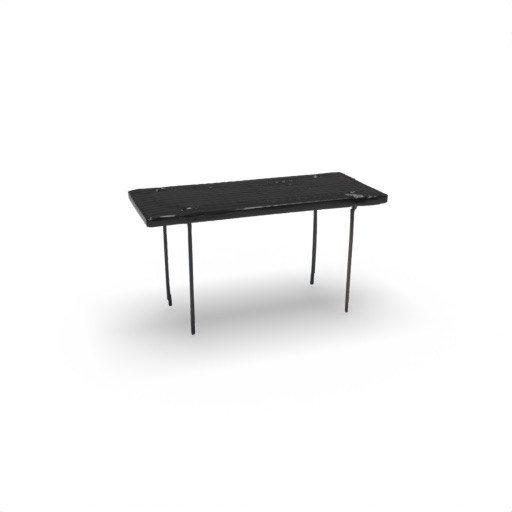} & 
        \includegraphics[width=0.122\linewidth, trim=3.0cm 2.0cm 3.0cm 4.0cm, clip]{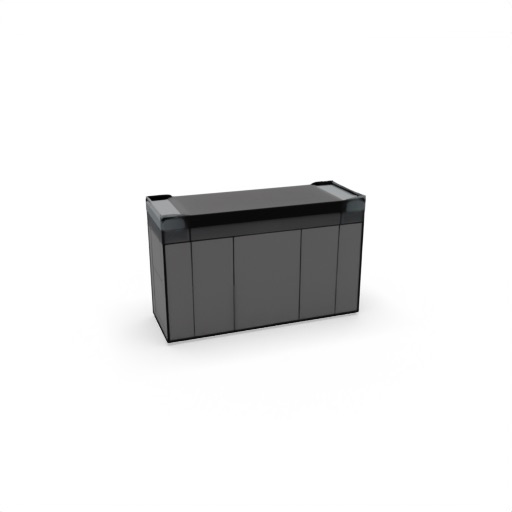} & 
        \includegraphics[width=0.122\linewidth, trim=3.0cm 2.0cm 3.0cm 4.0cm, clip]{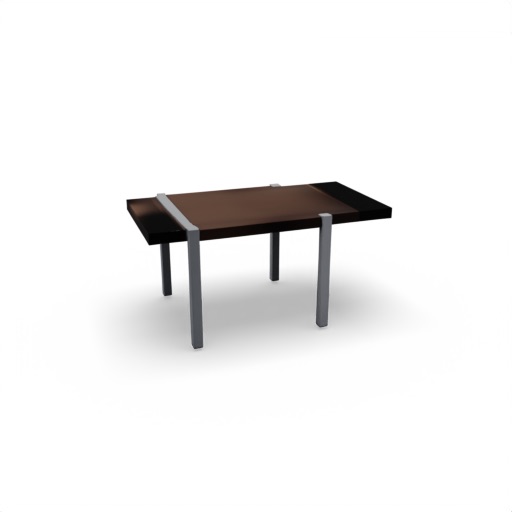} & 
        \includegraphics[width=0.122\linewidth, trim=3.0cm 2.0cm 3.0cm 4.0cm, clip]{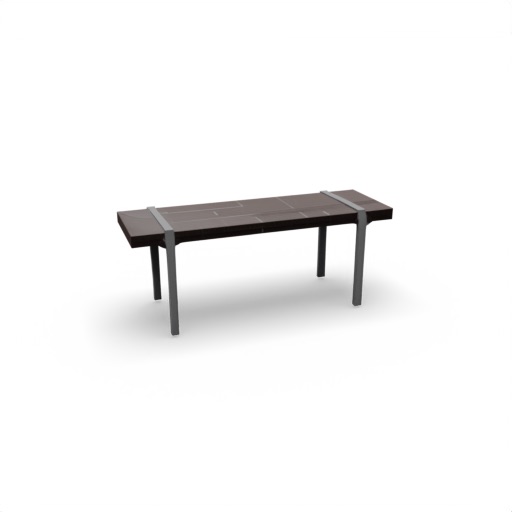} 
        \vspace{-6pt}
        \\
        \begin{tabular}{@{}c@{}} \footnotesize Input \end{tabular} & 
        \begin{tabular}{@{}c@{}} \footnotesize ChangeIt3D \end{tabular} &  
        \begin{tabular}{@{}c@{}} \footnotesize BlendedPC \end{tabular} & 
        \begin{tabular}{@{}c@{}} \footnotesize Spice-E \end{tabular} & 
        \begin{tabular}{@{}c@{}} \footnotesize EditP23 \end{tabular} & 
        \begin{tabular}{@{}c@{}} \footnotesize VoxHammer \end{tabular} & 
        \begin{tabular}{@{}c@{}} \footnotesize TRELLIS \end{tabular} & 
        \begin{tabular}{@{}c@{}} \footnotesize Ours \end{tabular} \\

    \end{tabular*}
    \vspace{-5pt}
    \caption{\new{\textbf{Qualitative comparisons on the ShapeTalk \cite{achlioptas2023shapetalk} benchmark.}} We compare our method against training based 3D editors such as ChangeIt3D \cite{achlioptas2022changeit3d} BlendedPC \cite{sella2025blended} and Spice-E \cite{sella2024spice}, single view editing based baselines such as VoxHammer \cite{li2025voxhammer} and TRELLIS \cite{xiang2025structured} (with FLUX Kontext \cite{labs2025flux1kontextflowmatching} edited image inputs) as well as the multi view editing based method EditP23 \cite{bar2025editp23}. }
    \label{fig:comp}
\end{figure*}

\begin{figure*}[t]
    \centering
    \setlength{\tabcolsep}{0pt} 
    
    \begin{tabular*}{\linewidth}{@{\extracolsep{\fill}}cccc c cccc}
        
        \multicolumn{4}{c}{\editprompt{``Make the tail longer''}} & & 
        \multicolumn{4}{c}{\editprompt{``Make it wear a red hat''}}
        \\
        
        \includegraphics[width=0.124\linewidth, trim=4.3cm 3.5cm 4.3cm 3.5cm, clip]{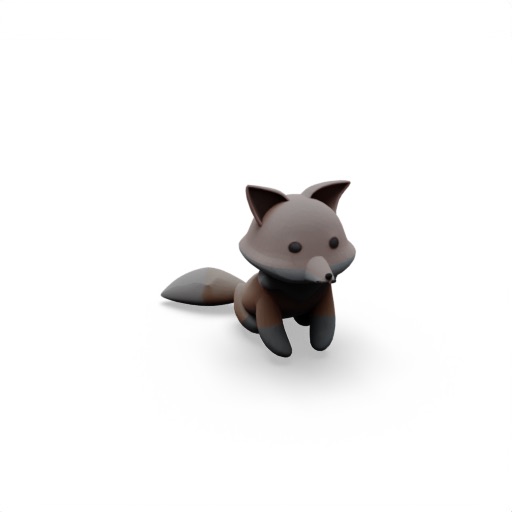} &
        \includegraphics[width=0.124\linewidth, trim=4.3cm 3.5cm 4.3cm 3.5cm, clip]{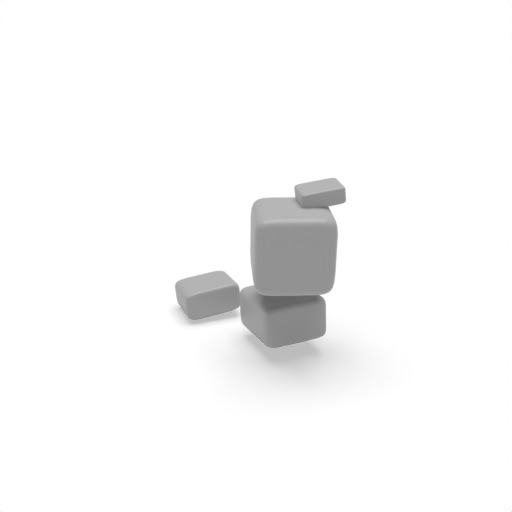} &
        \includegraphics[width=0.124\linewidth, trim=4.3cm 3.5cm 4.3cm 3.5cm, clip]{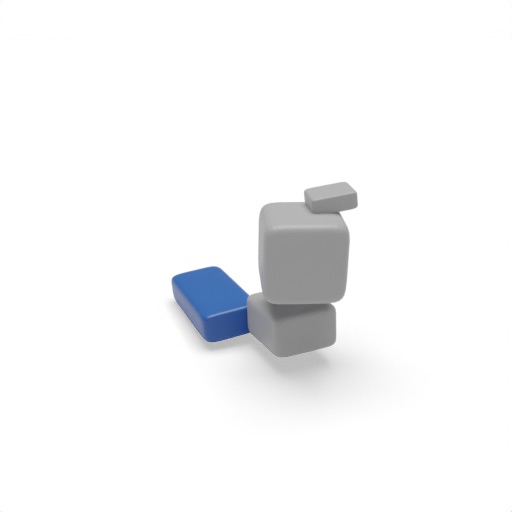} &
        \includegraphics[width=0.124\linewidth, trim=4.3cm 3.5cm 4.3cm 3.5cm, clip]{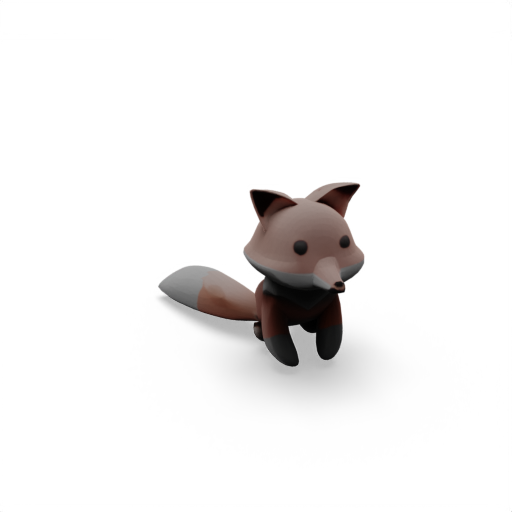} &
        
        \hspace{2pt} & %
        
        \includegraphics[width=0.124\linewidth, trim=4.3cm 3.5cm 4.3cm 3.5cm, clip]{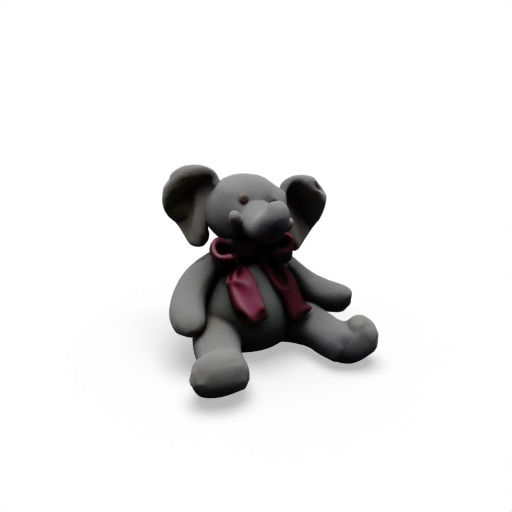} & 
        \includegraphics[width=0.124\linewidth, trim=4.3cm 3.5cm 4.3cm 3.5cm, clip]{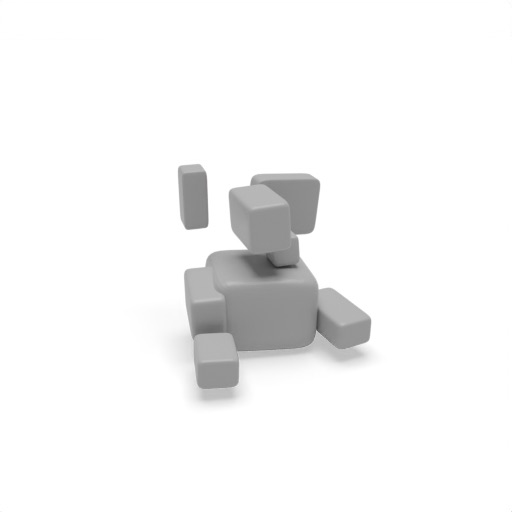} & 
        \includegraphics[width=0.124\linewidth, trim=4.3cm 3.5cm 4.3cm 3.5cm, clip]{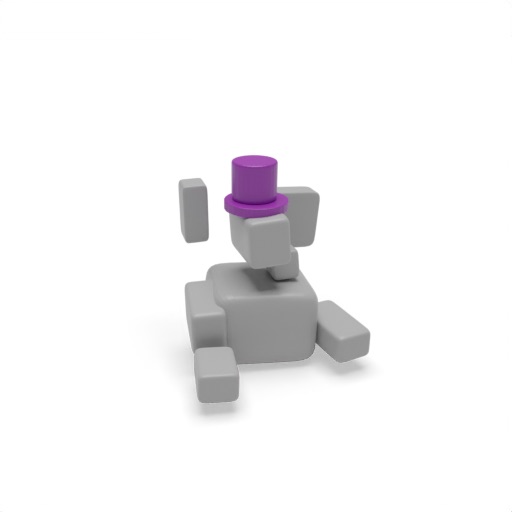} & 
        \includegraphics[width=0.124\linewidth, trim=4.3cm 3.5cm 4.3cm 3.5cm, clip]{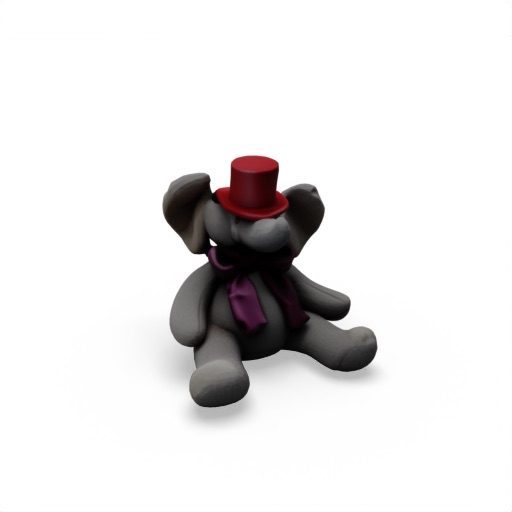} 
        \vspace{7pt}
        \\

        \multicolumn{4}{c}{\editprompt{``Make the pot cubic''}} & & 
        \multicolumn{4}{c}{\editprompt{``Make it into a windmill''}} 
        \\
        
        \includegraphics[width=0.124\linewidth, trim=4.3cm 3.5cm 4.3cm 3.5cm, clip]{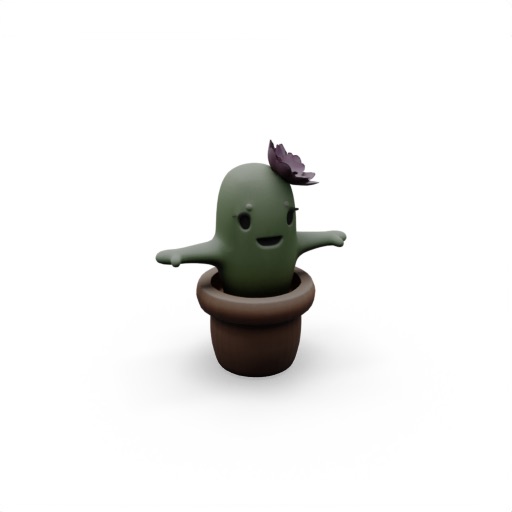} &
        \includegraphics[width=0.124\linewidth, trim=4.3cm 3.5cm 4.3cm 3.5cm, clip]{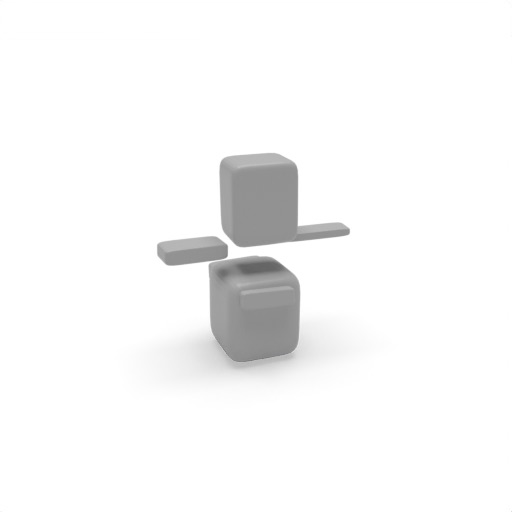} &
        \includegraphics[width=0.124\linewidth, trim=4.3cm 3.5cm 4.3cm 3.5cm, clip]{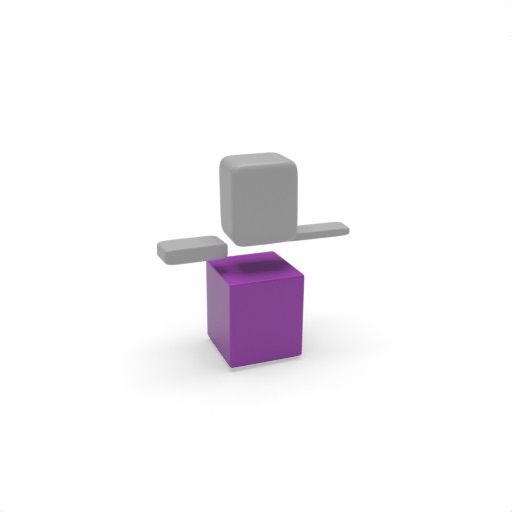} &
        \includegraphics[width=0.124\linewidth, trim=4.3cm 3.5cm 4.3cm 3.5cm, clip]{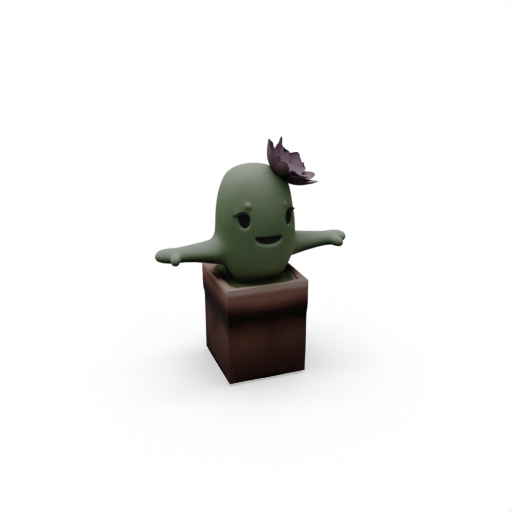} &
        
        \hspace{2pt} & %
        
        \includegraphics[width=0.124\linewidth, trim=4.3cm 3.5cm 4.3cm 3.5cm, clip]{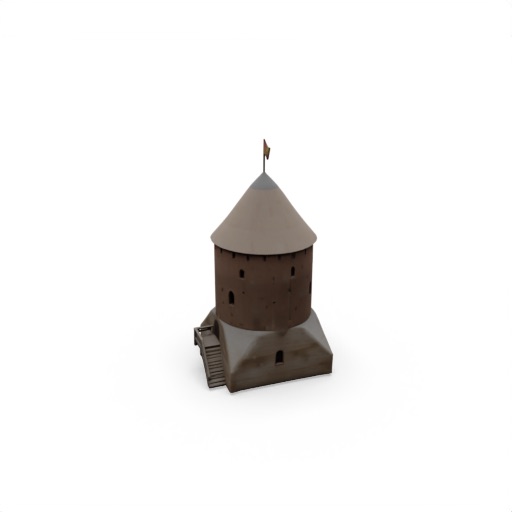} & 
        \includegraphics[width=0.124\linewidth, trim=4.3cm 3.5cm 4.3cm 3.5cm, clip]{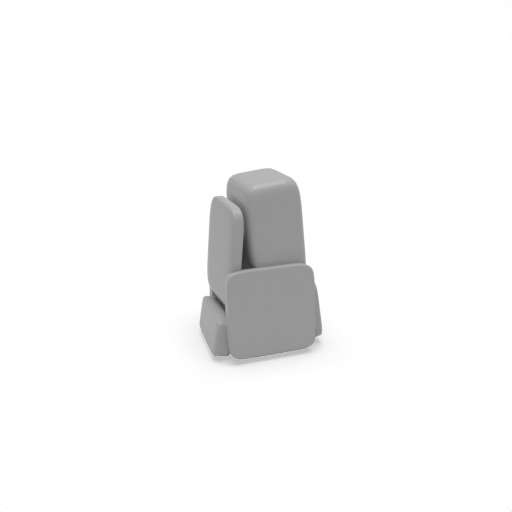} & 
        \includegraphics[width=0.124\linewidth, trim=4.3cm 3.5cm 4.3cm 3.5cm, clip]{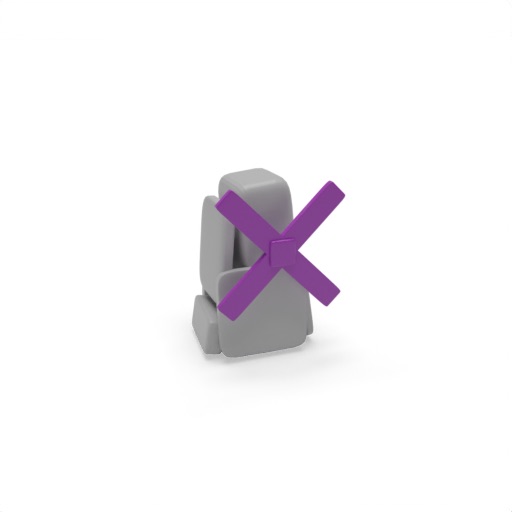} & 
        \includegraphics[width=0.124\linewidth, trim=4.3cm 3.5cm 4.3cm 3.5cm, clip]{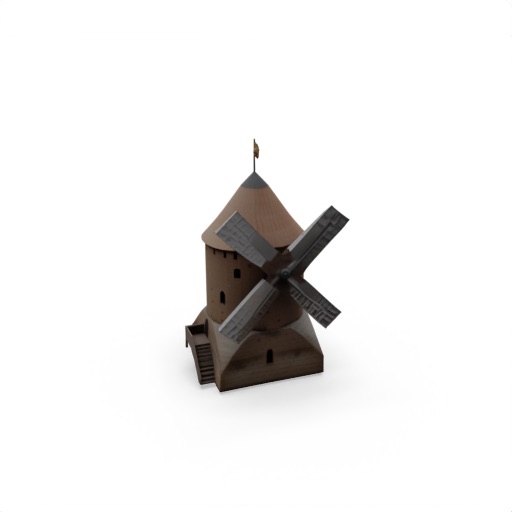} \\
        
        \begin{tabular}{@{}c@{}} \footnotesize Input \end{tabular} & 
        \begin{tabular}{@{}c@{}} \footnotesize Original Proxy \end{tabular} & 
        \begin{tabular}{@{}c@{}} \footnotesize Edited  Proxy \end{tabular} & 
        \begin{tabular}{@{}c@{}} \footnotesize Ours \end{tabular} & &
        \begin{tabular}{@{}c@{}} \footnotesize Input \end{tabular} & 
        \begin{tabular}{@{}c@{}} \footnotesize Original  Proxy \end{tabular} & 
        \begin{tabular}{@{}c@{}} \footnotesize Edited  Proxy \end{tabular} & 
        \begin{tabular}{@{}c@{}} \footnotesize Ours \end{tabular} \\[5mm]

    \end{tabular*}
    \vspace{-15pt}
    \caption{\new{\textbf{Qualitative results from the Edit3Dbench benchmark}}. In addition to the input shape and our method's output we also present the original and edited proxy shapes (edited primitives shown in \textcolor{blue}{blue}, added ones shown in \textcolor{violet}{purple}). Note that that the ``elephant'' and ``windmill'' examples require both structural and appearance modifications (\emph{e.g.}, generating the elephant's hat and then painting it red).}
    \label{fig:results_edit3dbench}
\end{figure*}

\begin{figure*}[t]
    \centering
    \setlength{\tabcolsep}{1.5pt} 
    
    \begin{tabular}{cccc c cccc}
        
        \multicolumn{4}{c}{\editprompt{"The legs of the chair are spaced further apart."}} 
        & & 
        \multicolumn{4}{c}{\editprompt{"The chair's backrest is curved."}} 
        \\
        
        \includegraphics[width=0.118\linewidth, trim=2cm 2cm 2cm 2cm, clip]{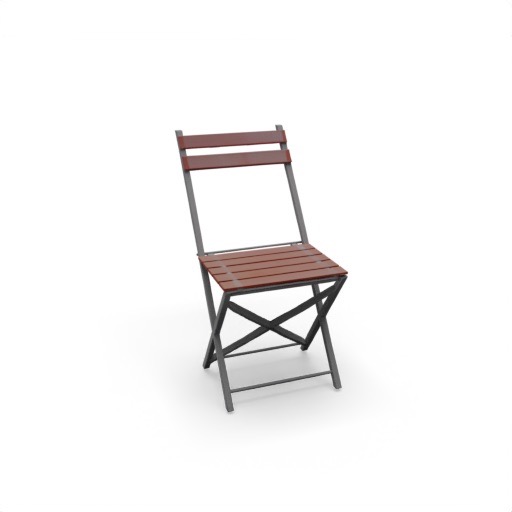} &
        \includegraphics[width=0.118\linewidth, trim=2cm 2cm 2cm 2cm, clip]{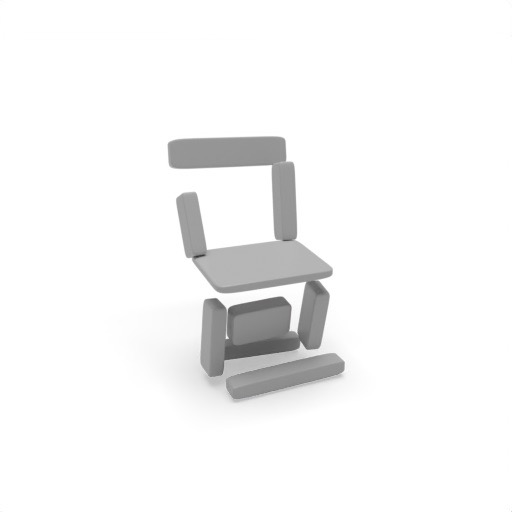} &
        \includegraphics[width=0.118\linewidth, trim=2cm 2cm 2cm 2cm, clip]{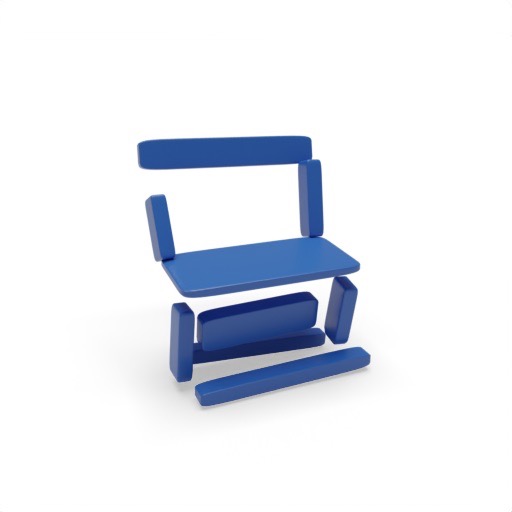} &
        \includegraphics[width=0.118\linewidth, trim=2cm 2cm 2cm 2cm, clip]{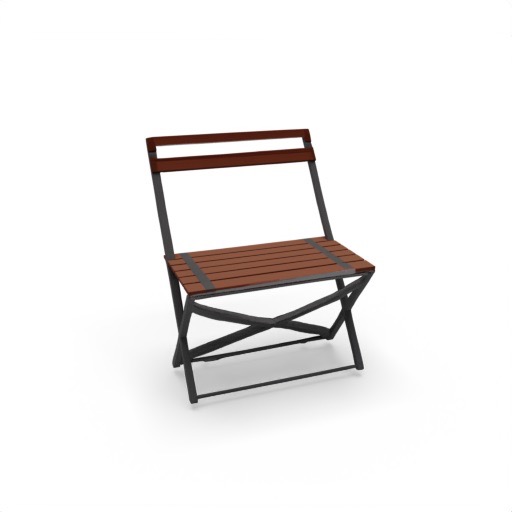} 
        & & %
        \includegraphics[width=0.118\linewidth, trim=2cm 2cm 2cm 2cm, clip]{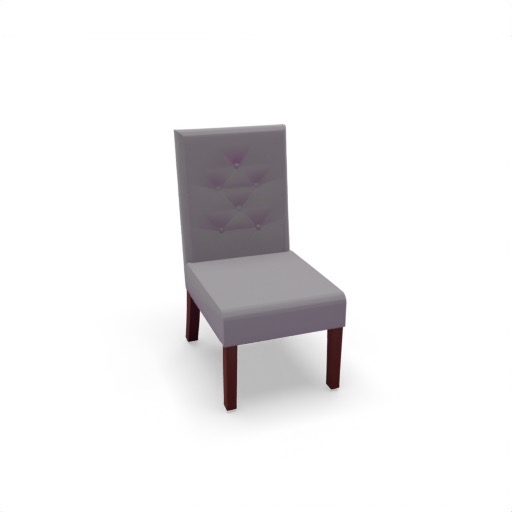} & 
        \includegraphics[width=0.118\linewidth, trim=2cm 2cm 2cm 2cm, clip]{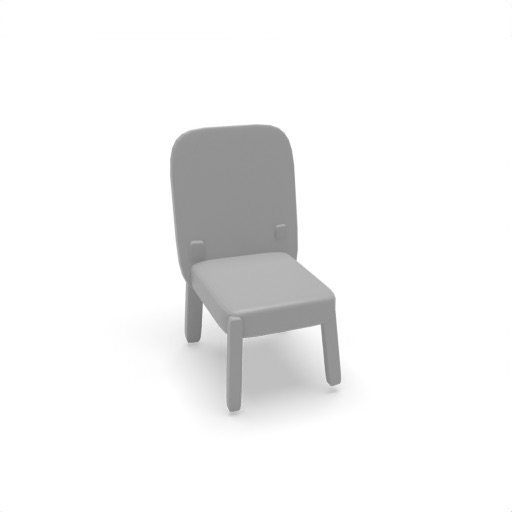} & 
        \includegraphics[width=0.118\linewidth, trim=2cm 2cm 2cm 2cm, clip]{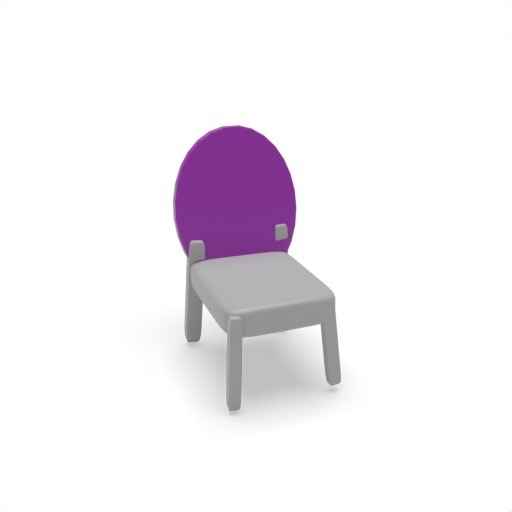} & 
        \includegraphics[width=0.118\linewidth, trim=2cm 2cm 2cm 2cm, clip]{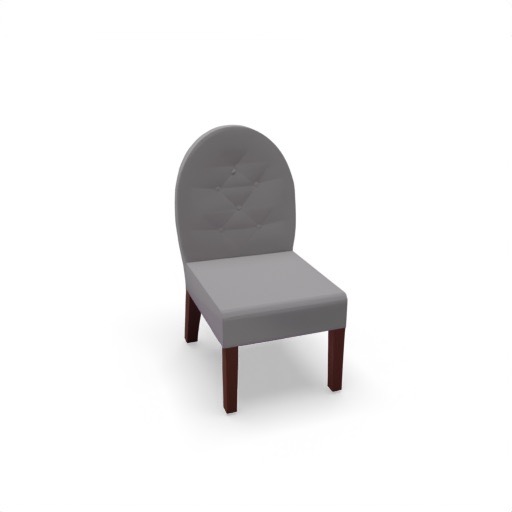}    
        \\
        
        \footnotesize{Input} &
        \footnotesize{Original Proxy} &
        \footnotesize{Edited Proxy} &
        \footnotesize{{Ours}} 
        & &
        \footnotesize{Input} &
        \footnotesize{Original Proxy} &
        \footnotesize{Edited Proxy} &
        \footnotesize{{Ours}} 
        \\[5mm] %
        
        \multicolumn{4}{c}{\editprompt{"The table's skirt is there."}} 
        & & 
        \multicolumn{4}{c}{\editprompt{"The table has a sub-table underneath."}} 
        \\
        
        \includegraphics[width=0.118\linewidth, trim=2cm 2cm 2cm 2cm, clip]{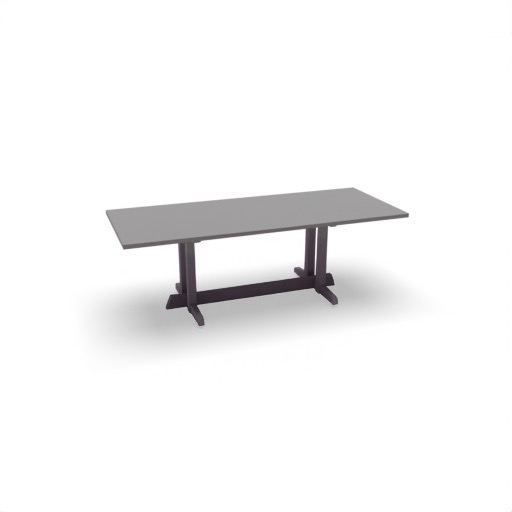} &
        \includegraphics[width=0.118\linewidth, trim=2cm 2cm 2cm 2cm, clip]{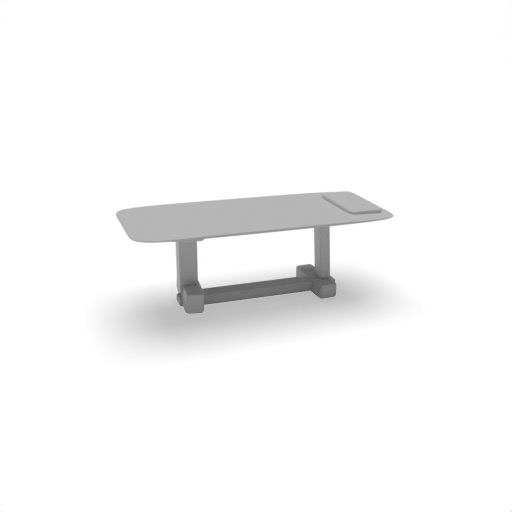} &
        \includegraphics[width=0.118\linewidth, trim=2cm 2cm 2cm 2cm, clip]{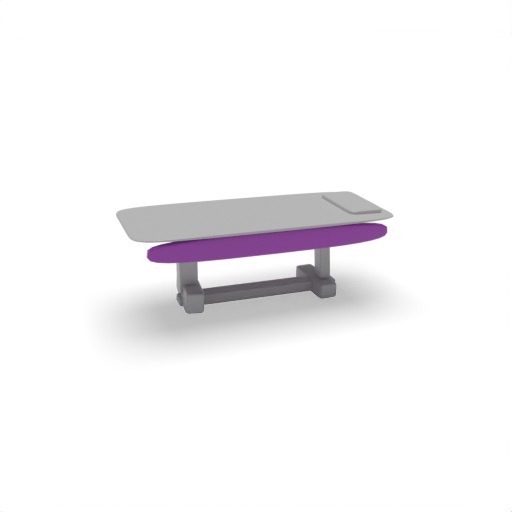} &
        \includegraphics[width=0.118\linewidth, trim=2cm 2cm 2cm 2cm, clip]{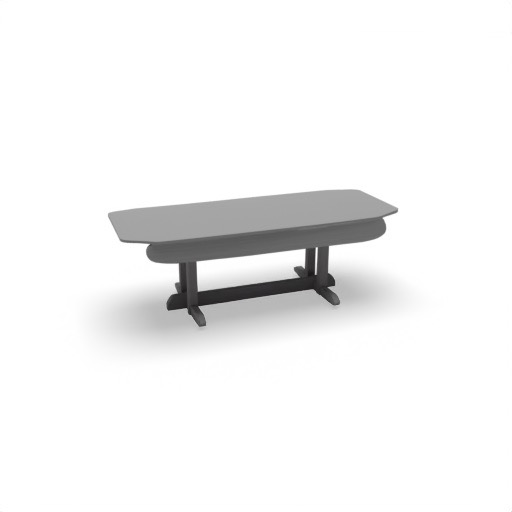} 
        & & %
        \includegraphics[width=0.118\linewidth, trim=2cm 2cm 2cm 2cm, clip]{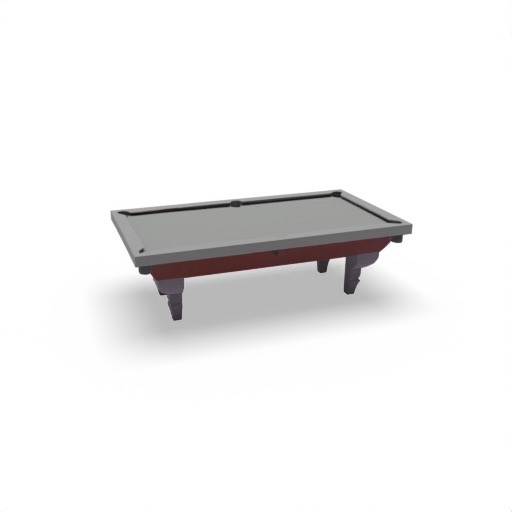} & 
        \includegraphics[width=0.118\linewidth, trim=2cm 2cm 2cm 2cm, clip]{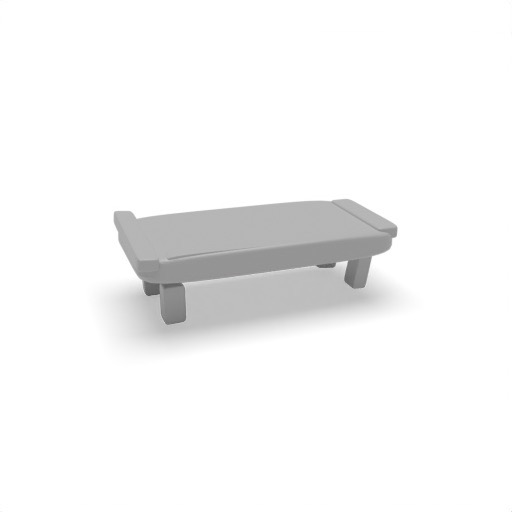} & 
        \includegraphics[width=0.118\linewidth, trim=2cm 2cm 2cm 2cm, clip]{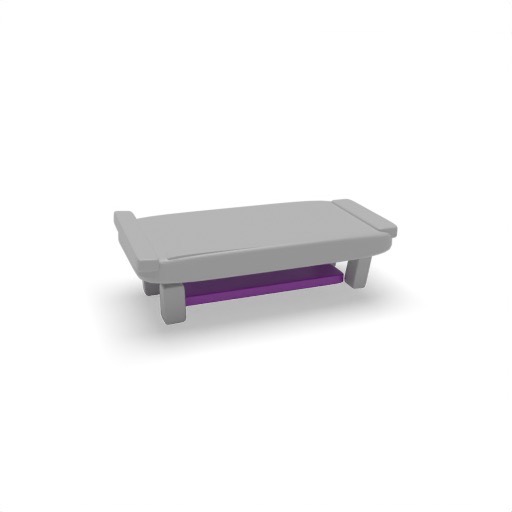} & 
        \includegraphics[width=0.118\linewidth, trim=2cm 2cm 2cm 2cm, clip]{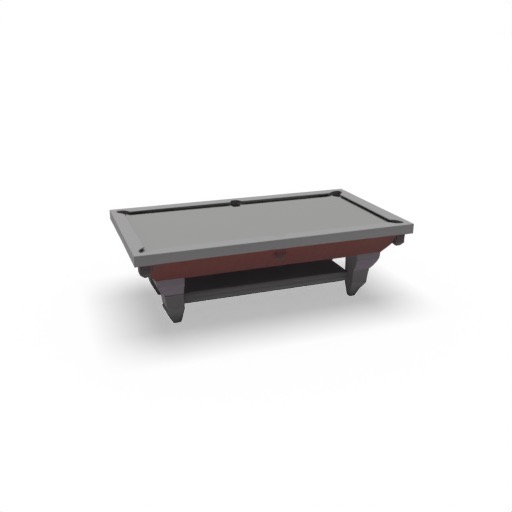}  
        \\
        \footnotesize{Input} &
        \footnotesize{Original Proxy} &
        \footnotesize{Edited Proxy} &
        \footnotesize{{Output}} 
        & &
        \footnotesize{Input} &
        \footnotesize{Original Proxy} &
        \footnotesize{Edited Proxy} &
        \footnotesize{{Output}} 
        \\[5mm] %
        
        \multicolumn{4}{c}{\editprompt{"The lamp has a thicker stem."}} 
        & & 
        \multicolumn{4}{c}{\editprompt{"The lamp has a bulbous shade."}} 
        \\
        \includegraphics[width=0.118\linewidth, trim=2cm 2cm 2cm 2cm, clip]{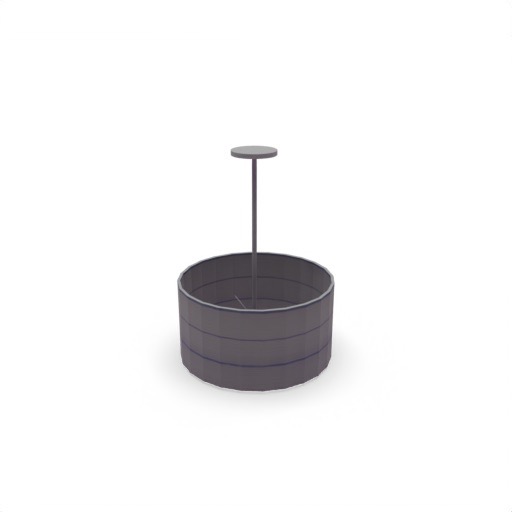} &
        \includegraphics[width=0.118\linewidth, trim=2cm 2cm 2cm 2cm, clip]{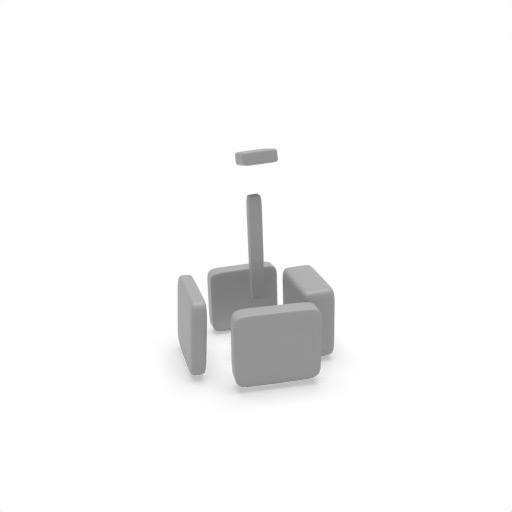} &
        \includegraphics[width=0.118\linewidth, trim=2cm 2cm 2cm 2cm, clip]{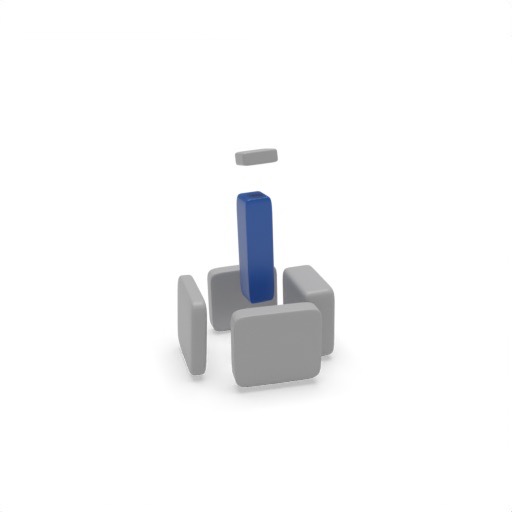} &
        \includegraphics[width=0.118\linewidth, trim=2cm 2cm 2cm 2cm, clip]{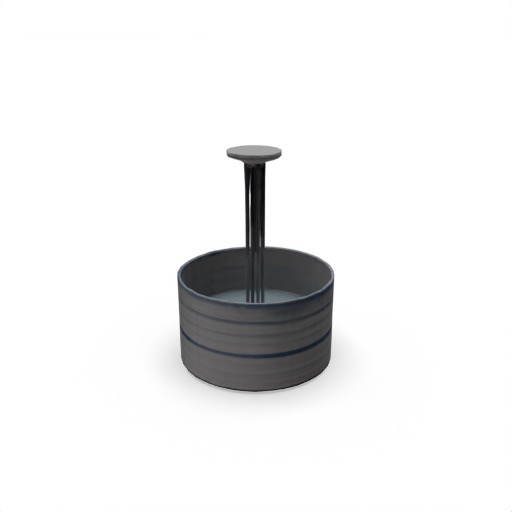} 
        & & %
        \includegraphics[width=0.118\linewidth, trim=2cm 2cm 2cm 2cm, clip]{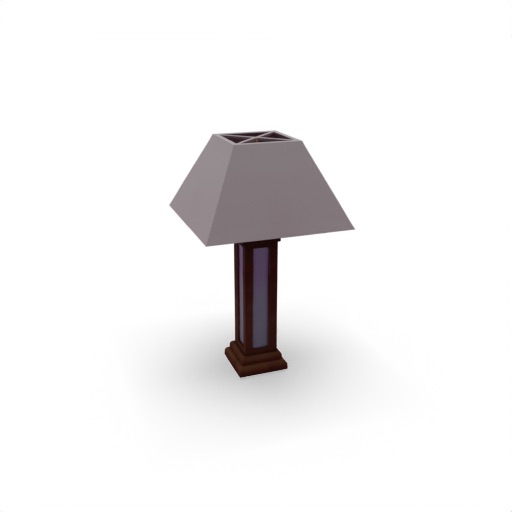} & 
        \includegraphics[width=0.118\linewidth, trim=2cm 2cm 2cm 2cm, clip]{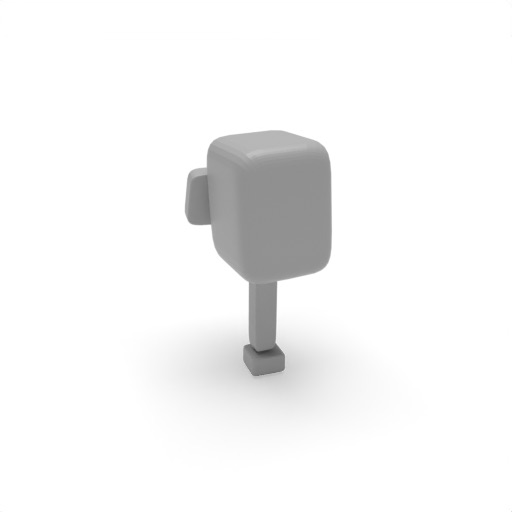} & 
        \includegraphics[width=0.118\linewidth, trim=2cm 2cm 2cm 2cm, clip]{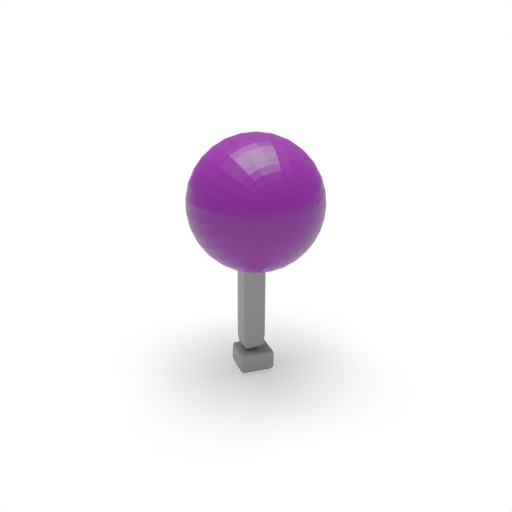} & 
        \includegraphics[width=0.118\linewidth, trim=2cm 2cm 2cm 2cm, clip]{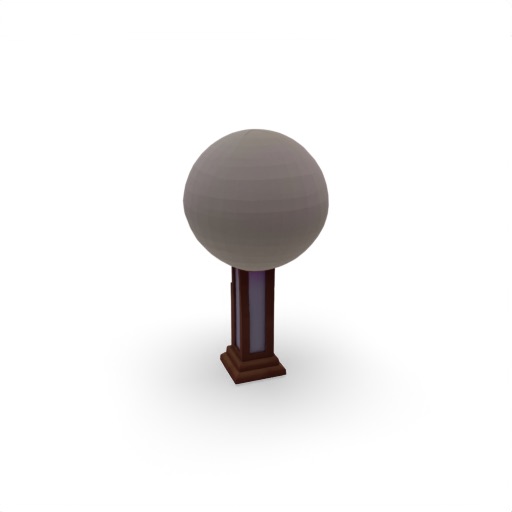}  
        \\ 

        \footnotesize{Input} &
        \footnotesize{Original Proxy} &
        \footnotesize{Edited Proxy} &
        \footnotesize{{Output}} 
        & &
        \footnotesize{Input} &
        \footnotesize{Original Proxy} &
        \footnotesize{Edited Proxy} &
        \footnotesize{{Output}} 
        \\[5mm] %

        \multicolumn{4}{c}{\editprompt{"The chair's back consists of multiple vertical spires and a heavy board on top."}} 
        & & 
        \multicolumn{4}{c}{\editprompt{"There is no stretcher connecting the front legs."}} 
        \\
        \includegraphics[width=0.118\linewidth, trim=2cm 2cm 2cm 2cm, clip]{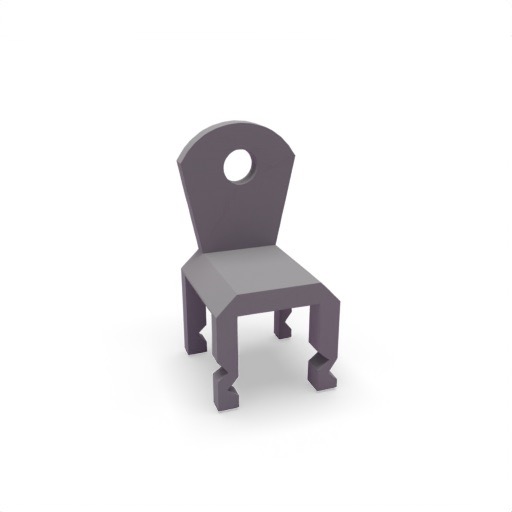} &
        \includegraphics[width=0.118\linewidth, trim=2cm 2cm 2cm 2cm, clip]{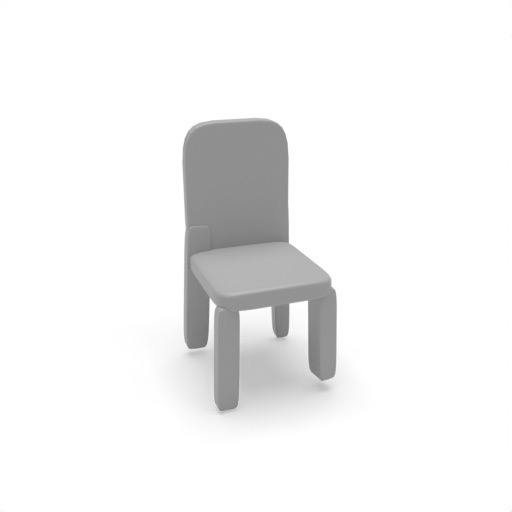} &
        \includegraphics[width=0.118\linewidth, trim=2cm 2cm 2cm 2cm, clip]{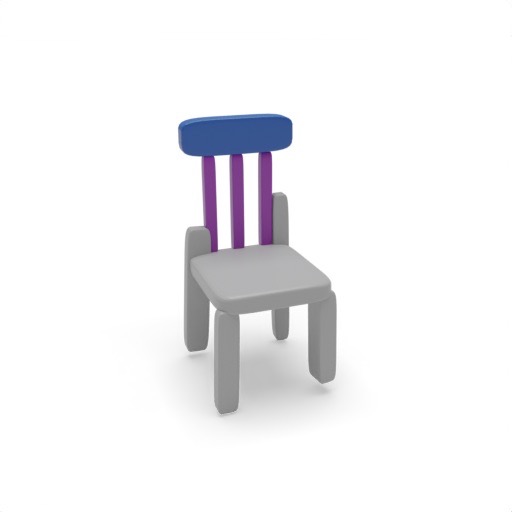} &
        \includegraphics[width=0.118\linewidth, trim=2cm 2cm 2cm 2cm, clip]{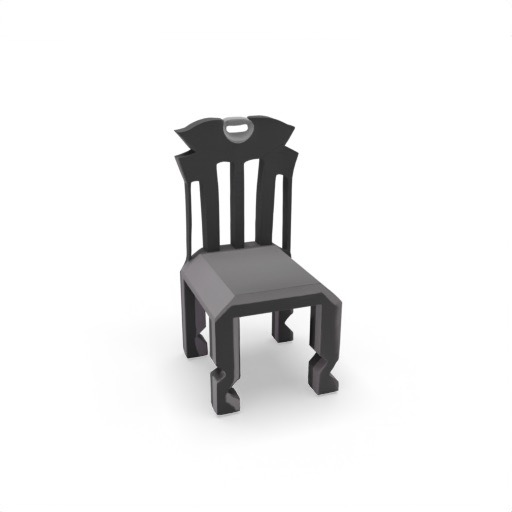} 
        & & %
        \includegraphics[width=0.118\linewidth, trim=2cm 2cm 2cm 2cm, clip]{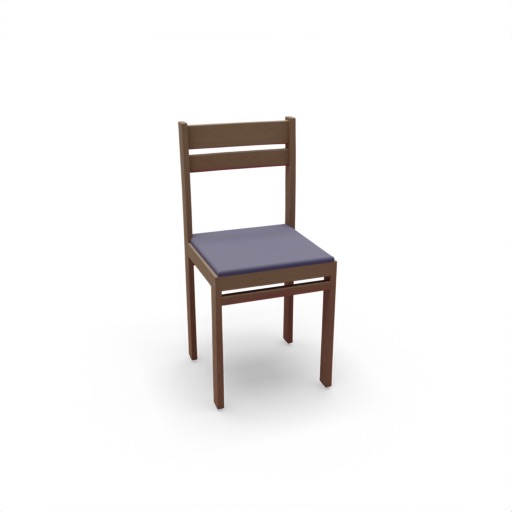} & 
        \includegraphics[width=0.118\linewidth, trim=2cm 2cm 2cm 2cm, clip]{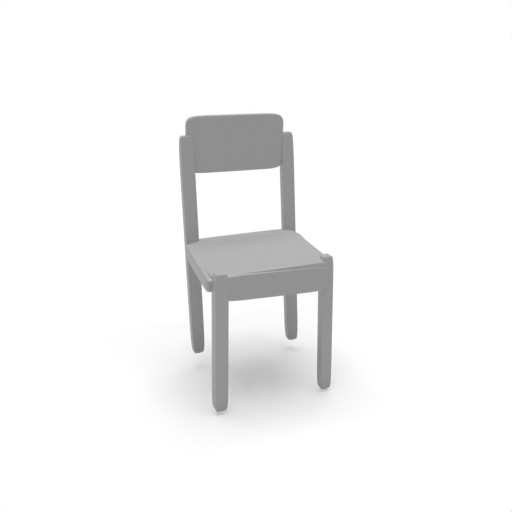} & 
        \includegraphics[width=0.118\linewidth, trim=2cm 2cm 2cm 2cm, clip]{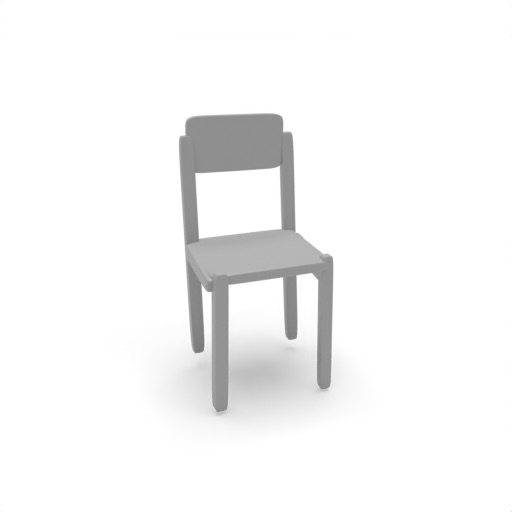} & 
        \includegraphics[width=0.118\linewidth, trim=2cm 2cm 2cm 2cm, clip]{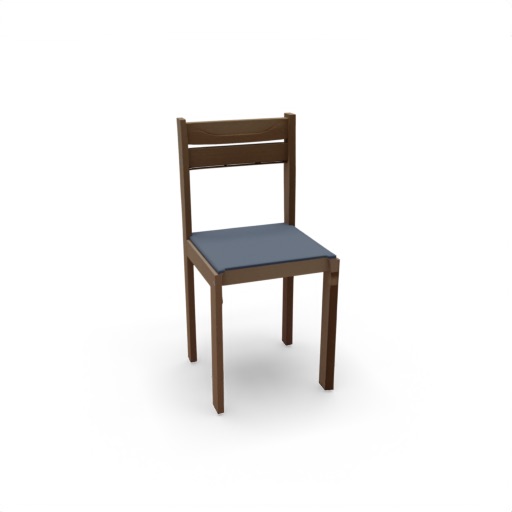}  
        \\ 

        \footnotesize{Input} &
        \footnotesize{Original Proxy} &
        \footnotesize{Edited Proxy} &
        \footnotesize{{Output}} 
        & &
        \footnotesize{Input} &
        \footnotesize{Original Proxy} &
        \footnotesize{Edited Proxy} &
        \footnotesize{{Output}} 
        \\[4mm] %

        \multicolumn{4}{c}{\editprompt{"There are two open shelves on the table."}} 
        & & 
        \multicolumn{4}{c}{\editprompt{"The table's top is more narrow."}} 
        \\
        \includegraphics[width=0.118\linewidth, trim=2cm 2cm 2cm 2cm, clip]{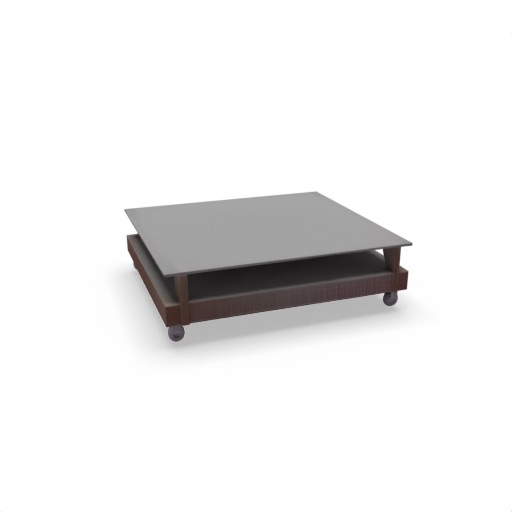} &
        \includegraphics[width=0.118\linewidth, trim=2cm 2cm 2cm 2cm, clip]{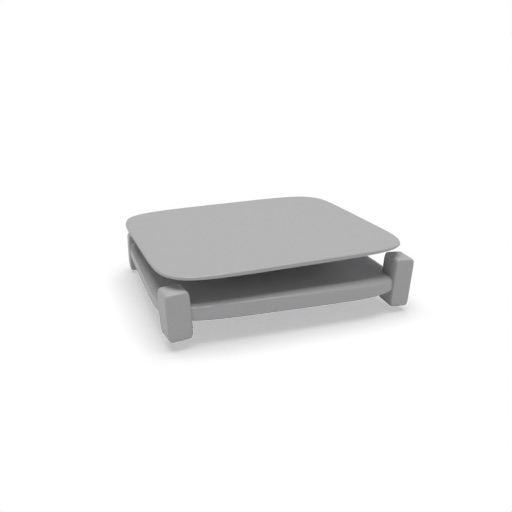} &
        \includegraphics[width=0.118\linewidth, trim=2cm 2cm 2cm 2cm, clip]{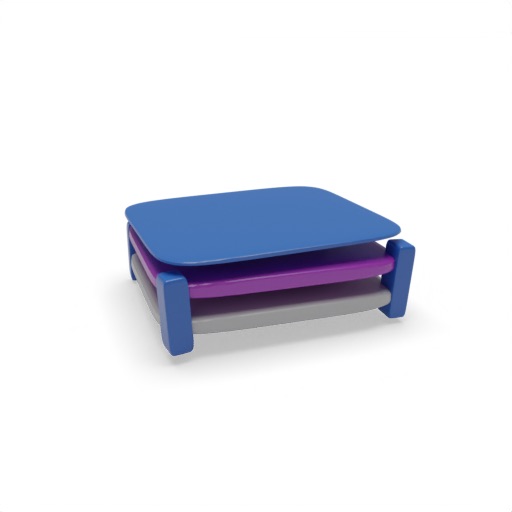} &
        \includegraphics[width=0.118\linewidth, trim=2cm 2cm 2cm 2cm, clip]{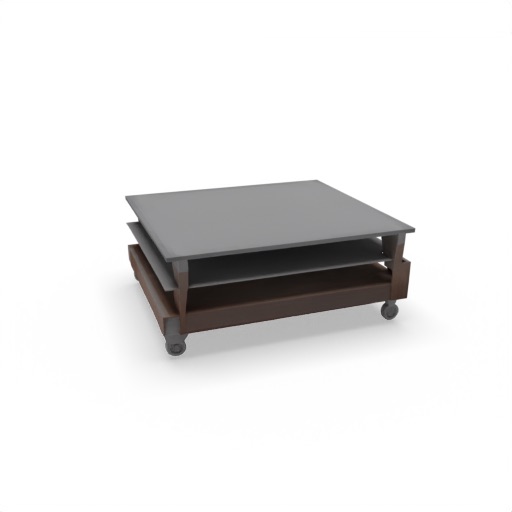} 
        & & %
        \includegraphics[width=0.118\linewidth, trim=2cm 2cm 2cm 2cm, clip]{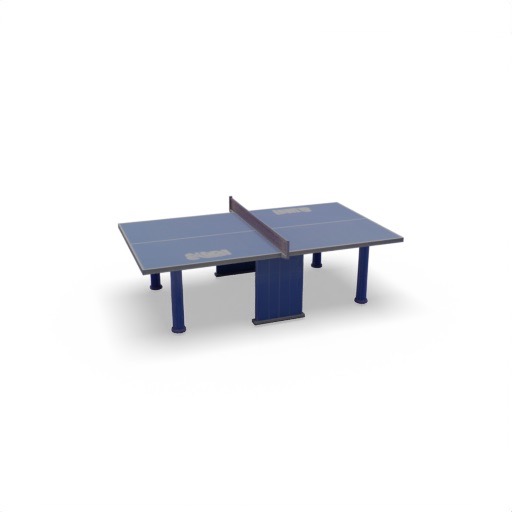} & 
        \includegraphics[width=0.118\linewidth, trim=2cm 2cm 2cm 2cm, clip]{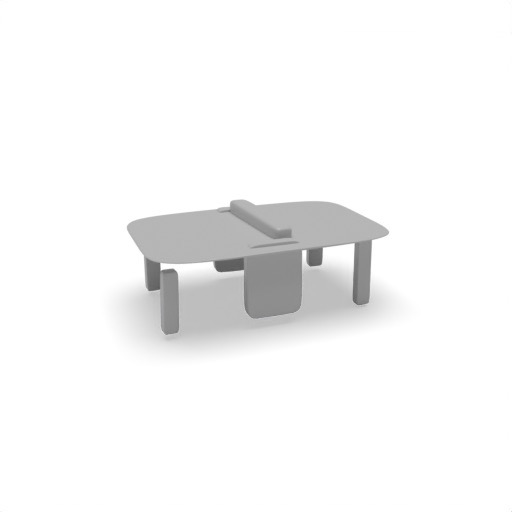} & 
        \includegraphics[width=0.118\linewidth, trim=2cm 2cm 2cm 2cm, clip]{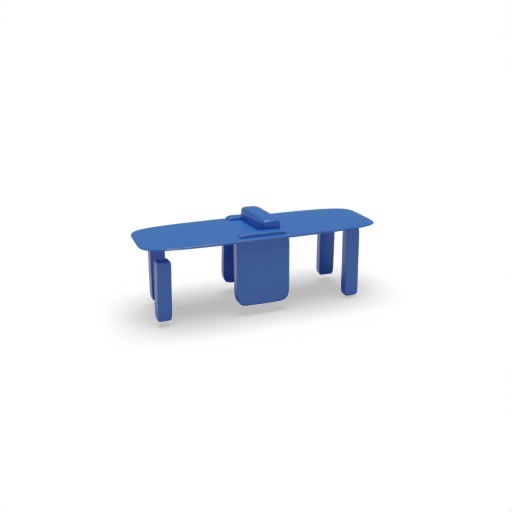} & 
        \includegraphics[width=0.118\linewidth, trim=2cm 2cm 2cm 2cm, clip]{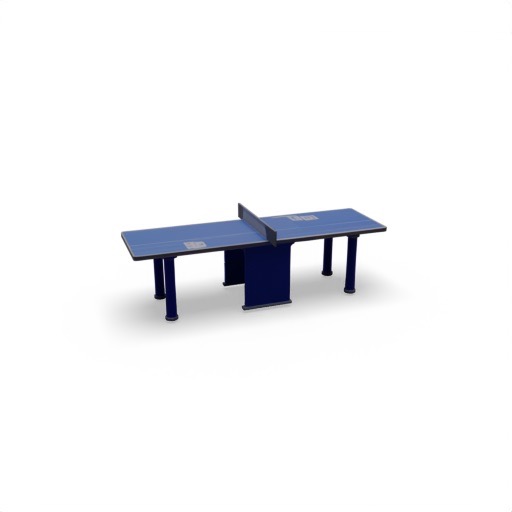}  
        \\ 

        \footnotesize{Input} &
        \footnotesize{Original Proxy} &
        \footnotesize{Edited Proxy} &
        \footnotesize{{Output}} 
        & &
        \footnotesize{Input} &
        \footnotesize{Original Proxy} &
        \footnotesize{Edited Proxy} &
        \footnotesize{{Output}} 
        \\[5mm] %
        
        \multicolumn{4}{c}{\editprompt{"The shade is bigger."}} 
        & & 
        \multicolumn{4}{c}{\editprompt{"The base has more layers."}} 
        \\
        \includegraphics[width=0.118\linewidth, trim=2cm 2cm 2cm 2cm, clip]{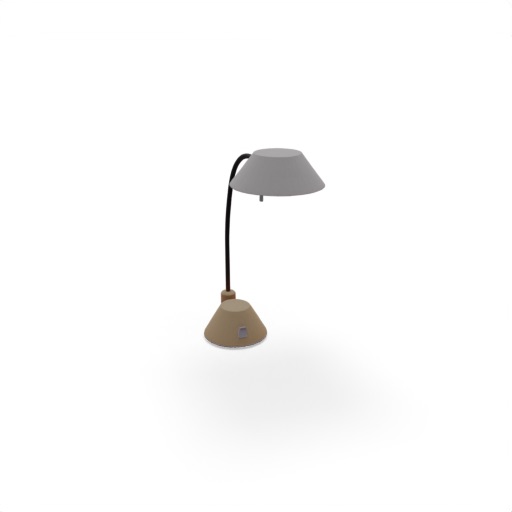} &
        \includegraphics[width=0.118\linewidth, trim=2cm 2cm 2cm 2cm, clip]{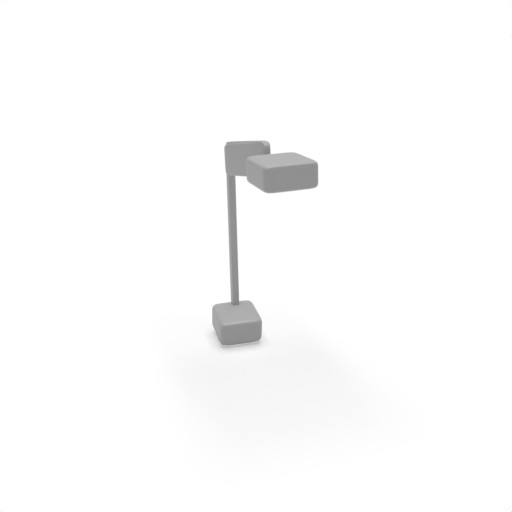} &
        \includegraphics[width=0.118\linewidth, trim=2cm 2cm 2cm 2cm, clip]{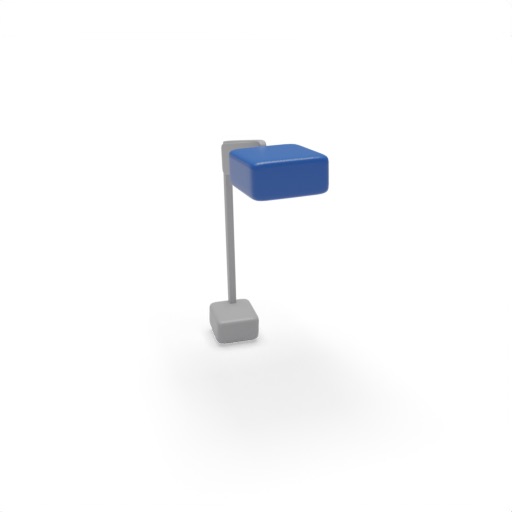} &
        \includegraphics[width=0.118\linewidth, trim=2cm 2cm 2cm 2cm, clip]{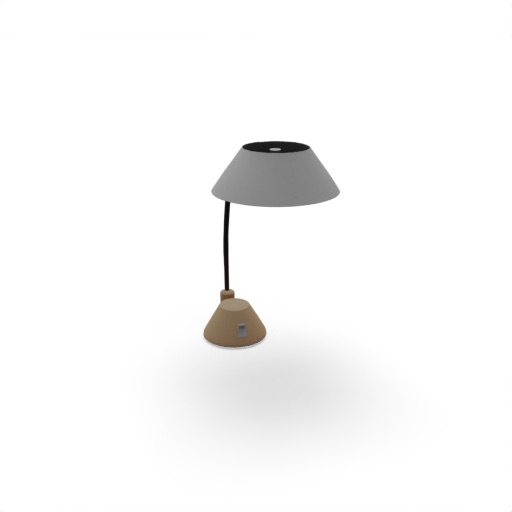} 
        & & %
        \includegraphics[width=0.118\linewidth, trim=2cm 2cm 2cm 2cm, clip]{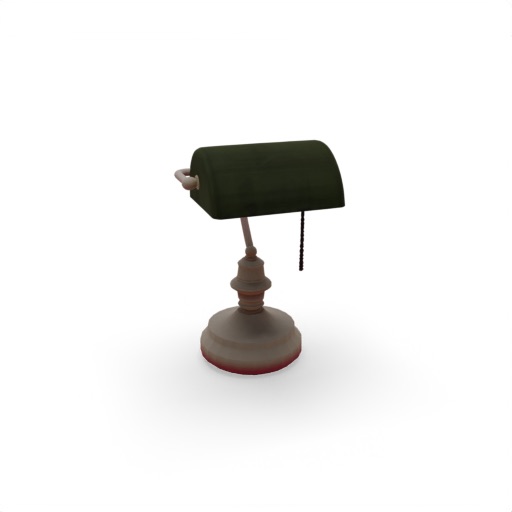} & 
        \includegraphics[width=0.118\linewidth, trim=2cm 2cm 2cm 2cm, clip]{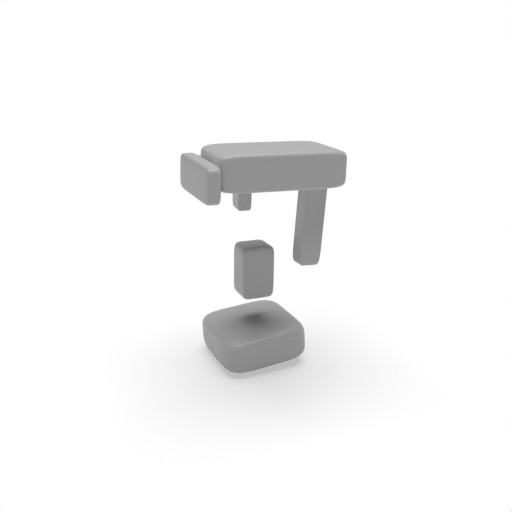} & 
        \includegraphics[width=0.118\linewidth, trim=2cm 2cm 2cm 2cm, clip]{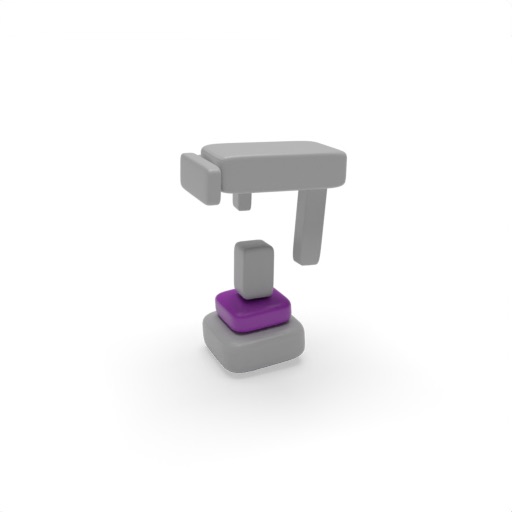} & 
        \includegraphics[width=0.118\linewidth, trim=2cm 2cm 2cm 2cm, clip]{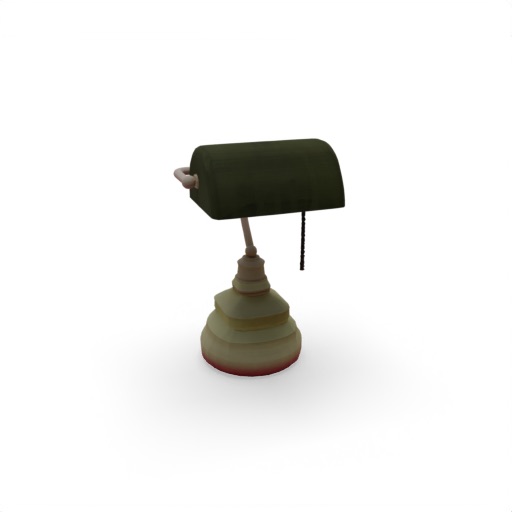}  
        \\ 

        \footnotesize{Input} &
        \footnotesize{Original Proxy} &
        \footnotesize{Edited Proxy} &
        \footnotesize{{Output}} 
        & &
        \footnotesize{Input} &
        \footnotesize{Original Proxy} &
        \footnotesize{Edited Proxy} &
        \footnotesize{{Output}} 
        \\[5mm] %
        
    \end{tabular}
    \caption{\textbf{Qualitative results from the ShapeTalk benchmark.} Above, we show input and output texture-based renderings, along with the original and edited proxy shape (middle columns). When presenting the edited proxy shapes we color the edited super-quadratics \textcolor{blue}{blue} and added super-quadratics in \textcolor{violet}{purple}.
   }
    \label{fig:results}
\end{figure*}

\clearpage

\renewcommand{\thesection}{\Alph{section}}

\twocolumn[
        \centering
        \Large
        \textbf{\ProxE: Fine-Grained 3D Shape Editing via Primitive-Based Abstractions}\\
        \vspace{0.5em}Supplementary Material \\
        \vspace{1.0em}
    ] %

In this document, we present implementation details (Section \ref{sec:details}) as well as additional results and experiments (Section \ref{sec:additional_results}).
\tableofcontents

\section{Additional Results and Information}
We refer readers to the interactive visualizations at \href{index.html}{index.html}. In this document, we provide  implementation details (Section \ref{sec:details}) for our  method and experiments. We also include all VLM instruction prompts used by our method in the ``\textit{vlm\_prompts}'' folder included alongside this document.
\section{Technical Details}
\label{sec:details}
\subsection{\methodName{}  Implementation Details}
This section details the implementation details of our proposed method, starting with the prompt parsing (Section. \ref{sec:preprocess_details}), followed by Editing Abstractions with a Vision-Language Model (Section. \ref{sec:abstractions_detail}), Structural Editing via an Edited Abstraction (Section. \ref{sec:structure_editing_details}) and finally, Appearance Refinement (Section. \ref{sec:appearance_refinement_details}). 

\subsubsection{Pre-process: Prompt Parsing}
\label{sec:preprocess_details}
When parsing the initial instruction prompt we use the \texttt{gemini-2.5-flash} VLM using the \href{https://aistudio.google.com/}{Google AI Studio API}. The instruction prompt given to the VLM at this stage alongside the editing instruction is included in the ``\textit{vlm\_prompts}'' folder (see ``\textit{analyze\_edit\_instruction.txt}''). In short, the message tells the VLM to break the instruction prompt in to two stand alone descriptions, one for appearance and one for structure. If the original prompt does not mention any structural or appearance changes, the VLM is instructed to return \textit{``a \{category\}''} as the structural or appearance description, signaling to the rest of the pipeline not to perform structural or appearance editing. 

\subsubsection{Editing Abstractions with a Vision-Language Model}
\label{sec:abstractions_detail}
To generate the abstracted proxy shape, we use the default \emph{SuperDec} implementation and model available on \href{https://github.com/elisabettafedele/superdec}{SuperDec's official github repo}. After obtaining the shape proxy, we render it from four different views (front, back, left, right) which are then combined into a single image. The parameters of the abstraction are then converted into a ``json'' format file and given alongside the combined image of the proxy, an image of the original shape, the structural description and the VLM's instruction prompt to the VLM. This instruction prompt is also available in the ``\textit{vlm\_prompts}'' (see ``\textit{VLM\_edit\_instruction.txt}''). 

\begin{figure*}[t]
    \centering
    \includegraphics[width=0.97\textwidth]{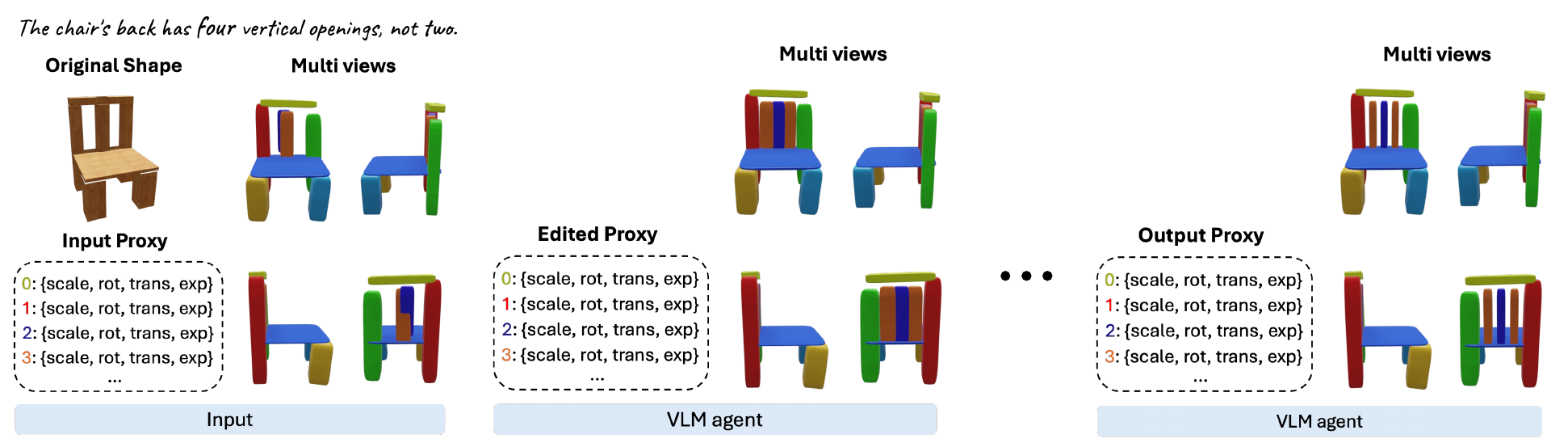}
    \caption{\new{\textbf{Editing Abstractions with a Vision-Language Model.} The VLM agent receives as input the proxy's JSON file—where each primitive is described by its scale, rotation, translation, and shape exponent parameters—a text editing instruction, the rendering of the original shape, and multi-view renderings of the proxy. It then produces an updated JSON file, which is used to generate new multi-view renderings of the edited proxy. This process repeats iteratively, with the outputs from the previous step fed back to the VLM agent, until the edit is confirmed or the maximum number of iterations is reached.}}
    \label{fig:vqa_editing}
\end{figure*}

The prompt instructs the VLM to work according to three steps (which are detailed in its reasoning text output). First the VLM describes the shape it is presented and its proxy, then it formulates an editing plan and finally it generates an updated ``json'' file. We then parse the updated ``json'' from the received textual output and re-render a VLM input. We give this new rendering along with the previous context back to the VLM alongside a \emph{feedback prompt} (see ``\textit{VLM\_feedback\_instruction.txt}''). The VLM is then prompted to either confirm the edit is correct or suggest new proxy parameters. In our implementation we repeat this process for a maximum of $3$ tries or until the VLM determines the edit is viable.
\new{We illustrate this process in Figure \ref{fig:vqa_editing}. }

\subsubsection{Structural Editing via an Edited Abstraction}
\label{sec:structure_editing_details}
We build on top of TRELLIS and use the official implementation and models available on the \href{https://github.com/microsoft/TRELLIS}{official TRELLIS GitHub repository}. After obtaining $\mathcal{S}_{orig}$, $\mathcal{S}_{warp}$, and $\mathcal{P}_{edit}$ we use the inversion process introduced by VoxHammer available on \href{https://github.com/Nelipot-Lee/VoxHammer}{their official GitHub repository} to invert their structure and for $\mathcal{S}_{orig}$ the appearance as well. For inversion and inference in both structure and appearance we use $25$ time-steps and the default diffusion hyperparameters. Structure is inverted with a text prompt (``\emph{a \{category\}''}) as guidance. For appearance we use view rendered from the original shape. To extract the dino features required to encode the original shape into TRELLIS' SLAT space we render $75$ images of the original shape from varying directions in a $360$ sphere.

After inversion, we run the structure proxy-induced denoising process as described in Section 3.3 of the main paper. The time-step hyper-parameters used in this process are as follows $t_{init} = T - 12$, $t_{warp}=T - 16$, $t_{uc}=T-20$ with $T=25$. We start with ``Original Shape Injection'', then perform ``Warped Shape Injection'' and finally perform ``Proxy Injection''.

\subsubsection{Appearance Refinement}
\label{sec:appearance_refinement_details}
After obtaining the edited structure, we move on to the appearance refinement stage as described in Section 3.4 of the main paper. We edit the same view used to invert the SLAT features in the inversion stage only when the appearance description $c_{\text{txt}}^{app}$ is something other than ``\emph{a \{category\}''}. When this is indeed the case we use \texttt{FLUX.1-Kontext-dev} using the \href{https://huggingface.co/black-forest-labs/FLUX.1-Kontext-dev}{HuggingFace API}. The prompt we give to this model is \emph{``make this \{category\} into a $c_{\text{txt}}^{app}$ ''}. We run this model in it's default settings. When an edit is requested we set $t_{app} = T - 4$, when it is not requested we set $t_{app} = T - 16$ again with $T=25$.

\section{Evaluation Details}
\label{sec:eval_details}
\subsection{Result rendering.}
\minihead{Point-based rendering.} This rendering method serves as the primary means for visually comparing different baselines, encompassing both point cloud generation methods--such as Changeit3D and BlendedPC--and mesh-based approaches, such as Spice-E, EditP23, VoxHammer, and our method.

\minihead{Texture-based rendering.} Beyond shape editing, our method generates colored textures—a feature lacking in point-based approaches like ChangeIt3D and BlendedPC. We render these textured results to facilitate direct comparison with texture-supporting methods such as VoxHammer and TRELLIS.

\subsection{Metrics}
\label{subsec:metrics}
\noindent 

\minihead{Identity preservation.} This includes three following metrics:

\textit{localized-Geometric Distance (l-GD).} We the official implementation of Changeit3D \cite{achlioptas2023shapetalk} and the improved modification of BlendedPC on using a stronger point segmentation model from Point-E\cite{nichol2022point} to help identify the binary mask for unedited regions. 

Additionally, since point-based methods such as Changeit3D and BlendedPC directly output point clouds, whereas shape-based methods produce meshes from which point clouds must be sampled. This sampling process introduces inherent spatial misalignment that is absent in directly predicted point clouds. 
This sampling process can cause spatial shifts in the resulting point cloud relative to the input, whereas directly predicted point clouds remain spatially consistent.
To ensure a fair comparison, we apply the Iterative Closest Point (ICP) algorithm to align points outside the binary mask with the corresponding input points before computing the l-GD metric.

\textit{LPIPS and DINO-I.} We adopt the evaluation tool from VoxHammer~\cite{li2025voxhammer} to compute the scores on the rendering outputs of mesh objects. We first takes output object files from different baselines (excluding Changeit3D and BlendedPC, since they produces 3D point clouds, not texture meshes) to get texture-rendering outputs. Here, we consider only one rendering view for most experiments.

\minihead{3D Quality.} For P-FID, we utilize the evaluation protocol from Point-E \cite{nichol2022point}, performing uniform sampling to extract 2,048 points from the edited outputs. This ensures a consistent setting for baseline comparisons. For FID, we adopt the VoxHammer implementation to compute the distribution divergence between the extracted features of renderings of the input shapes and the output objects.

\begin{table}[t]
\caption{\new{\textbf{Performance comparison across methods and VLMs.} The table lists the base VQA score and the enhanced score with Chain-of-Thought (CoT) prompting.}}
\centering

\label{tab:vqa-cot}
\resizebox{0.85\columnwidth}{!}{%
\begin{tabular}{llcc}
\toprule
\textbf{Method} & \textbf{VLM} & \textbf{VQA} & \textbf{VQA$+$CoT} \\
\midrule
\multirow{4}{*}{EditP23} 
    & Qwen2.5-VL-7b & 0.18 & 0.62 \\
    & SAIL-VL-8B & 0.36 & 0.81 \\
    & SAIL-VL-8B-Thinking & 0.44 & 0.56 \\
\midrule
\multirow{4}{*}{Spice-E} 
    & Qwen2.5-VL-7b & 0.14 & 0.58 \\
    & SAIL-VL-8B & 0.34 & 0.78 \\
    & SAIL-VL-8B-Thinking & 0.41 & 0.48 \\
\midrule
\multirow{4}{*}{VoxHammer} 
    & Qwen2.5-VL-7b & 0.15 & 0.55 \\
    & SAIL-VL-8B & 0.32 & 0.75 \\
    & SAIL-VL-8B-Thinking & 0.40 & 0.47 \\
\midrule
\multirow{4}{*}{TRELLIS} 
    & Qwen2.5-VL-7b & 0.28 & 0.65 \\
    & SAIL-VL-8B & 0.48 & 0.77 \\
    & SAIL-VL-8B-Thinking & 0.48 & 0.58 \\
\midrule
\multirow{4}{*}{Ours} 
    & Qwen2.5-VL-7b & 0.28 & 0.71 \\
    & SAIL-VL-8B & 0.54 & 0.89 \\
    & SAIL-VL-8B-Thinking & 0.46 & 0.67 \\
\bottomrule
\end{tabular}%
}
\end{table}

\begin{figure*}[t]
\noindent\rule{\textwidth}{1pt}
\begin{lstlisting}[style=promptstyle]
You are an automated grading system. You will be asking to assess the difference between two input images based on the text prompt.

1. Visual Context Analysis
Observe the images and the question. Provide a concise description of the specific visual elements (objects, attributes, spatial relations, or actions) that are directly relevant to this query.

2. Reasoning Plan
Identify the logical steps required to verify the answer. Break this down into 2-3 specific "checkpoints" or observations (e.g., identifying a specific object, then verifying its attribute, then checking its relation to others).

3. Step-by-Step Execution
Systematically address each checkpoint from your plan. Provide a brief, evidence-based reasoning trace for each step based solely on the visual data.

4. Final Conclusion
Based on the reasoning above, provide a definitive answer with this format
Final Answer:
Yes/No

Note: You must end with "Final Answer:\nYes" or "Final Answer:\nNo". Before providing your answer, you must explicitly write out your reasoning, starting with the phrase '1. Visual Context Analysis:'.
\end{lstlisting}
\noindent\rule{\textwidth}{1pt}
\caption{\new{System prompt with CoT integration for improving VQA evaluation}, as further detailed in section \ref{subsec:metrics}.}
\label{lst:vlm_prompt}
\end{figure*}

\begin{figure*}[t]
    \centering
    \includegraphics[width=0.48\textwidth]{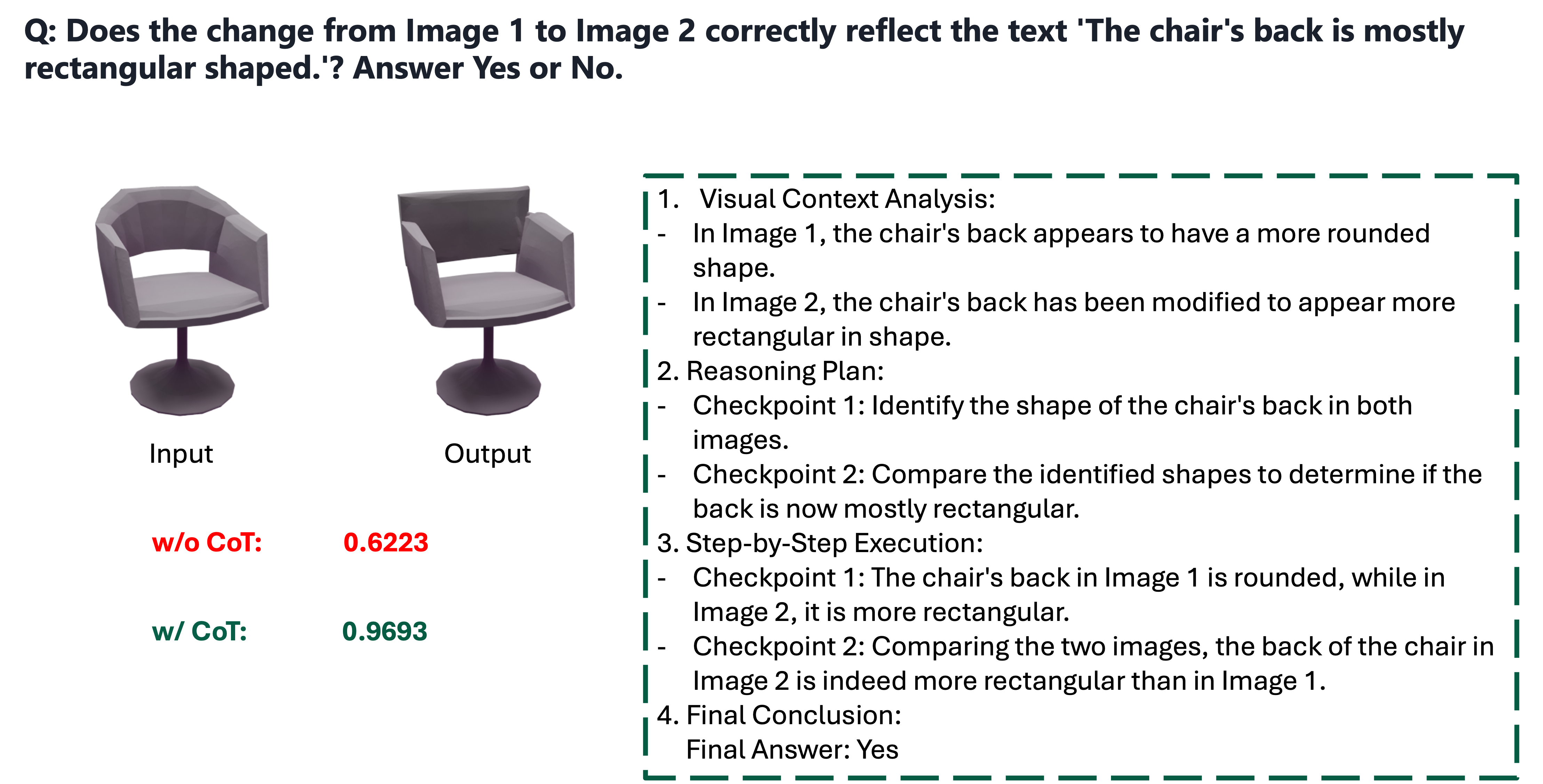}
    \hfill
    \includegraphics[width=0.48\textwidth]{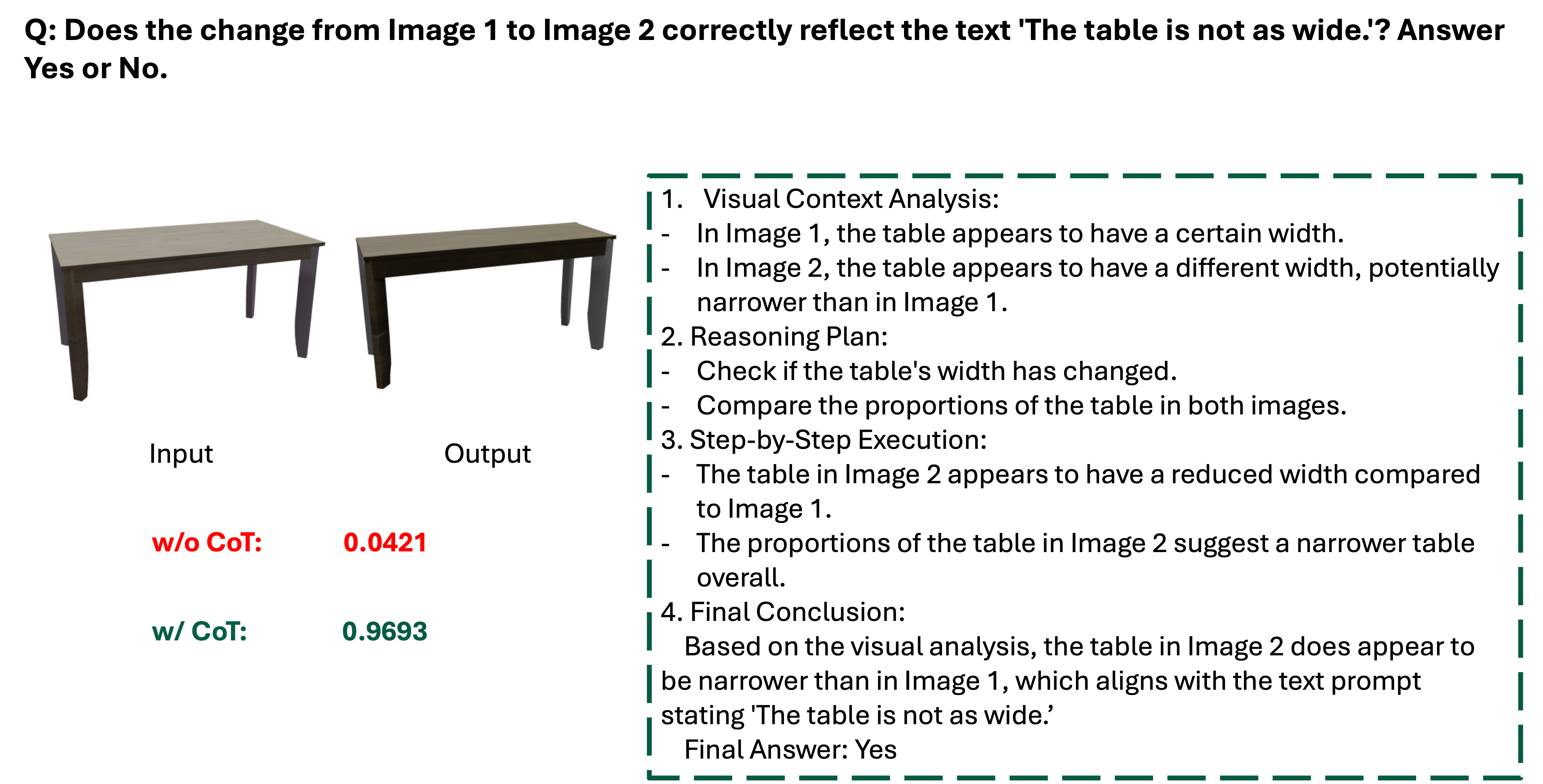}

    \vspace{0.5em}

    \includegraphics[width=0.48\textwidth]{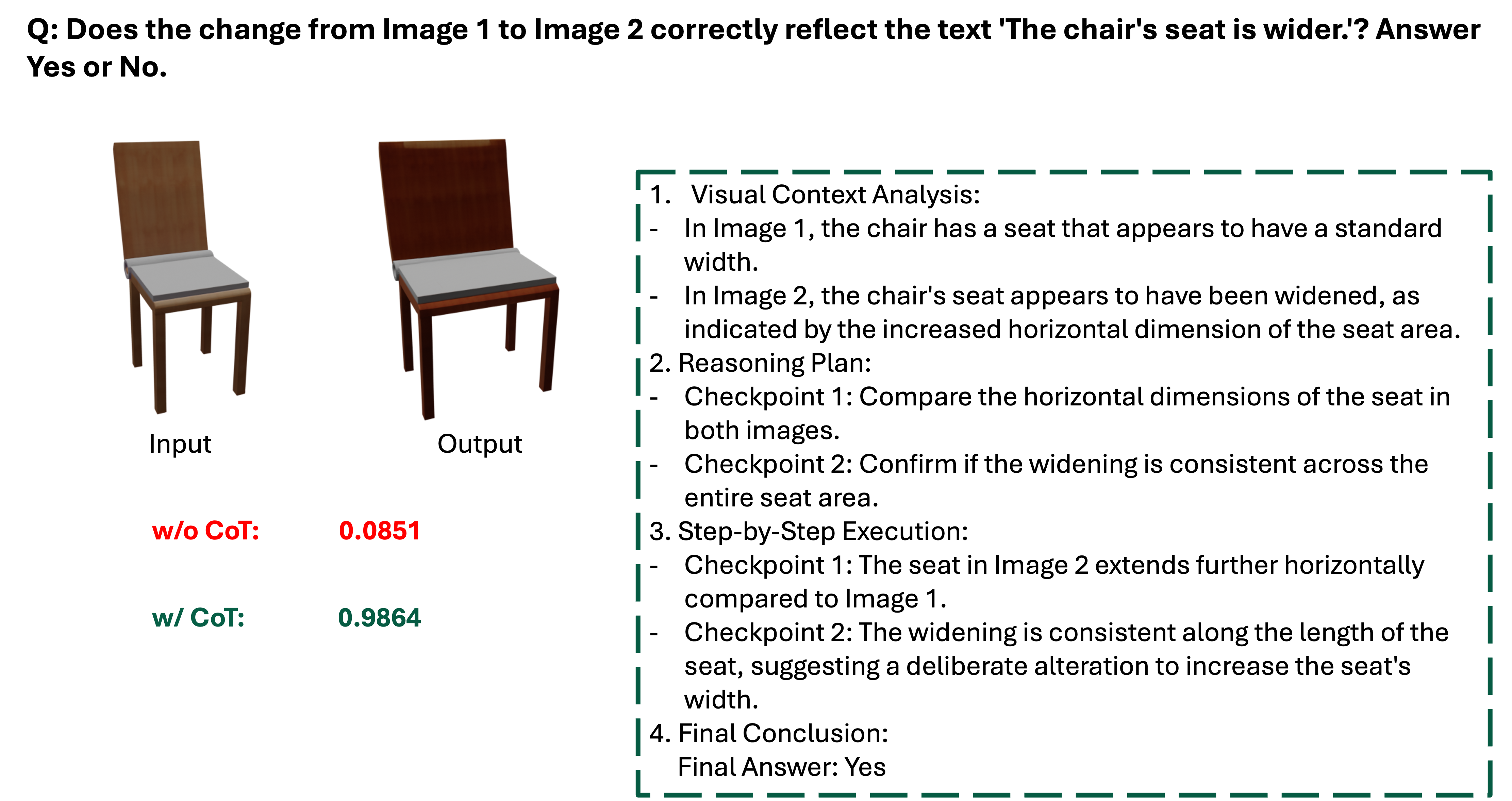}
    \hfill
    \includegraphics[width=0.48\textwidth]{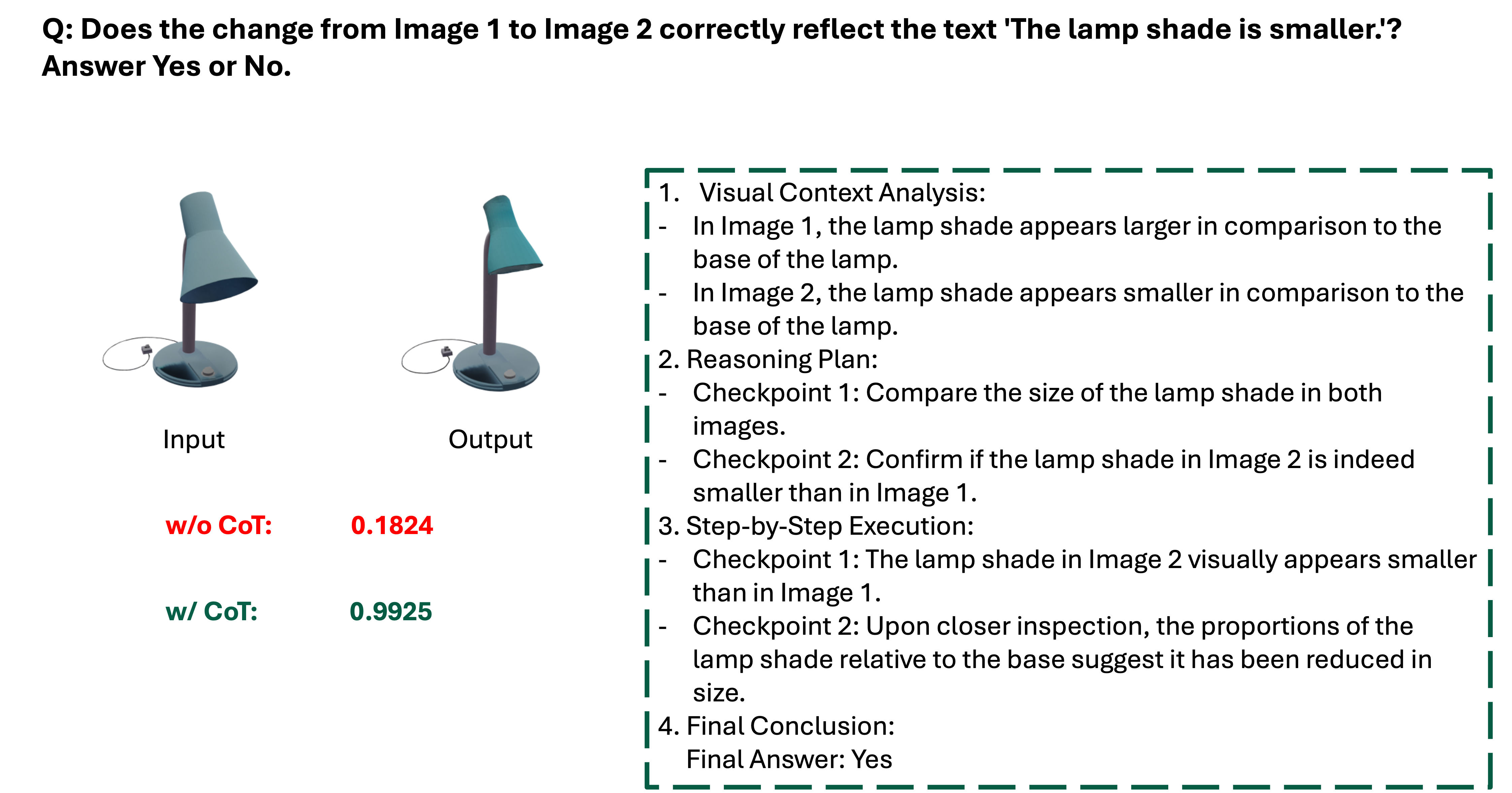}

    \caption{\new{Qualitative comparisons between the vanilla VQA and our VQA with CoT prompting, demonstrating the benefit of integrating CoT reasoning into the VQA evaluation.}}
    \label{fig:vqa_cot}
\end{figure*}

\minihead{Edit Fidelity.} For the CLIP score, we utilize a pretrained CLIP model (ViT-B/32) to extract features from renderings of the edited results and the corresponding text descriptions of the edits.

For the VQA score, we base our evaluation on the official implementation of VQAScore \cite{lin2024evaluating}. Instead of querying the model regarding a single output, we require the Vision-Language Model (VLM) to analyze a pair of images: the rendering of the input shape and the rendering of the edited output. We phrase the prompt as follows: 'Image 1 is the original and Image 2 is the edited version. Does the change from Image 1 to Image 2 reflect the text ‘[input prompt]’? Answer Yes or No.' We then utilize the output probability of the 'Yes' token to quantify the success of the edit.

\new{As mentioned in the main paper, we adopt a Chain-of-Thought (CoT) prompting to explicitly require the VLMs to produce reasoning traces before returning a "Yes/No" answer. This helps avoid the ambiguity of the black-box answers from original VQAScore while making the evaluation more interpretable through detailed justifications.  Specifically, we impose a structured response format that requires the VLM to first analyze the visual inputs, then formulate a reasoning plan with 2-3 checkpoints. For each checkpoint, the VLM must provide a brief judgment with supporting evidence before arriving at a final answer. As before, we use the output probability of the 'Yes' token as the final score. The full CoT prompt is provided in Figure \ref{lst:vlm_prompt}. To illustrate the benefit of CoT, we compare the original VQA score with our CoT-augmented variant in Figure \ref{fig:vqa_cot}. CoT prompting substantially improves evaluation accuracy across diverse editing scenarios, yielding scores that more reliably reflect the actual success of the requested edit. As shown in the qualitative examples, the VLM produces plausible, structured reasoning that grounds its final judgment.}

To fairly evaluate editing fidelity across both shape and texture, we adapt different rendering strategies to the output format of each baseline. For point-based methods (e.g., ChangeIt3D, BlendedPC), we utilize point-based rendering for CLIP and VQA evaluation. For the remaining methods, which generate textured meshes, we employ standard texture-based rendering. This distinction ensures that the metrics accurately reflect the true quality of each model type.

Additionally, it is important to note that since Changeit3D and BlendedPC generates point clouds rather than textured meshes, they are omitted from these specific metrics, FID, LPIPS, and DINO-I.

\subsection{Baselines}
\minihead{Changeit3D.} We use the official checkpoint of Changeit3D, point-cloud-based autoencoder with decoupling the magnitude of the-edit (namely, "idpen\_0.1\_sc\_True" model) to perform inference on our conducted ShapeTalk subset.   

\minihead{BlendedPC.} Since BlendedPC is a category-specific pretrained model, we use the corresponding checkpoints for individual categories (i.e. chair, table, and lamp) to perform inference editing on the same ShapeTalk subset.

\minihead{Spice-E.} Spice-E is a 3D diffusion method that requires a input shape condition in addition to the text prompt to guide the editing process. We use pretrained semantic checkpoints for individual categories (i.e. chair, table, lamp) to obtain output editing results.

\minihead{EditP23.} The EditP23 \cite{bar2025editp23} pipeline follows a multi-stage image-to-3D editing workflow. First, source meshes are rendered into a multi-view grid using the EditP23 helper script . The resulting conditioning view is then edited via FLUX Kontext \cite{labs2025flux1kontextflowmatching} (guidance: 2.5, steps: 24) using LLaMA-3 rephrased prompts. 

Next, the EditP23 editing process propagates these 2D modifications across all views to ensure consistency (target guidance: 21.0, $n_{\max}=39$, $T_{\text{steps}}=50$), producing an edited multi-view image. Finally, 3D reconstruction is performed using the EditP23 reconstruction script with the \texttt{instant-mesh-large.yaml} configuration, leveraging InstantMesh \cite{xu2024instantmesh} to generate the final mesh.

\minihead{VoxHammer.} To address VoxHammer's~\cite{li2025voxhammer} requirement for a user-provided mask, we automate the process by utilizing a pre-trained PointNet~\cite{qi2016pointnet} to identify the target region and constructing a solid bounding box around it to ensure robust downstream editing. Using this generated mask, we execute the standard three-stage VoxHammer pipeline—multi-view rendering, diffusion-based 2D inpainting, and 3D inference propagation—while maintaining a constant editing view azimuth; notably, this automated masking exhibited a failure rate of $\sim$6\%, primarily in cases where instructions referenced components absent from the source geometry (e.g., adding armrests).

\begin{table}[t]
\caption{\new{\textbf{User Study}. We report win rates of our method compared against baseline techniques. As illustrated below, our method is consistently preferred both in edit quality and identity preservation.}
}
\footnotesize
\centering
\begin{tabular}{l c c c c}
\toprule
& EditP23 & Spice-E & VoxHammer & TRELLIS \\

\midrule

Edit Quality & 86.6 & 92.7 & 91.7 &  78.8  \\
Identity Preservation & 86.3 & 88.9 & 85.8 & 59.5 \\

\bottomrule
\end{tabular}

\label{tab:vqascore}
\end{table}

\ignorethis{
\begin{table}[t]
\footnotesize
\centering
\begin{tabular}{l cc cc c}
\toprule
 & \multicolumn{2}{c}{Qwen2.5-VL-7B} & \multicolumn{2}{c}{SenseNova-SI} & Human \\
\cmidrule(lr){2-3} \cmidrule(lr){4-5} \cmidrule(lr){6-6}
& Full & Sub & Full & Sub & Sub [\emph{Diff.}]\\
\midrule
TRELLIS & 57.5 & 70.0 & 66.4 & 48.01 & 66.6 $[\textit{-3.4,+18.6}]$ \\
VoxHammer       & 66.4 & 75.0 & 61.2 & 65.0 & 78.1 $[\textit{+3.1,+16.9}]$ \\
Spice-E         & 63.4 & 65.0 & 59.7 & 60.0 & 78.2 $[\textit{+13.2,+18.5}]$ \\
EditP23         & 60.7 & 70.0 & 57.2 & 50.0 & 76.3 $[\textit{+6.3,+16.3}]$ \\
\bottomrule
\end{tabular}
\caption{User study reports win rates of our method compared to baseline models on the selected subset. As seen, our method shows a favorable trend compared to the baseline models.
\vspace{-8pt}

}

\label{tab:vqascore}
\end{table}
}

\begin{figure}[t]
    \centering
    \includegraphics[height=0.7\textheight]{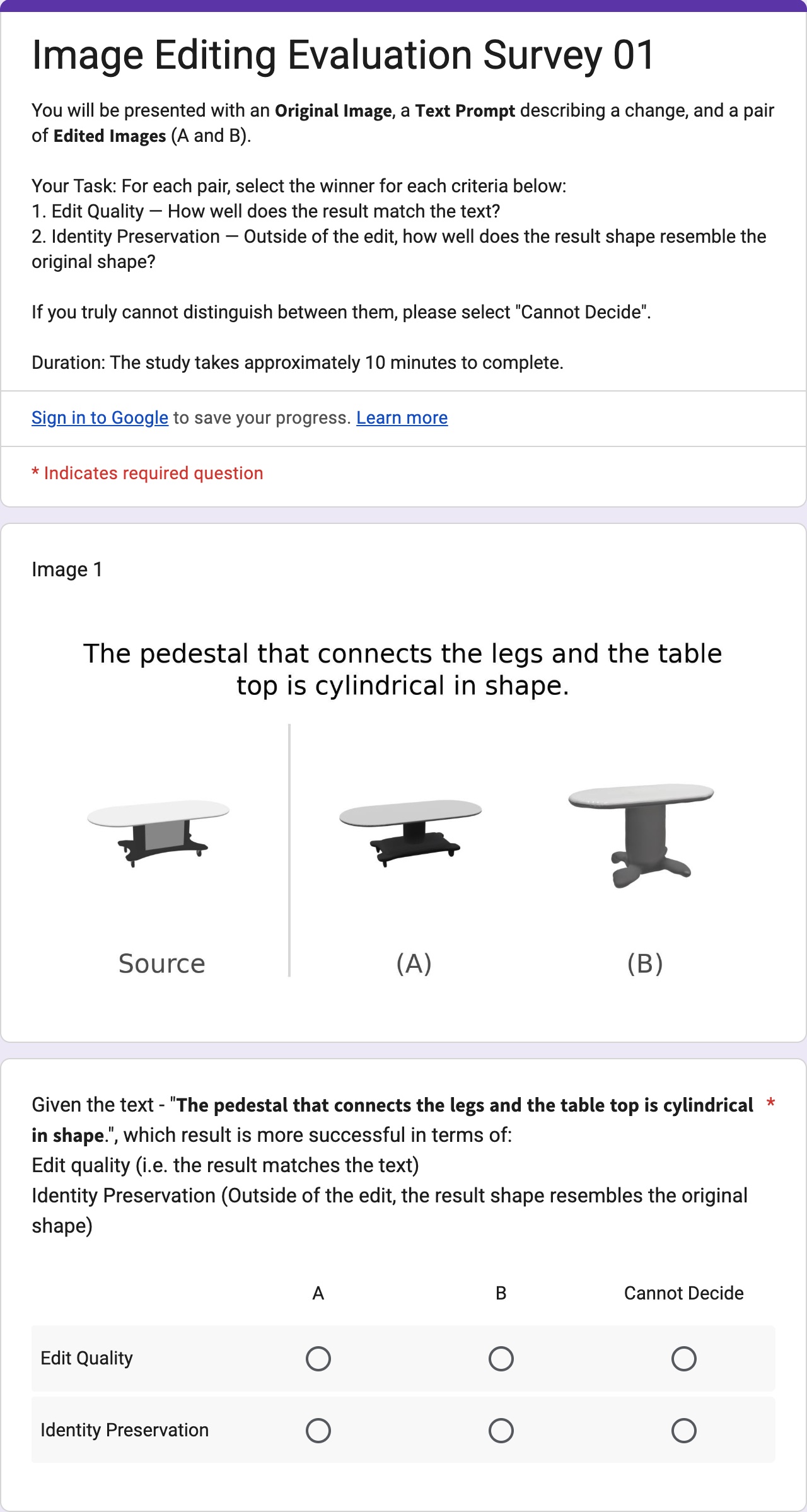}
    \caption{\new{\textbf{User study instruction and example question.} Users were presented with an input shape, an editing prompts and two editing results: one produced by our method and one by a competing method. They were then asked to separately select their preferred output shape in terms of edit quality and identity preservation.}}
    \label{fig:user_study_form}
\end{figure}

\begin{figure*}[t]
    \centering
    \setlength{\tabcolsep}{1pt} %

    \begin{tabular}{c c c | c c c c c c}
        \multicolumn{3}{c}{} & 
        \multicolumn{2}{c}{\footnotesize \editprompt{Remove the well}} & 
        \multicolumn{2}{c}{\footnotesize \editprompt{Make the tallest house smaller}} & 
        \multicolumn{2}{c}{\footnotesize \editprompt{Add a tree}} \\
        
        \noalign{\smallskip} %

        \includegraphics[width=0.12\textwidth, trim=4cm 4cm 4cm 4cm, clip]{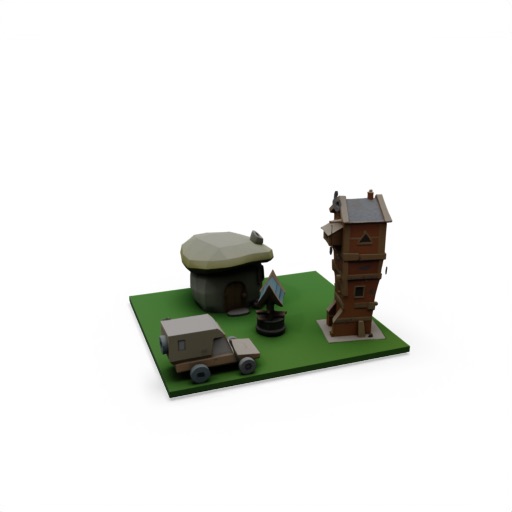} &
        \includegraphics[width=0.12\textwidth, trim=4cm 4cm 4cm 4cm, clip]{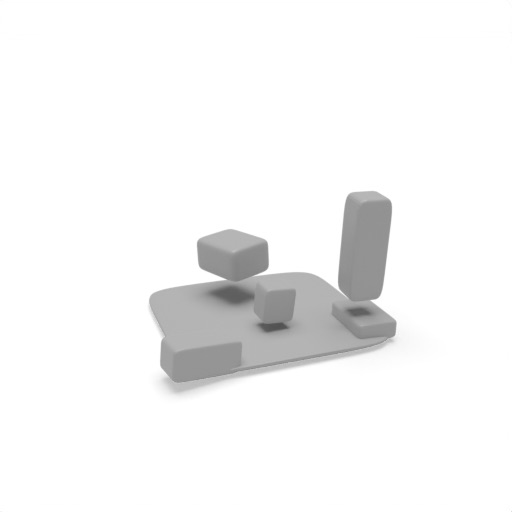} &
        &
        \includegraphics[width=0.12\textwidth, trim=4cm 4cm 4cm 4cm, clip]{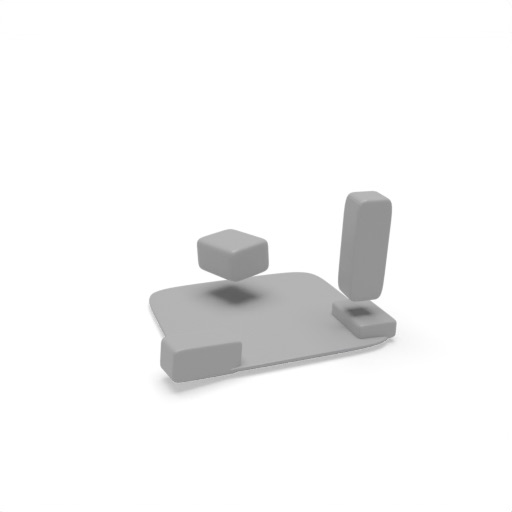} &
        \includegraphics[width=0.12\textwidth, trim=4cm 4cm 4cm 4cm, clip]{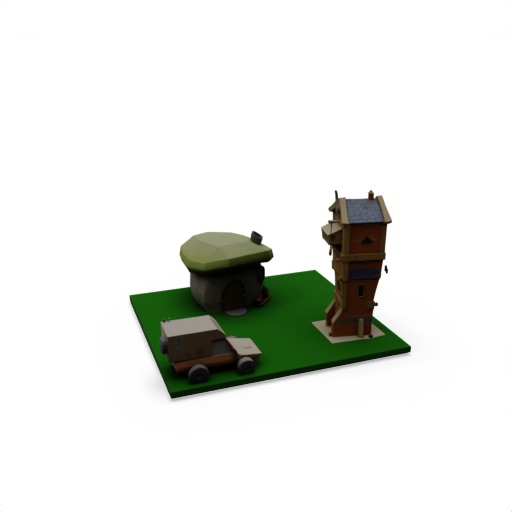} &
        \includegraphics[width=0.12\textwidth, trim=4cm 4cm 4cm 4cm, clip]{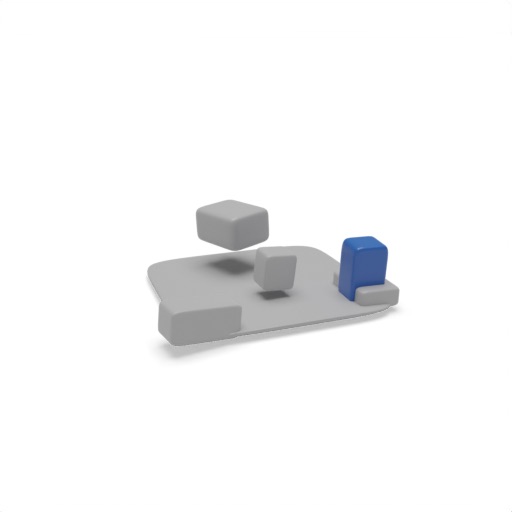} &
        \includegraphics[width=0.12\textwidth, trim=4cm 4cm 4cm 4cm, clip]{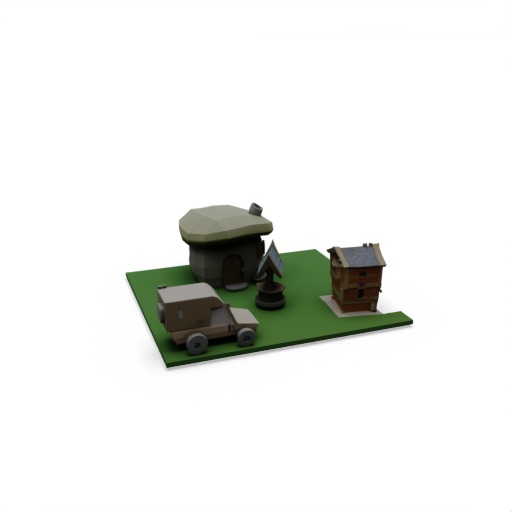} &
        \includegraphics[width=0.12\textwidth, trim=4cm 4cm 4cm 4cm, clip]{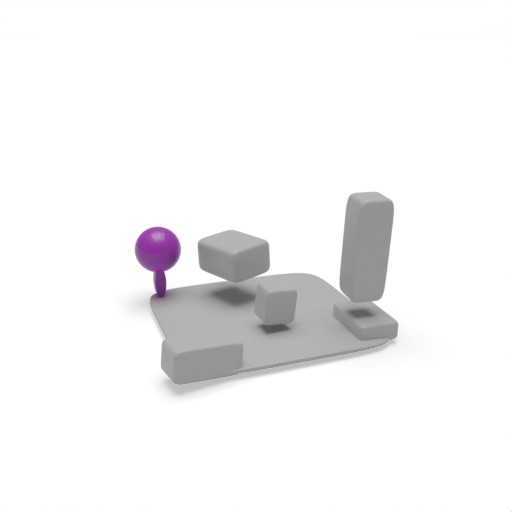} &
        \includegraphics[width=0.12\textwidth, trim=4cm 4cm 4cm 4cm, clip]{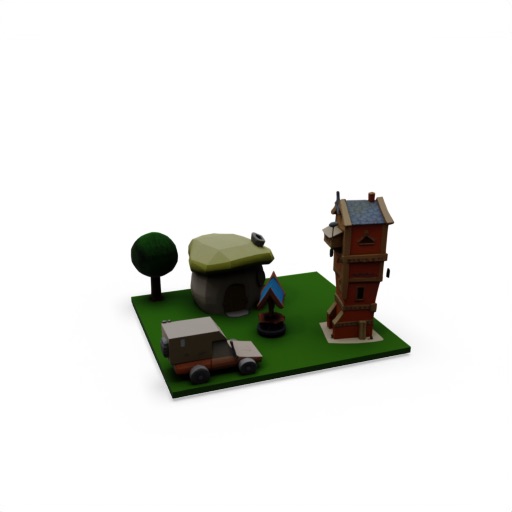} \\
        
        \footnotesize Input Scene &
        \footnotesize Original Proxy &
        &
        \footnotesize Edited Proxy & \footnotesize Ours &
        \footnotesize Edited Proxy & \footnotesize Ours &
        \footnotesize Edited Proxy & \footnotesize Ours \\
    \end{tabular}

    \vspace{-3pt}
    \caption{\new{\textbf{Scene editing examples.} To test our method's ability to edit scenes as opposed to objects exclusively, we composed a scene out of Edit3Dbench objects and performed various edits, including object removal (third and fourth columns), object modification (fifth and sixth columns) and new object generation (seventh and eight columns).}}
    \label{fig:scene}
\end{figure*}

\section{Additional Results and Discussions}
\label{sec:additional_results}
\subsection{User study}
\new{We compute the win rate of our method against multiple baselines based on votes from 26 participants. Each participant was shown the original rendered shapes, the edit instruction, and two generated results. Users were asked to select the output that better reflects each of these two criteria: edit quality and identity preservation. If both outputs are of equal quality, users may select `Cannot decide'. We evaluated on 80 samples, obtained by randomly sampling from the full set and manually filtering out ones with unclear instructions. These 80 samples were randomly split into two surveys of 40 samples each. The final win rate against a compared baseline is the proportion of votes where our method was preferred, averaged across the two survey splits. A screenshot of the study interface is shown in Figure \ref{fig:user_study_form}.}

\new{As shown in Table \ref{tab:vqascore}, our method achieves the highest win rates in both categories: edit quality and identity preservation. This further demonstrates the advantage of our method over all baselines, consistent with the quantitative results in the main paper.}

\begin{table}[t]
\caption{\new{\textbf{Runtime Comparison}. Average runtime per sample for baseline methods and the individual components of \methodName{}. All runtimes were measured on a single NVIDIA A100 80GB GPU.}}
\centering
\small
\setlength{\tabcolsep}{8pt}
\begin{tabular}{lc}
\toprule
\textbf{Method / Component} & \textbf{Runtime} \\
\midrule
ChangeIt3D & \textbf{2s} \\
BlendedPC & 48s \\
Spice-E & 32s \\
EditP23 & 1m 18s \\
VoxHammer & 9m 7s \\
TRELLIS (+ Flux Kontext) & 1m 27s \\
\midrule
\methodName{} - Proxy Editing (VLM) & 3m 28s \\
\methodName{} - Structure Inversion & 51s  \\
\methodName{} - SLAT Inversion & 4m 18s \\
\methodName{} - Structure Editing & 25s \\
\methodName{} - Appearance Refinement & 48s \\
\midrule
\textbf{\methodName{} - Total} & 10m 28s \\
\bottomrule
\end{tabular}
\label{tab:runtime}
\end{table}

\subsection{Method runtimes}

\new{
Table~\ref{tab:runtime} reports the runtime of our method and the compared baselines, measured on a single NVIDIA A100 80GB GPU. Overall, \methodName{} requires approximately 10m 28s per edit. As in VoxHammer, a substantial portion of the runtime is spent on 3D inversion and encoding steps, which together account for nearly half of the total runtime. In particular, SLAT inversion introduces a non-negligible overhead, but is only required for the appearance refinement stage and can be omitted in settings where only structural edits are desired or in consecutive editing scenarios. As shown in Table 2 of the main paper, our method remains competitive even without this component.}

\new{
While our method is slower than some highly specialized baselines, it is important to note that methods such as Spice-E, ChangeIt3D, and BlendedPC rely on dedicated training procedures that can take days and are often limited in their ability to generalize beyond the distributions they were trained on. In contrast, \methodName{} is entirely training-free and can be applied directly to arbitrary input shapes, offering a favorable trade-off between runtime, flexibility, and generalization.
}

\subsection{Scene editing}
\new{We additionally evaluated our method on simple scene-level editing by composing four Edit3D-Bench assets into a single scene and applying a range of edits (Figure~\ref{fig:scene}). These experiments suggest that our framework can already support meaningful scene manipulations, such as removing objects or modifying individual scene elements. At the same time, more fine-grained, part-level scene edits would likely require further modifications to the pipeline, such as first segmenting the scene into individual objects and applying SuperDec separately to each one. Moreover, our approach is currently limited by the voxel resolution of TRELLIS, making it less suitable for very large or highly detailed scenes without additional partitioning or hierarchical processing, which we leave for future work.}

\begin{figure}[t]
    \centering

    {\editprompt{"The lamp's shade is bulbous"}} \par\vspace{1pt}
    
     \includegraphics[width=0.24\linewidth, trim=3.3cm 3.3cm 3.3cm 3.3cm, clip]{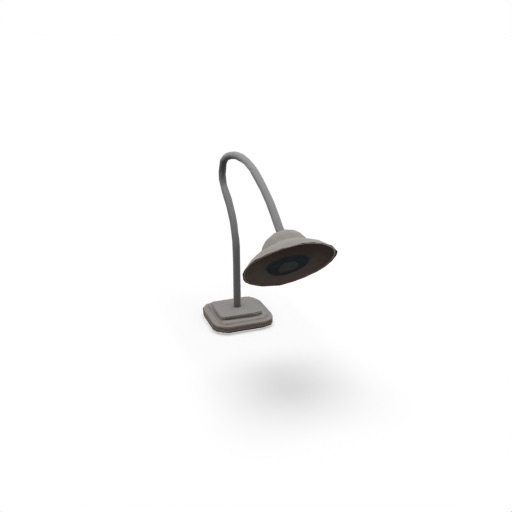}\hfill
    \includegraphics[width=0.24\linewidth, trim=3.3cm 3.3cm 3.3cm 3.3cm, clip]{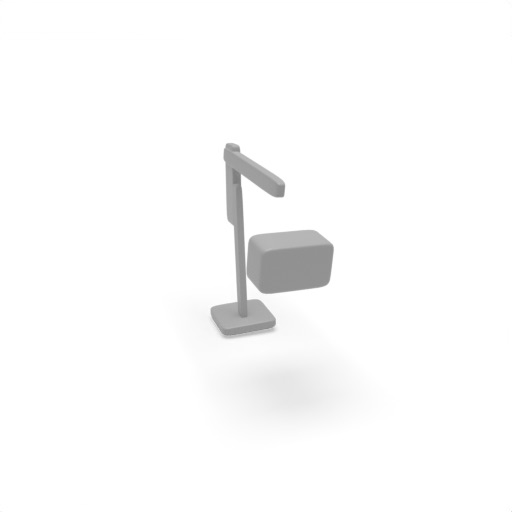}\hfill
    \includegraphics[width=0.24\linewidth, trim=3.3cm 3.3cm 3.3cm 3.3cm, clip]{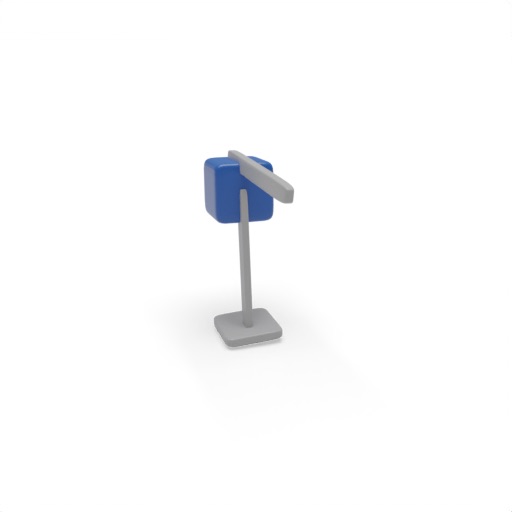}\hfill
    \includegraphics[width=0.24\linewidth, trim=3.3cm 3.3cm 3.3cm 3.3cm, clip]{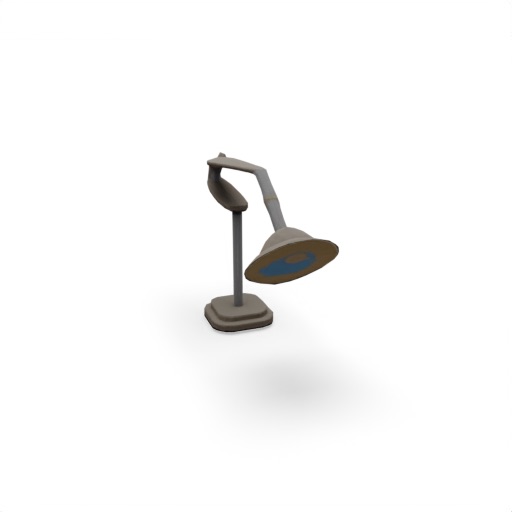}

    {\editprompt{"The lamp's shade is made from concentric thin rings"}} \par\vspace{1pt}
    
     \includegraphics[width=0.24\linewidth, trim=3cm 3cm 3cm 3cm, clip]{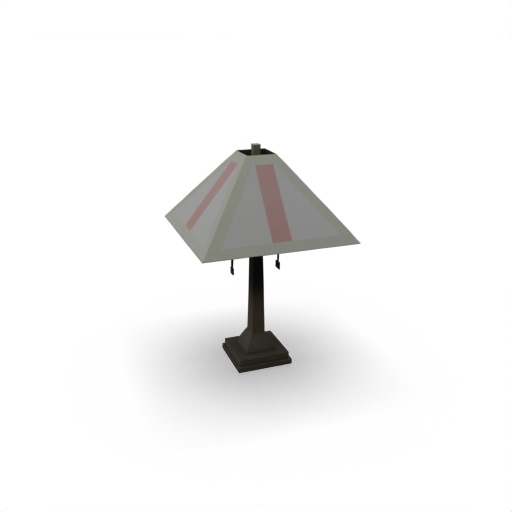}\hfill
    \includegraphics[width=0.24\linewidth, trim=3cm 3cm 3cm 3cm, clip]{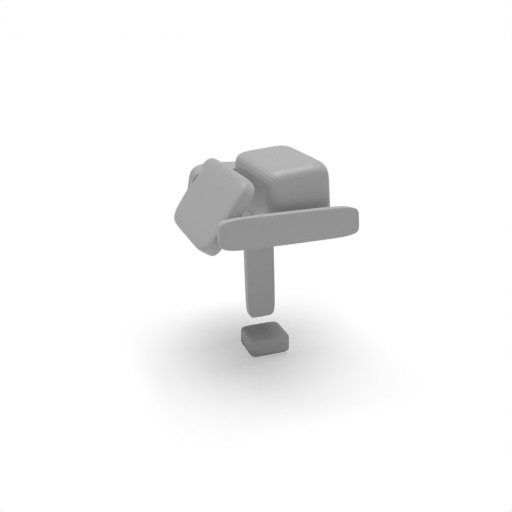}\hfill
    \includegraphics[width=0.24\linewidth, trim=3cm 3cm 3cm 3cm, clip]{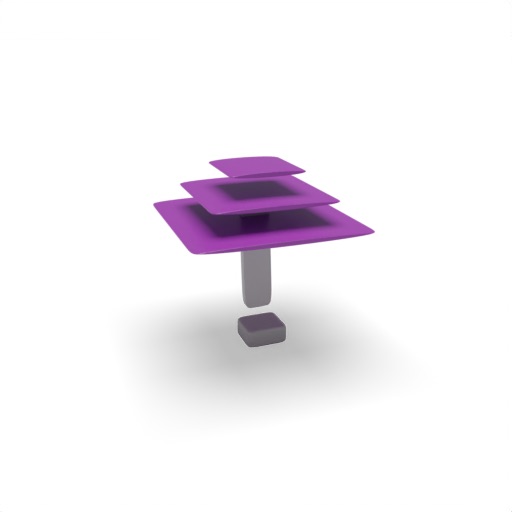}\hfill
    \includegraphics[width=0.24\linewidth, trim=3cm 3cm 3cm 3cm, clip]{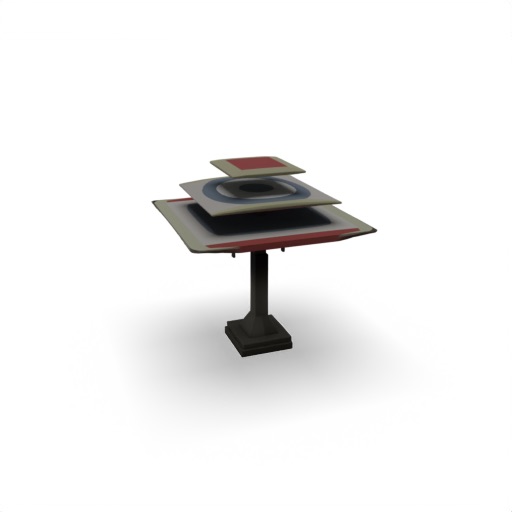}

    {\editprompt{"The chair sits closer to the ground"}} \par\vspace{1pt}
    
    \includegraphics[width=0.24\linewidth, trim=3cm 3cm 3cm 3cm, clip]{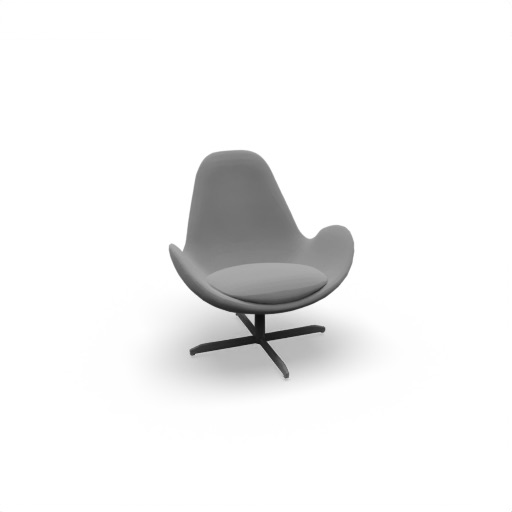}\hfill
    \includegraphics[width=0.24\linewidth, trim=3cm 3cm 3cm 3cm, clip]{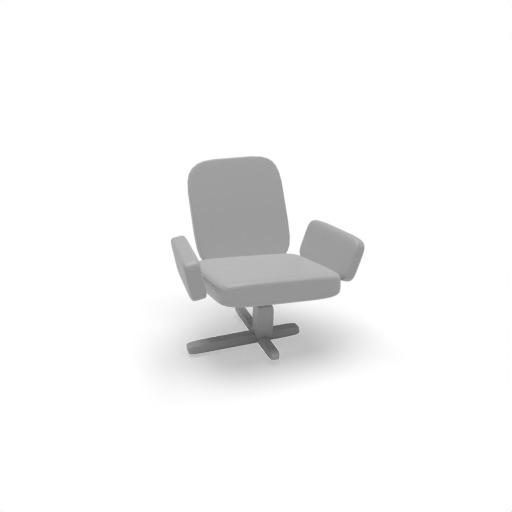}\hfill
    \includegraphics[width=0.24\linewidth, trim=3cm 3cm 3cm 3cm, clip]{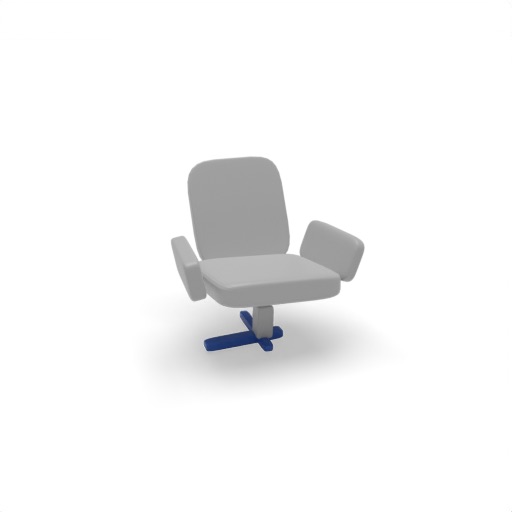}\hfill
    \includegraphics[width=0.24\linewidth, trim=3cm 3cm 3cm 3cm, clip]{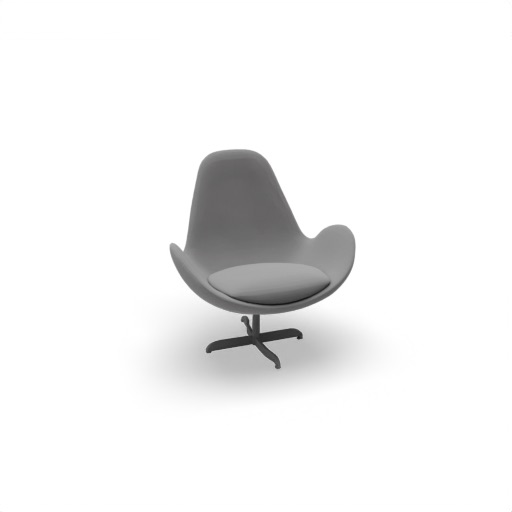}
    
    \makebox[0.24\linewidth][c]{\footnotesize Original Shape}\hfill
    \makebox[0.24\linewidth][c]{\footnotesize Original Proxy}\hfill
    \makebox[0.24\linewidth][c]{\footnotesize Edited Proxy}\hfill
    \makebox[0.24\linewidth][c]{\footnotesize Ours}
\vspace{-3pt}
    \caption{\new{\textbf{VLM / LLM failures.} We present failure cases in which the VLM failed to correctly edit the proxy according to the edit prompts (top and middle rows), and a failure example in which the LLM did not correctly parse the edit prompt (bottom row). In the latter case, the LLM converted the editing prompt ``The chair sits closer to the ground'' to ``a chair with shorter legs'', thereby causing the legs to shorten horizontally. Note that in the $90$ randomly sampled results included in the supplementary material (see \texttt{index.html} there were only two cases of VLM editing failures and no LLM parsing failures.
    }}
    \label{fig:vlm_failures}
\end{figure}

\subsection{Robustness of VLM/LLM based components}
\new{Systematically quantifying failures in our multi-stage pipeline is challenging, as the task lacks ground-truth intermediate representations such as proxy abstractions and edited proxies. To nevertheless assess robustness, we manually analyzed the 90 non-curated, randomly selected results included in the supplementary material (see index.html). Excluding minor artifacts, we identified 14 failures. Of these, only two cases (examples 23 and 84) were caused by incorrect proxy editing; for instance, in example 84, the VLM generated square elements instead of discs. We did not observe any failures caused by the LLM parsing stage. Overall, we found the LLM parsing to be robust, and when errors do occur, they are typically better characterized as misinterpretations rather than hallucinations. For example, the prompt ``the target sits closer to the ground'' was translated into ``a chair with shorter legs,'' which in that case led to shortening horizontal base supports rather than lowering the seat vertically. We present the two VLM editing failures (examples 23 and 84) as well as the LLM parsing error example we previously discussed in Figure \ref{fig:vlm_failures}.}

\clearpage

\end{document}